\documentclass[11pt, a4paper]{article}

\usepackage[a4paper, margin=2.5cm]{geometry}
\usepackage[utf8]{inputenc}
\usepackage[T1]{fontenc}
\usepackage[english]{babel}
\usepackage{lmodern}
\usepackage{microtype}

\usepackage{amsmath}
\usepackage{amssymb}
\usepackage{amsthm}
\usepackage{booktabs}
\usepackage{hyperref}

\hypersetup{
    colorlinks=true,
    linkcolor=blue,
    citecolor=blue,
    urlcolor=blue
}

\numberwithin{equation}{section}
\newtheorem{theorem}{Theorem}[section]
\newtheorem{lemma}[theorem]{Lemma}
\newtheorem{proposition}[theorem]{Proposition}
\newtheorem{corollary}[theorem]{Corollary}
\theoremstyle{definition}
\newtheorem{definition}[theorem]{Definition}

\newtheorem{remark}[theorem]{Remark}

\begin{document}

\begin{center}
{\LARGE \textbf{Algebra-Valued Yang-Mills-Type Theories: Polynomial Obstructions and Holonomy Reductions}}\\[0.6cm]
{\large Yuri Ximenes Lecca$^{a,}$\footnote{Formerly publishing as Yuri Ximenes Martins.} \quad and \quad Willian Goulart Gomes Velasco$^{b}$}\\[0.25cm]
{\small $^{a}$\itshape Lemma Group, Brazil}\\[0.1cm]
{\small $^{b}$\itshape UTFPR, Curitiba, Brazil}\\
{\small $^{a}$\texttt{ximenesyuri@gmx.ie}}\\
{\small $^{b}$\texttt{willianvelasco@utfpr.edu.br}}
\end{center}

\vspace{0.3cm}
\hrule
\vspace{0.3cm}
\noindent\textbf{Abstract}\\[0.15cm]
{\small We study algebra-valued Yang-Mills-type ($A$-YMT) theories over a smooth $n$-manifold $M$, where connections take values in a general (not necessarily associative, commutative, or Lie) $\mathfrak{m}$-graded algebra $(A, *)$ and descend to an associated algebra bundle $E_A = P \times_\rho A \to M$ over a principal $G$-bundle $\pi : P \to M$. We first derive the variational field equations, infinitesimal gauge transformations, and generalized Bianchi identities, showing that affine Ehresmann compatibility forces the vertical image of $\mathfrak{g}$ into the Lie nucleus $\mathcal{N}_{\mathrm{Lie}}(A)$ of the commutator $[x,y]_* = x*y - y*x$, and proving that non-linear gauge covariance and the differential Bianchi identity are obstructed precisely by the Jacobiator $\mathrm{Jac}_{[\cdot,\cdot]_*}$ (leaving principal $G$-gauge transformations universally anomaly-free). We then establish polynomial and geometric obstructions in three settings. First, we prove a vanishing theorem for internal-product Chern-Weil and self-dual $A$-YMT functionals in weak $(2,s)$-solvable algebras, invariant under $(Q,S)$-splitting extensions. Second, in reductive Cartan-Yang-Mills theories valued in $A = \mathbb{R}^k \oplus \mathfrak{h}$, we compare quartic-block against recursive binary-tree parenthesization, proving a strict dimensional cutoff at $n \ge 8$ for quartic blocks and an exact Jacobsthal-number cutoff ($n - 4 \notin \bigcup_{r \ge 3}\{2 J_r, J_{r+1}\}$, plus coframe-rank vanishing for $n \ge k + 5$) for binary trees. We also catalog the admissible Berger holonomy reductions for $4 \le n \le 7$, establishing an on-shell vanishing (torsion-free) versus off-shell survival (torsionful) dichotomy both in 4D for $U(2)$ and $SU(2)$ connections---via $e \curlywedge_\rtimes (e \curlywedge_\rtimes \Omega_\omega) = \frac{1}{2}(e \curlywedge_\rtimes e) \curlywedge_\rtimes \Omega_\omega$ and the first Bianchi identity---and in 7D for $G_2$-connections. Third, we prove an associator-quotient collapse and Jacobiator vertex-dichotomy theorem for odd-dimensional internal $A$-Chern-Simons theories, with applications to $A$-$BF$ theories, octonionic and reductive Yang-Mills-Higgs Proca mass spectra, and reductive dualities.}

\vspace{0.3cm}
\hrule
\vspace{0.5cm}

\section{Introduction}

Classical Yang-Mills (YM) theory is formulated on a principal $G$-bundle $\pi : P \to M$ over an orientable $n$-dimensional semi-Riemannian manifold $(M, g)$, where $G$ is a compact semi-simple Lie group with Lie algebra $\mathfrak{g}$ \cite{KobayashiNomizu, DonaldsonKronheimer}. Given a connection 1-form $D \in \mathrm{Conn}(P; \mathfrak{g}) \subset \Omega^1(P; \mathfrak{g})$, its curvature (or field strength) is the horizontal, $G$-equivariant $\mathfrak{g}$-valued 2-form $F_D = dD + \frac{1}{2} D [\wedge]_{\mathfrak{g}} D \in \Omega^2_{\mathrm{heq}}(P; \mathfrak{g})$, where $[\wedge]_{\mathfrak{g}}$ denotes the exterior product induced by the Lie bracket of $\mathfrak{g}$. Using the canonical Weil isomorphism $j : \Omega^2_{\mathrm{heq}}(P; \mathfrak{g}) \xrightarrow{\sim} \Omega^2(M; E_{\mathfrak{g}})$ onto 2-forms with values in the adjoint bundle $E_{\mathfrak{g}} = P \times_{\mathrm{Ad}} \mathfrak{g}$, the classical Yang-Mills action functional is given by
\begin{equation}\label{eq:classical_YM}
S_{\mathrm{YM}}[D] = \int_M \langle F_D, F_D \rangle_{\mathfrak{g}}\, \mathrm{dvol}_g = \int_M \mathrm{tr}\left( j(F_D) \wedge *_g j(F_D) \right),
\end{equation}
where $*_g$ is the Hodge star operator of $g$ and $\mathrm{tr}$ represents the Killing form $B_{\mathfrak{g}}$ of $\mathfrak{g}$.

While physically foundational, the classical action \eqref{eq:classical_YM} relies on three structural assumptions:
\begin{enumerate}
    \item The base manifold $M$ must admit a semi-Riemannian metric $g$ of prescribed signature (which is subject to topological obstructions, such as $\chi(M) = 0$ for compact Lorentzian manifolds \cite{HawkingEllis, Beem});
    \item The internal symmetry algebra must be a Lie algebra $\mathfrak{g}$ admitting a non-degenerate $\mathrm{Ad}$-invariant quadratic form (typically requiring $G$ to be semi-simple or compact, or belong to specific classes of quadratic solvable Lie algebras \cite{Ovando});
    \item The gauge potential $D$ takes values in a Lie algebra rather than a more general (possibly non-associative or $\mathfrak{m}$-graded) algebra $(A, *)$ modeling internal, worldsheet, supergeometric, Malcev, or reductive Cartan symmetries \cite{BaezMuniain, Wise, Quillen, Coimbra}.
\end{enumerate}

In the context of first-order tetradic gravity, it was shown in \cite{MartinsBiezunerEHP} that the Einstein-Hilbert-Palatini (EHP) action functional can be detached from Lorentzian Cartan geometry and formulated algebraically over an arbitrary $\mathfrak{m}$-graded algebra $(A, *)$ decomposed as $A \simeq A_0 \oplus A_1$. Doing so revealed that the non-triviality of the $A$-valued EHP functional is governed by the graded polynomial identities of $(A, *)$: specifically, if $A_0$ is weak $(k,s)$-solvable, the cosmological constant decouples for $n \ge k + s + 1$ and the entire $A$-EHP functional vanishes identically for $n \ge k + s + 3$ \cite{MartinsBiezunerEHP}.

The purpose of the present paper is to formulate \emph{algebra-valued Yang-Mills-type ($A$-YMT) theories} directly in a concrete differential-geometric and algebraic setting (without requiring category-theoretic functorial bundle systems), establish their variational and gauge-theoretic foundations, and determine the algebraic, parenthesization, and holonomy obstructions that govern them. Given a principal $G$-bundle $\pi : P \to M$ and a representation $\rho : G \to \mathrm{Aut}_{\mathrm{Alg}}(A)$ of $G$ by algebra automorphisms of an $\mathfrak{m}$-graded algebra $(A, *)$, horizontal $\rho$-equivariant $A$-valued differential forms $\Omega^k_{\mathrm{heq},\rho}(P; A)$ descend via the canonical Weil algebra isomorphism $j : \Omega^\bullet_{\mathrm{heq},\rho}(P; A) \xrightarrow{\sim} \Omega^\bullet(M; E_A)$ to forms valued in the associated algebra bundle $E_A = P \times_\rho A \to M$. We simultaneously decouple the quadratic curvature pairing from both the metric Hodge star $*_g$ and the Killing form $B_{\mathfrak{g}}$ by allowing a general $\mathbb{R}$-bilinear or $C^\infty(M)$-bilinear pairing $\langle \cdot, \cdot \rangle_A$ on $\Omega^2(M; E_A)$.

A fundamental structural contrast immediately emerges between $A$-EHP gravity and $A$-YMT gauge theory:
\begin{itemize}
    \item In $A$-EHP theory, the Hilbert-Palatini $n$-form $\alpha_{n,\Lambda} = \wedge_*^{n-2} e \wedge_* \Omega + \frac{\Lambda}{(n-1)!} \wedge_*^n e$ contains a number of factors of $e$ that grows linearly with $\dim M = n$, requiring either quotienting by associativity-like polynomial ideals or fixing a parenthesization scheme whenever $(A, *)$ is non-associative, and converting $(k,s)$-solvability directly into upper bounds on the spacetime dimension $n$ \cite{MartinsBiezunerEHP}.
    \item In pure $A$-YMT theory, the curvature 2-form $F_D = dD + \frac{1}{2} D \wedge_* D = dD + \frac{1}{4} D \wedge_{[\cdot,\cdot]_*} D$ involves only a single binary product $D \wedge_* D$, which automatically projects onto the skew-symmetric commutator bracket $[x,y]_* = x*y - y*x$ while annihilating the symmetric Jordan product $x \circ_* y = \frac{1}{2}(x*y + y*x)$. Consequently, $F_D$ is unambiguously defined for \emph{any} non-associative algebra $(A, *)$ without choosing parenthesizations or quotienting by associativity ideals, and the action $S^A[D] = \int_M \langle j(F_D), j(F_D) \rangle_A$ is universally quartic in $D$ across all spacetime dimensions $n \ge 2$. However, whenever higher curvature powers $\wedge_*^r F_D$, constitutive soldering powers $\wedge_*^{n-4} e$, or differential identities such as the Bianchi identity $dF_D + \frac{1}{2}[D, F_D]_*$ and Chern-Simons transgression are formed, the symmetric product $\circ_*$, the parenthesization scheme, the associator $\mathrm{Assoc}_*$, and the commutator Jacobiator $\mathrm{Jac}_{[\cdot,\cdot]_*}$ play decisive roles.
\end{itemize}

Specifically, the main results of this paper are organized as follows:
\begin{enumerate}
    \item \textbf{Ehresmann Admissibility, the Lie Nucleus Constraint, Variational Field Equations, and the Jacobiator Bianchi/Gauge Obstructions (Proposition \ref{prop:Cartan_Ehresmann} and Theorem \ref{thm:Bianchi_gauge_EL}):} On a principal $G$-bundle $\pi : P \to M$ with automorphism representation $\rho : G \to \mathrm{Aut}_{\mathrm{Alg}}(A)$, we prove that an affine Ehresmann connection $D \in \Omega^1_{\mathrm{eq},\rho}(P; A)$ with vertical map $D(\underline{X}) = \eta(X)$ has horizontal curvature $F_D \in \Omega^2_{\mathrm{heq},\rho}(P; A)$ if and only if $d\rho(X)(y) = \frac{1}{2}[\eta(X), y]_*$, which in turn forces the vertical image $\eta(\mathfrak{g})$ to lie in the \emph{Lie nucleus} $\mathcal{N}_{\mathrm{Lie}}(A)$ of $(A, [\cdot,\cdot]_*)$. We then derive the Euler-Lagrange equations of a general $A$-YMT theory and prove that under an infinitesimal gauge transformation $\delta_\epsilon D = d\epsilon + \frac{1}{2} D \wedge_{[\cdot,\cdot]_*} \epsilon$, the failure of homogeneous curvature covariance $\delta_\epsilon F_D - \frac{1}{2} F_D \wedge_{[\cdot,\cdot]_*} \epsilon$ and the failure of the differential Bianchi identity $dF_D + \frac{1}{2} D \wedge_{[\cdot,\cdot]_*} F_D$ are governed purely by the horizontal Jacobiator 2-form $-\frac{1}{8}\mathrm{Jac}_{[\cdot,\cdot]_*}(D, D, \epsilon)$ and 3-form $-\frac{1}{24}\mathrm{Jac}_{[\cdot,\cdot]_*}(D, D, D)$. In particular, principal $G$-gauge transformations $\epsilon \in \Omega^0_{\mathrm{heq},\rho}(P; \eta(\mathfrak{g})) \subset \Omega^0_{\mathrm{heq},\rho}(P; \mathcal{N}_{\mathrm{Lie}}(A))$ are universally anomaly-free for any algebra $(A, *)$, whereas extended $A$-gauge transformations $\epsilon \in \Omega^0_{\mathrm{heq},\rho}(P; A)$ outside $\mathcal{N}_{\mathrm{Lie}}(A)$ are obstructed by $\mathrm{Jac}_{[\cdot,\cdot]_*}$. Using a rigorous pointwise localization argument relative to the basepoint connection $D_0 = \eta \circ \omega_0$, we also establish the exact converse when $\langle \cdot, \cdot \rangle_{A,\mathrm{sym}}$ is $C^\infty(M)$-bilinear (tensorial), pointwise $\mathrm{ad}_{[\cdot,\cdot]_*}$-invariant, and non-degenerate.
    \item \textbf{Internal-Product Topological and Self-Dual $A$-YMT Obstructions (Theorem \ref{thm:topological_obstruction} and Proposition \ref{prop:splitting_invariance}):} After recording the commutator projection, abelianization of weak $(1,1)$-solvable algebras, and derived-filtration localization (Proposition \ref{prop:abelianization}), we study internal-trace topological and self-dual pairings constructed purely from the internal multiplication $\wedge_*$ and a bundle linear functional $\mathrm{tr}_\xi : E_A \to M \times \mathbb{R}$ (as opposed to an external symmetric bilinear invariant form $\kappa \in \mathrm{inv}^2(\mathfrak{g})$). We prove that weak $(2,s)$-solvability forces the $r$-th degree internal Chern-Weil action $\int_M \mathrm{tr}_\xi(j(\wedge_*^r F_D))$ in dimension $n = 2r$ to vanish for $n \ge 2(s+1)$, and we show that these obstructions are preserved under $(Q,S)$-splitting extensions. For $s = 1$, this internal-product vanishing encompasses all skew-commutative algebras (including Lie algebras under $[\wedge]_{\mathfrak{g}}$, Berger holonomy algebras, Hitchin's generalized complex geometry, and Cayley-Dickson unitary algebras $\mathfrak{u}(k; \mathrm{CD}^l(A))$ under the skew-symmetric matrix product $\curlywedge$).
    \item \textbf{Parenthesization Schemes, 4D First-Bianchi Contractions, and Berger Holonomy Reductions in Cartan-Yang-Mills / MacDowell-Mansouri Theories (Lemma \ref{lem:skew_comm_powers}, Lemma \ref{lem:4D_Bianchi_dichotomy}, Theorem \ref{thm:Berger_CYM}, and Proposition \ref{prop:G2_survival}):} For reductive connections $\nabla = e + \omega$ valued in $A = \mathbb{R}^k \oplus \mathfrak{h}$ with $\mathfrak{h} \subset \mathfrak{so}(k_1) \oplus \dots \oplus \mathfrak{so}(k_r)$, we formulate the internal product $\curlywedge_\rtimes$ via an $H$-equivariant soldering map $\sigma : \mathbb{R}^k \wedge \mathbb{R}^k \to \mathfrak{h}$ (including the $G_2$-equivariant vector cross product on $\mathbb{R}^7$ when $n=k=7$). We distinguish between two non-associative parenthesization schemes for exterior powers of $e$: (a) \emph{quartic-block parenthesization} $\wedge_{*, \mathrm{blk}}^m e$ (which factors out $\wedge_{*, \mathrm{bal}}^4 e = (e \wedge_* e) \wedge_* (e \wedge_* e) = 0$ and vanishes for all $m \ge 4$), and (b) \emph{recursive binary-tree halving} $\wedge_{*, \mathrm{bal}}^m e$ (which we prove vanishes algebraically for all $m \ge 4$ outside the consecutive Jacobsthal pairs $m \in \bigcup_{r \ge 3}\{2 J_r, J_{r+1}\} = \{5,6\} \cup \{10,11\} \cup \{21,22\} \cup \{42,43\} \cup \dots$ where $J_r = \frac{2^r - (-1)^r}{3}$, and vanishes by coframe rank for all $m > k$, so for $k \le 9$ it vanishes for all $m \ge 7$). Under quartic-block parenthesization, the cosmological term vanishes for $n \ge 4$, the EHP term vanishes for $n \ge 6$, and the quadratic Yang-Mills term survives up to $n = 7$ before vanishing for $n \ge 8$ (while under recursive binary halving the YM term vanishes algebraically for all $n \ge 8$ with $n - 4 \notin \bigcup_{r \ge 3}\{2 J_r, J_{r+1}\}$ and by coframe rank for all $n \ge k + 5$, so for $k \le 9$ it vanishes at $n = 8, 9$ and for all $n \ge 11$). Furthermore, at $n = k = 4$, we prove an exact proportionality and on-shell/off-shell dichotomy for $U(2)$ and $SU(2)$ connections (Lemma \ref{lem:4D_Bianchi_dichotomy}): for any $\mathfrak{h} \subset \mathfrak{so}(k)$ with $\times_H = 0$, the right-nested EHP 4-form satisfies $e \curlywedge_\rtimes (e \curlywedge_\rtimes \Omega_\omega) = (0, -\frac{1}{2}\sigma(e \wedge (\Omega_\omega \cdot e))) = \frac{1}{2}(e \curlywedge_\rtimes e) \curlywedge_\rtimes \Omega_\omega = (0, \frac{1}{4}[\sigma(e\wedge e), \Omega_\omega])$; at $n = k = 4$, because the anti-self-dual basis 2-forms $\omega_a^- \in \Lambda^2_- \mathbb{R}^4$ satisfy $\omega_a^- \wedge \omega_b^- = -2\delta_{ab}\mathrm{vol}_4$ and the algebraic first Bianchi identity forces the curvature matrix $R_{ab} = R_{ba}$ on $\mathfrak{su}(2)_-$ to be symmetric, \emph{both} $(e \curlywedge_\rtimes e) \curlywedge_\rtimes \Omega_\omega$ and $e \curlywedge_\rtimes (e \curlywedge_\rtimes \Omega_\omega)$ vanish identically on-shell for every torsion-free $U(2)$ and $SU(2)$ connection on K\"ahler and $K3$ surfaces while remaining non-zero off-shell when torsion $\Theta_{e,\omega} \neq 0$ is present (whereas in dimension $n = k = 6$, the pure quadratic YM term $\Omega_\omega \curlywedge_\rtimes ((e \curlywedge_\rtimes e) \curlywedge_\rtimes \Omega_\omega)$ is non-vanishing on-shell for torsion-free $U(3)$ and $SU(3)$ connections!). At the critical dimension $n = k = 7$, we prove a complete representation-theoretic structure proposition (Proposition \ref{prop:G2_survival}) showing that for $H = G_2 \subset SO(7)$, Schur's lemma forces the $\mathfrak{g}_2$-component $B_e^{(3)} = \sigma((e \wedge_{\times_{G_2}} e) \wedge e) \in \mathrm{Hom}_{G_2}(\Lambda^3 \mathbb{R}^7, \Lambda^2_{14})$ of $\curlywedge_{\rtimes, \mathrm{bal}}^3 e$ to vanish identically, while the $\mathbb{R}^7$-component $(v^{(3)})_a = -7\, i_{\partial_a}(*\varphi_0) \neq 0$ satisfies $(\Omega_\omega \cdot v^{(3)})_b = 7\sum_{c,d=1}^7 R_{bc,cd}\, e^d \wedge *\varphi_0$ whenever $\Omega_\omega \in \Lambda^2_{14} \otimes \mathfrak{g}_2$ by the identity $\Lambda^2_{14} \wedge *\varphi_0 = 0$; consequently, just as at $n = k = 4$, $\mathcal{L}_7^{\mathrm{CYM}}$ vanishes on-shell on the torsion-free $G_2$-curvature space $\mathcal{W}_{77} \subset S^2(\Lambda^2_{14})$ (where the first Bianchi identity and $G_2$-holonomy force $\mathrm{Ric}_{bd} = \sum_{c=1}^7 R_{bc,cd} = 0$) and is non-vanishing off-shell on $\Omega^2(P; \mathfrak{g}_2)$ whenever torsion $\Theta_{e,\omega} \neq 0$ excites the $\Lambda^2_7 \otimes \mathfrak{g}_2$ or non-Bianchi curvature sectors.
    \item \textbf{Associator-Quotient Collapse and Jacobiator Dichotomy in Internal Chern-Simons Theories, with Applications to $BF$, Higgs, and Dual Theories (Theorem \ref{thm:CS_obstructions} and Propositions \ref{prop:BF_obstructions}--\ref{prop:dual_obstructions}):} We extend our framework to odd-dimensional internal $A$-valued Chern-Simons ($A$-CS) theories on linear connection spaces. We prove a three-way structural dichotomy for skew-commutative algebras $(A, \curlywedge)$: (a) in the associative quotient, all $\ge 3$-fold products vanish identically; (b) in the unquotiented grouped formulation, if $(A, \curlywedge)$ is a Lie algebra ($\mathrm{Jac}_\curlywedge = 0$), the Jacobi identity forces $(D \curlywedge D) \curlywedge D = \frac{1}{3}\mathrm{Jac}_\curlywedge(D,D,D) = 0$, annihilating the 3D cubic vertex and killing all $\ge 5$D internal $A$-CS theories under \emph{both} quartic-block and recursive binary-tree parenthesizations; and (c) if $(A, \curlywedge)$ is a non-Lie skew-commutative algebra ($\mathrm{Jac}_\curlywedge \neq 0$), the 3D cubic and 5D quartic Jacobiator vertices survive under quartic-block grouping (with an additional 5D quintic Jacobsthal vertex under recursive binary halving), while for $n \ge 7$ all block-grouped $A$-CS theories vanish identically and binary-tree $A$-CS theories vanish outside the sparse odd Jacobsthal pairs $n \in \bigcup_{m \ge 4}\{J_m, J_m + 2\} \cap [7, \infty) = \{7\} \cup \{11, 13\} \cup \{21, 23\} \cup \dots$ (and for $n \ge \dim A + 3$ by rank). We conclude with applications to Cattaneo-Plebanski $A$-$BF$ equivalences, explicit octonionic and reductive Proca mass spectra in the $A$-valued Yang-Mills-Higgs mechanism, and reductive dualities.
\end{enumerate}

\section{Geometric-Algebraic Foundations of \texorpdfstring{$A$}{A}-YMT Theories}\label{sec:foundations}

\subsection{Non-Associative Algebras, Parenthesization, and \texorpdfstring{$(k,s)$}{(k,s)}-Solvability}

Let $R$ be a commutative unital ring of characteristic zero (throughout, $R = \mathbb{R}$ or $\mathbb{C}$) and let $(A, *)$ be an $R$-algebra, i.e., an $R$-module $A$ equipped with an $R$-bilinear multiplication $* : A \times A \to A$, without assuming associativity, commutativity, or a unit element \cite{SchaferBook}. Every multiplication $*$ decomposes canonically into its skew-symmetric commutator bracket and its symmetric Jordan product:
\begin{equation}\label{eq:comm_jordan_decomp}
x * y = \frac{1}{2}[x, y]_* + x \circ_* y, \quad \text{where} \quad [x, y]_* := x * y - y * x, \quad x \circ_* y := \frac{1}{2}(x * y + y * x).
\end{equation}
We denote the \emph{associator} of $(A, *)$ and the \emph{Jacobiator} of $(A, [\cdot,\cdot]_*)$ for $x, y, z \in A$ by
\begin{equation}\label{eq:assoc_jac_def}
\mathrm{Assoc}_*(x, y, z) := (x * y) * z - x * (y * z), \quad \mathrm{Jac}_{[\cdot,\cdot]_*}(x, y, z) := [[x, y]_*, z]_* + [[y, z]_*, x]_* + [[z, x]_*, y]_*.
\end{equation}
Thus $(A, [\cdot,\cdot]_*)$ is a Lie algebra if and only if $\mathrm{Jac}_{[\cdot,\cdot]_*}(x, y, z) = 0$ for all $x, y, z \in A$ (in which case $(A, *)$ is called \emph{Lie-admissible} \cite{SchaferBook}; every associative algebra is Lie-admissible since $\mathrm{Jac}_{[\cdot,\cdot]_*}(x,y,z) = \sum_{\sigma \in S_3} \mathrm{sign}(\sigma)\, \mathrm{Assoc}_*(x_{\sigma(1)}, x_{\sigma(2)}, x_{\sigma(3)})$). More generally, the \emph{Lie nucleus} (or \emph{Jacobi nucleus}) of $(A, [\cdot,\cdot]_*)$ is the Lie subalgebra
\begin{equation}\label{eq:Lie_nucleus}
\mathcal{N}_{\mathrm{Lie}}(A) := \left\{ x \in A \;\middle|\; \mathrm{Jac}_{[\cdot,\cdot]_*}(x, y, z) = 0 \text{ for all } y, z \in A \right\}.
\end{equation}
By definition of $\mathrm{Jac}_{[\cdot,\cdot]_*}$, the inner adjoint map $\mathrm{ad}^{[\cdot,\cdot]_*}_x(y) := [x, y]_*$ is a derivation of the commutator bracket $[y, z]_*$ if and only if $x \in \mathcal{N}_{\mathrm{Lie}}(A)$.

Given a smooth manifold $P$, let $\Omega(P; A) = \bigoplus_{k \ge 0} \Omega^k(P; A)$ denote the $\mathbb{N}$-graded $R$-module of smooth $A$-valued differential forms on $P$, where $\Omega^k(P; A) = \Gamma(\Lambda^k T^*P \otimes A)$. The bilinear multiplication $* : A \otimes A \to A$ induces a canonical exterior product
\begin{equation}\label{eq:wedge_star}
\wedge_* : \Omega^k(P; A) \times \Omega^l(P; A) \longrightarrow \Omega^{k+l}(P; A),
\end{equation}
defined pointwise at $p \in P$ on tangent vectors $v_1, \dots, v_{k+l} \in T_p P$ by
\begin{equation}\label{eq:wedge_star_pointwise}
(\alpha \wedge_* \beta)_p(v_1, \dots, v_{k+l}) = \frac{1}{k!\, l!} \sum_{\sigma \in S_{k+l}} \mathrm{sign}(\sigma)\, \alpha_p(v_{\sigma(1)}, \dots, v_{\sigma(k)}) * \beta_p(v_{\sigma(k+1)}, \dots, v_{\sigma(k+l)}).
\end{equation}
In terms of the decomposition \eqref{eq:comm_jordan_decomp}, for any $\alpha \in \Omega^k(P; A)$ and $\beta \in \Omega^l(P; A)$ we have
\begin{equation}\label{eq:graded_comm_symmetry}
\alpha \wedge_{[\cdot,\cdot]_*} \beta = \alpha \wedge_* \beta - (-1)^{kl} \beta \wedge_* \alpha, \qquad \alpha \wedge_{\circ_*} \beta = \frac{1}{2}\big(\alpha \wedge_* \beta + (-1)^{kl} \beta \wedge_* \alpha\big).
\end{equation}
Setting $\alpha = \beta \in \Omega^k(P; A)$ in \eqref{eq:graded_comm_symmetry} yields the basic parity dichotomy of exterior squares:
\begin{itemize}
    \item \textbf{Odd-Degree Forms ($k$ odd, e.g., connection 1-forms $D \in \Omega^1(P; A)$):} $(-1)^{k^2} = -1$, so $\alpha \wedge_{\circ_*} \alpha = 0$ identically and
    \begin{equation}\label{eq:odd_form_comm_projection}
    \alpha \wedge_* \alpha = \frac{1}{2} \alpha \wedge_{[\cdot,\cdot]_*} \alpha \quad (\alpha \in \Omega^{\mathrm{odd}}(P; A)).
    \end{equation}
    Hence the exterior square of any odd-degree form depends \emph{solely} on the commutator bracket $[x,y]_*$ and vanishes identically for all odd forms (when $\dim P \ge k$) if and only if $(A, *)$ is commutative ($[x,y]_* = 0$).
    \item \textbf{Even-Degree Forms ($k$ even, e.g., curvature 2-forms $F_D \in \Omega^2(P; A)$):} $(-1)^{k^2} = +1$, so $\alpha \wedge_{[\cdot,\cdot]_*} \alpha = 0$ identically and
    \begin{equation}\label{eq:even_form_jordan_projection}
    \alpha \wedge_* \alpha = \alpha \wedge_{\circ_*} \alpha \quad (\alpha \in \Omega^{\mathrm{even}}(P; A)).
    \end{equation}
    Hence the exterior square of any even-degree form depends \emph{solely} on the symmetric Jordan product $x \circ_* y$ and vanishes identically whenever $(A, *)$ is skew-commutative ($x \circ_* y = 0$).
\end{itemize}

Notice that when passing from algebra elements $x_i \in A$ to differential forms $\alpha_i \in \Omega^{k_i}(P; A)$, permuting odd-degree forms introduces Koszul signs $(-1)^{k_i k_j}$ that twist symmetric algebra identities into alternating ones (and vice versa). Thus, whereas a non-permuting multilinear identity $p(x_1, \dots, x_m) = 0$ (such as a parenthesized nilpotency identity $x_1 * \dots * x_m = 0$) or an identity evaluated on even-degree forms lifts directly by multilinearity \cite{Kaplansky, Drensky}, polynomial nilpotency conditions on $k$-forms must be defined directly on $\Omega^k(P; A)$ \cite{MartinsBiezunerEHP}.

When $(A, *)$ is non-associative, products of $m \ge 3$ forms depend essentially on the choice of parenthesization. To avoid any ambiguity in higher-degree powers $\wedge_*^m \alpha$, we distinguish three explicit parenthesization conventions for an $A$-valued form $\alpha \in \Omega^k(P; A)$:
\begin{enumerate}
    \item \textbf{Full parenthesization:} All non-associative monomial parenthesizations $p_m(\alpha, \dots, \alpha)$ of degree $m$.
    \item \textbf{Recursive binary-tree parenthesization ($\wedge_{*, \mathrm{bal}}^m \alpha$):} Defined inductively for $m \ge 1$ by $\wedge_{*, \mathrm{bal}}^1 \alpha := \alpha$ and by exact binary halving
    \begin{equation}\label{eq:balanced_power}
    \wedge_{*, \mathrm{bal}}^m \alpha := \big(\wedge_{*, \mathrm{bal}}^{\lceil m/2 \rceil} \alpha\big) \wedge_* \big(\wedge_{*, \mathrm{bal}}^{\lfloor m/2 \rfloor} \alpha\big) \quad (m \ge 2).
    \end{equation}
    In particular, for $1 \le m \le 8$, recursive binary halving produces:
    \begin{align}
    \wedge_{*, \mathrm{bal}}^2 \alpha &= \alpha \wedge_* \alpha, \qquad \wedge_{*, \mathrm{bal}}^3 \alpha = (\alpha \wedge_* \alpha) \wedge_* \alpha, \qquad \wedge_{*, \mathrm{bal}}^4 \alpha = (\alpha \wedge_* \alpha) \wedge_* (\alpha \wedge_* \alpha), \nonumber\\
    \wedge_{*, \mathrm{bal}}^5 \alpha &= \big((\alpha \wedge_* \alpha) \wedge_* \alpha\big) \wedge_* (\alpha \wedge_* \alpha), \qquad \wedge_{*, \mathrm{bal}}^6 \alpha = \big((\alpha \wedge_* \alpha) \wedge_* \alpha\big) \wedge_* \big((\alpha \wedge_* \alpha) \wedge_* \alpha\big), \label{eq:binary_tree_5_to_8}\\
    \wedge_{*, \mathrm{bal}}^7 \alpha &= \big(\wedge_{*, \mathrm{bal}}^4 \alpha\big) \wedge_* \big(\wedge_{*, \mathrm{bal}}^3 \alpha\big), \qquad \wedge_{*, \mathrm{bal}}^8 \alpha = \big(\wedge_{*, \mathrm{bal}}^4 \alpha\big) \wedge_* \big(\wedge_{*, \mathrm{bal}}^4 \alpha\big).\nonumber
    \end{align}
    \item \textbf{Quartic-block parenthesization ($\wedge_{*, \mathrm{blk}}^m \alpha$):} Defined so that it agrees with $\wedge_{*, \mathrm{bal}}^m \alpha$ for $1 \le m \le 4$, and for $m \ge 5$ factors out leading quartic blocks $\wedge_{*, \mathrm{bal}}^4 \alpha$ recursively:
    \begin{equation}\label{eq:quartic_block_power}
    \wedge_{*, \mathrm{blk}}^m \alpha := \begin{cases}
    \wedge_{*, \mathrm{bal}}^m \alpha, & \text{if } 1 \le m \le 4, \\[0.1cm]
    \big(\wedge_{*, \mathrm{bal}}^4 \alpha\big) \wedge_* \big(\wedge_{*, \mathrm{blk}}^{m-4} \alpha\big), & \text{if } m \ge 5.
    \end{cases}
    \end{equation}
    By convention, when $m = 0$, $\big(\wedge_{*, \mathrm{bal}}^0 \alpha\big) \wedge_* \beta = \big(\wedge_{*, \mathrm{blk}}^0 \alpha\big) \wedge_* \beta := \beta$ omits the $\alpha$-factor.
\end{enumerate}

\begin{lemma}[Comparison of Parenthesization Schemes and All-Orders Jacobsthal Characterization on 1-Forms in Skew-Commutative Algebras]\label{lem:skew_comm_powers}
Let $(A, *)$ be a skew-commutative algebra ($x * y = -y * x$, so $x \circ_* y = 0$), and let $e \in \Omega^1(P; A)$ be an $A$-valued 1-form.
\begin{enumerate}
    \item \textbf{Quartic-Block Vanishing:} Because $e \wedge_* e \in \Omega^2(P; A)$ is an even-degree form, \eqref{eq:even_form_jordan_projection} gives
    \begin{equation}\label{eq:balanced_four_zero}
    \wedge_{*, \mathrm{bal}}^4 e = (e \wedge_* e) \wedge_* (e \wedge_* e) = 0.
    \end{equation}
    Consequently, under \emph{quartic-block parenthesization} \eqref{eq:quartic_block_power}, $\wedge_{*, \mathrm{blk}}^m e = 0$ identically for all $m \ge 4$.
    \item \textbf{Recursive Binary-Tree Halving and Jacobsthal Survivor Pairs:} Under \emph{recursive binary-tree parenthesization} \eqref{eq:balanced_power}, let $J_r := \frac{2^r - (-1)^r}{3}$ ($r \ge 1$) denote the $r$-th Jacobsthal number ($J_1 = 1, J_2 = 1, J_3 = 3, J_4 = 5, J_5 = 11, J_6 = 21, J_7 = 43, \dots$, satisfying $J_{r+1} = 2 J_r + (-1)^r$), and define the set of binary-tree survivor degrees
    \begin{equation}\label{eq:Jacobsthal_survivor_set}
    \mathcal{S}_{\mathrm{bal}} := \{1, 2, 3\} \cup \bigcup_{r=3}^\infty \big\{2 J_r,\, J_{r+1}\big\} = \{1, 2, 3\} \cup \{5, 6\} \cup \{10, 11\} \cup \{21, 22\} \cup \{42, 43\} \cup \dots
    \end{equation}
    \begin{enumerate}
        \item[(a)] For every integer $m \notin \mathcal{S}_{\mathrm{bal}}$ (in particular, for $m = 4$, $7 \le m \le 9$, $12 \le m \le 20$, $23 \le m \le 41$, etc.), $\wedge_{*, \mathrm{bal}}^m e = 0$ holds identically in $\Omega^m(P; A)$. Moreover, if $\wedge_{*, \mathrm{bal}}^3 e = 0$ as well (which holds whenever $(A, *)$ is a Lie algebra), then $\wedge_{*, \mathrm{bal}}^m e = 0$ for \emph{all} $m \ge 3$; more generally, if $\wedge_{*, \mathrm{bal}}^{J_r} e = 0$ for some $r \ge 3$, then $\wedge_{*, \mathrm{bal}}^m e = 0$ for all $m \ge 2 J_r - 1$.
        \item[(b)] Conversely, at each dyadic level $r \ge 3$, the consecutive pair $\{2 J_r,\, J_{r+1}\} \subset \mathcal{S}_{\mathrm{bal}}$ consists of the exterior square of the odd-degree $J_r$-form $\wedge_{*, \mathrm{bal}}^{J_r} e \in \Omega^{J_r}(P; A)$,
        \[
        \wedge_{*, \mathrm{bal}}^{2 J_r} e = \big(\wedge_{*, \mathrm{bal}}^{J_r} e\big) \wedge_* \big(\wedge_{*, \mathrm{bal}}^{J_r} e\big) = \frac{1}{2}\big(\wedge_{*, \mathrm{bal}}^{J_r} e\big) \wedge_{[\cdot,\cdot]_*} \big(\wedge_{*, \mathrm{bal}}^{J_r} e\big),
        \]
        which by \eqref{eq:odd_form_comm_projection} projects onto the commutator bracket $[x,y]_* = 2\, x * y$ rather than the Jordan product, together with the odd-degree product of $\wedge_{*, \mathrm{bal}}^{J_r} e$ and its even neighbor $\wedge_{*, \mathrm{bal}}^{J_r + (-1)^r} e = \wedge_{*, \mathrm{bal}}^{2 J_{r-1}} e$:
        \[
        \wedge_{*, \mathrm{bal}}^{J_{r+1}} e = \begin{cases}
        \big(\wedge_{*, \mathrm{bal}}^{J_r} e\big) \wedge_* \big(\wedge_{*, \mathrm{bal}}^{2 J_{r-1}} e\big), & \text{if } r \text{ is odd } (J_{r+1} = 2 J_r - 1), \\[0.1cm]
        \big(\wedge_{*, \mathrm{bal}}^{2 J_{r-1}} e\big) \wedge_* \big(\wedge_{*, \mathrm{bal}}^{J_r} e\big), & \text{if } r \text{ is even } (J_{r+1} = 2 J_r + 1).
        \end{cases}
        \]
        \item[(c)] Furthermore, if $e$ takes values in a $k$-dimensional subspace $V \subset A$ (or has pointwise rank $\le k$, so that $e = \sum_{a=1}^k e^a \otimes v_a$ is spanned by $k$ scalar 1-forms $e^1, \dots, e^k \in \Omega^1(P)$), then $\wedge_{*, \mathcal{P}}^m e = 0$ holds identically on $P$ for all $m > k$ and every parenthesization scheme $\mathcal{P}$ (since $\Lambda^m(\mathbb{R}^k)^* = 0$ for $m > k$). In particular, if $\mathrm{rk}(e) \le 9$, then $\wedge_{*, \mathrm{bal}}^m e = 0$ for all $m \ge 7$.
    \end{enumerate}
    \item \textbf{Left-Normed Products:} In contrast to both \eqref{eq:balanced_power} and \eqref{eq:quartic_block_power}, left-normed (iterated adjoint) products $(((\dots(e \wedge_* e) \wedge_* e)\dots) \wedge_* e)$ do not vanish at degree $4$ in general Lie algebras $\mathfrak{so}(k)$ ($k \ge 4$).
\end{enumerate}
\end{lemma}

\begin{proof}
Since $e \wedge_* e$ has even degree $2$ and $(A, *)$ is skew-commutative, \eqref{eq:even_form_jordan_projection} yields $\wedge_{*, \mathrm{bal}}^4 e = (e \wedge_* e) \wedge_* (e \wedge_* e) = 0$. By \eqref{eq:quartic_block_power}, $\wedge_{*, \mathrm{blk}}^m e = (\wedge_{*, \mathrm{bal}}^4 e) \wedge_* (\wedge_{*, \mathrm{blk}}^{m-4} e) = 0$ for all $m \ge 4$, proving (1).

For (2a)--(2b), note that under recursive binary halving \eqref{eq:balanced_power}, an even degree $m = 2q$ satisfies $\wedge_{*, \mathrm{bal}}^{2q} e = (\wedge_{*, \mathrm{bal}}^q e) \wedge_* (\wedge_{*, \mathrm{bal}}^q e)$, which vanishes identically by \eqref{eq:even_form_jordan_projection} whenever $q$ is even, and can survive only when $q$ is odd and $\wedge_{*, \mathrm{bal}}^q e \neq 0$. Meanwhile, an odd degree $m = 2q + 1$ splits into $\lceil m/2 \rceil = q + 1$ and $\lfloor m/2 \rfloor = q$, so $\wedge_{*, \mathrm{bal}}^{2q+1} e = (\wedge_{*, \mathrm{bal}}^{q+1} e) \wedge_* (\wedge_{*, \mathrm{bal}}^q e)$ can survive only if the consecutive pair $\{q, q+1\}$ both survive at the previous level. We now prove by induction on the dyadic level $r \ge 2$ that in the dyadic interval $I_{r+1} := [2^r + 1,\, 2^{r+1}]$ (for $r \ge 3$), the only surviving degrees $m \ge 4$ are the consecutive pair $\{2 J_r, J_{r+1}\} \subset I_{r+1}$:
\begin{itemize}
    \item At base level $r = 2$, the surviving pair in $\{1, 2, 3, 4\}$ of degree $\ge 2$ is $\{2, 3\} = \{2 J_2, J_3\}$, where $2$ is even and $J_3 = 3$ is odd (since $m = 4 = 2 \cdot 2$ vanishes).
    \item Inductively, suppose for $r \ge 2$ that the only surviving degrees at level $r$ are the consecutive pair $\{2 J_{r-1},\, J_r\}$ (which equals $\{J_r - 1, J_r\}$ when $r$ is odd and $\{J_r, J_r + 1\}$ when $r$ is even), where $2 J_{r-1}$ is even and $J_r$ is odd. At level $r + 1$, any surviving even degree $m = 2q$ requires $q \in \{2 J_{r-1}, J_r\}$ to be odd, which uniquely forces $q = J_r$ and $m = 2 J_r$ (whereas $m = 2(2 J_{r-1}) = 4 J_{r-1}$ vanishes by \eqref{eq:even_form_jordan_projection} because $2 J_{r-1}$ is even!). Any surviving odd degree $m = 2q + 1$ requires both $q$ and $q + 1$ to survive at level $r$, which uniquely forces $\{q, q+1\} = \{2 J_{r-1}, J_r\}$ and hence $m = 2 J_{r-1} + J_r = J_r + (-1)^{r-1} + J_r = 2 J_r + (-1)^r = J_{r+1}$.
\end{itemize}
Since $\{2 J_r, J_{r+1}\}$ is again a consecutive pair consisting of one even integer $2 J_r$ and one odd integer $J_{r+1}$, the induction closes for all $r \ge 3$, proving that $\wedge_{*, \mathrm{bal}}^m e = 0$ identically for all $m \notin \mathcal{S}_{\mathrm{bal}}$. Moreover, if $\wedge_{*, \mathrm{bal}}^{J_r} e = 0$ for some $r \ge 2$, then both $\wedge_{*, \mathrm{bal}}^{2 J_r} e$ and $\wedge_{*, \mathrm{bal}}^{J_{r+1}} e$ vanish, which inductively annihilates all higher levels.

For (2c), expanding $e = \sum_{a=1}^k e^a \otimes v_a$ shows by multilinearity that every $m$-fold exterior product $\wedge_{*, \mathcal{P}}^m e$ is a linear combination of $m$-forms $e^{a_1} \wedge \dots \wedge e^{a_m}$, all of which vanish identically for $m > k$.
\end{proof}

\begin{definition}[$(k,s)$-Nil and $(k,s)$-Solvable Algebras \cite{MartinsBiezunerEHP}]\label{def:ks_solv}
Let $(A, *)$ be an $R$-algebra and let $k, s \ge 1$ be integers. Fix a parenthesization scheme $\mathcal{P} \in \{\mathrm{full}, \mathrm{bal}, \mathrm{blk}\}$.
\begin{enumerate}
    \item We say that $(A, *)$ is \emph{$\mathcal{P}$-$(k,s)$-nilpotent of index $\le s$} (or simply \emph{$(k,s)$-nil} when $\mathcal{P} = \mathrm{full}$, \emph{binary-balanced $(k,s)$-nil} when $\mathcal{P} = \mathrm{bal}$, and \emph{block-balanced $(k,s)$-nil} when $\mathcal{P} = \mathrm{blk}$) if for every smooth manifold $P$ and every $A$-valued $k$-form $\alpha \in \Omega^k(P; A)$, we have $\wedge_{*, \mathcal{P}}^m \alpha = 0$ for all $m \ge s+1$. (For $s = 1$, $\wedge_*^2 \alpha = \alpha \wedge_* \alpha$ is a single binary product, so all three schemes coincide: $(1,1)$-nilpotency is equivalent to commutativity of $(A, *)$ and $(2,1)$-nilpotency is equivalent to skew-commutativity of $(A, *)$. By Lemma \ref{lem:skew_comm_powers}(1), every skew-commutative algebra is block-balanced $(1,3)$-nil since $\wedge_{*, \mathrm{blk}}^m \alpha = 0$ for all $\alpha \in \Omega^1(P; A)$ and $m \ge 4$.)
    \item A submodule $V \subset A$ is a \emph{$\mathcal{P}$-$(k,s)$-nilpotent submodule} if every $V$-valued $k$-form $\alpha \in \Omega^k(P; V)$ satisfies $\wedge_{*, \mathcal{P}}^m \alpha = 0$ for all $m \ge s+1$ in $\Omega(P; A)$.
    \item We say that $A$ (resp.\ $V \subset A$) is \emph{$\mathcal{P}$-$(k,s)$-solvable}---denoted \emph{$(k,s)$-solv}---if it decomposes as a finite direct sum $A = \bigoplus_{i=1}^r A_i$ of mutually orthogonal $\mathcal{P}$-$(k,s)$-nilpotent submodules satisfying $A_i * A_j = 0$ for $i \neq j$. More generally, if $A = \bigoplus_{i=1}^r A_i$ is a vector space direct sum of $\mathcal{P}$-$(k,s)$-nilpotent subalgebras without $A_i * A_j = 0$, a $k$-form $\alpha = \sum_{i=1}^r \alpha^{(i)} \in \Omega^k(P; A)$ is called \emph{algebraically split-adapted} if $\alpha^{(i)} \wedge_* \alpha^{(j)} = 0$ for all $i \neq j$; we denote the set of algebraically split-adapted $k$-forms on $P$ by $\mathcal{C}^k_{\mathrm{split}}(P; A)$.
    \item Suppose $A = \bigoplus_{m \in \mathfrak{m}} A^m$ is an $\mathfrak{m}$-graded algebra for an abelian group (or monoid) $(\mathfrak{m}, +)$. A graded submodule $V = \bigoplus_{m \in \mathfrak{m}} V^m$ is \emph{homogeneous weak $\mathcal{P}$-$(k,s)$-nilpotent} if $\wedge_{*, \mathcal{P}}^r \alpha^m = 0$ for all $r \ge s+1$ and each individual homogeneous degree $m \in \mathfrak{m}$ \cite{MartinsBiezunerEHP}, and \emph{(multidegree) weak $\mathcal{P}$-$(k,s)$-nilpotent} if $p_r(\alpha^{m_1}, \dots, \alpha^{m_r}) = 0$ for all $\mathcal{P}$-parenthesizations of degree $r \ge s+1$, every $V$-valued $k$-form $\alpha = \sum_{m \in \mathfrak{m}} \alpha^m$, and all $m_1, \dots, m_r \in \mathfrak{m}$. (Notice that for finite-dimensional algebras or finitely supported graded sums, multidegree weak $\mathcal{P}$-$(k,s)$-nilpotency of $V$ is equivalent by multilinearity to ungraded $\mathcal{P}$-$(k,s)$-nilpotency of $V$ and differs only for infinite-dimensional $\mathfrak{m}$-gradings, whereas homogeneous weak $\mathcal{P}$-$(k,s)$-nilpotency requires vanishing only on single-degree forms $\alpha^m \in \Omega^k(P; V^m)$ and is strictly weaker even in finite dimensions.) We say that $V$ (and in particular $A$) is \emph{weak $\mathcal{P}$-$(k,s)$-solvable} (resp.\ \emph{homogeneous weak $\mathcal{P}$-$(k,s)$-solvable}) if it decomposes as a finite direct sum $V = \bigoplus_{i=1}^r V_i$ of mutually orthogonal weak (resp.\ homogeneous weak) $\mathcal{P}$-$(k,s)$-nilpotent graded submodules ($V_i * V_j = 0$ for $i \neq j$).
\end{enumerate}
\end{definition}

Recall that the \emph{derived series} of an algebra $A$ is defined inductively by $A^{(0)} = A$ and $A^{(r)} = A^{(r-1)} * A^{(r-1)}$ \cite{SchaferBook, Sweedler}; $A$ is solvable of derived length $s$ if $A^{(s+1)} = 0$. For any form $\alpha \in \Omega^k(P; A)$, defining the iterated squaring sequence $\alpha^{[0]} := \alpha$ and $\alpha^{[m+1]} := \alpha^{[m]} \wedge_* \alpha^{[m]}$ immediately implies by induction on $m$ that $\alpha^{[m]} \in \Omega^{2^m k}(P; A^{(m)})$, and hence $\alpha^{[s+1]} = 0$ whenever $A^{(s+1)} = 0$.

\subsection{Associated Algebra Bundles, \texorpdfstring{$A$}{A}-Connections, and the Lie Nucleus Constraint}

Rather than introducing category-theoretic functorial bundle systems, we formulate our gauge-theoretic framework directly in terms of principal $G$-bundles and associated algebra bundles \cite{KobayashiNomizu, MichorGauge}.

Let $M$ be a smooth $n$-manifold, let $\pi : P \to M$ be a principal $G$-bundle with structure Lie group $G$ and Lie algebra $\mathfrak{g}$, and let $(A, *)$ be a finite-dimensional real (or complex) algebra equipped with a smooth representation by algebra automorphisms:
\begin{equation}\label{eq:rho_aut_rep}
\rho : G \longrightarrow \mathrm{Aut}_{\mathrm{Alg}}(A) \subset \mathrm{GL}(A), \qquad \rho(g)(x * y) = \rho(g)(x) * \rho(g)(y) \quad (\forall g \in G,\, x, y \in A).
\end{equation}
Differentiating $\rho$ at the identity element $e \in G$ yields a Lie algebra representation by derivations of $(A, *)$:
\begin{equation}\label{eq:drho_derivation}
d\rho : \mathfrak{g} \longrightarrow \mathrm{Der}(A), \qquad d\rho(X)(x * y) = d\rho(X)(x) * y + x * d\rho(X)(y) \quad (\forall X \in \mathfrak{g},\, x, y \in A).
\end{equation}
Because $G$ acts on $A$ by algebra automorphisms, the associated vector bundle
\begin{equation}\label{eq:associated_algebra_bundle}
E_A := P \times_\rho A \longrightarrow M
\end{equation}
is a smooth \emph{algebra bundle} over $M$ with typical fiber $(A, *)$, whose fiberwise multiplication is well-defined by $[p, x] * [p, y] := [p, x * y]$. Moreover, for every $G$-invariant subspace $V \subset A$ (in particular, for every $\mathrm{Aut}_{\mathrm{Alg}}(A)$-invariant subspace such as the commutator subspace $[A, A]_*$, the derived terms $A^{(r)}$, or the Lie nucleus $\mathcal{N}_{\mathrm{Lie}}(A)$), $E_V := P \times_\rho V \subset E_A$ is a canonical smooth vector subbundle of $E_A$.

For each $k \ge 0$, let $\Omega^k_{\mathrm{eq},\rho}(P; A)$ denote the space of $\rho$-equivariant $A$-valued $k$-forms $\alpha \in \Omega^k(P; A)$ satisfying $R_g^* \alpha = \rho(g^{-1}) \circ \alpha$ for all $g \in G$, and let
\begin{equation}\label{eq:heq_forms}
\Omega^k_{\mathrm{heq},\rho}(P; A) := \left\{ \alpha \in \Omega^k_{\mathrm{eq},\rho}(P; A) \;\middle|\; i_{\underline{X}} \alpha = 0 \text{ for all } X \in \mathfrak{g} \right\}
\end{equation}
denote the subspace of \emph{horizontal}, $\rho$-equivariant $A$-valued $k$-forms on $P$, where $\underline{X}_p := \left.\frac{d}{dt}\right|_{t=0} p \cdot \exp(tX)$ is the fundamental vertical vector field generated by $X \in \mathfrak{g}$. Because $\rho(g) \in \mathrm{Aut}_{\mathrm{Alg}}(A)$ and $i_{\underline{X}}(\alpha \wedge_* \beta) = (i_{\underline{X}}\alpha) \wedge_* \beta + (-1)^{\deg\alpha}\alpha \wedge_* (i_{\underline{X}}\beta)$, the direct sum $\Omega^\bullet_{\mathrm{heq},\rho}(P; A) = \bigoplus_{k=0}^n \Omega^k_{\mathrm{heq},\rho}(P; A)$ is a $\mathbb{Z}$-graded subalgebra of $(\Omega^\bullet(P; A), \wedge_*)$. The canonical \emph{Weil isomorphism} \cite{KobayashiNomizu, MichorGauge}
\begin{equation}\label{eq:Weil_isomorphism}
j : \Omega^\bullet_{\mathrm{heq},\rho}(P; A) \xrightarrow{\;\sim\;} \Omega^\bullet(M; E_A), \qquad j(\alpha)_{\pi(p)}\big(d\pi_p(v_1), \dots, d\pi_p(v_k)\big) := \big[p,\, \alpha_p(v_1, \dots, v_k)\big],
\end{equation}
is an isomorphism of $\mathbb{Z}$-graded algebras: $j(\alpha \wedge_* \beta) = j(\alpha) \wedge_* j(\beta)$ for all $\alpha, \beta \in \Omega^\bullet_{\mathrm{heq},\rho}(P; A)$, and for every $G$-invariant subspace $V \subset A$, $j$ maps $\Omega^k_{\mathrm{heq},\rho}(P; V)$ isomorphically onto $\Omega^k(M; E_V)$.

A smooth \emph{bundle trace functional} on $E_A$ is a fiberwise linear bundle morphism $\mathrm{tr}_\xi : E_A \to M \times \mathbb{R}$ determined by a smooth global section $\xi \in \Gamma(E_A^*)$ of the dual bundle $E_A^* = P \times_{\rho^*} A^*$ via $\mathrm{tr}_\xi(\Psi) := \langle \xi, \Psi \rangle \in \Omega^\bullet(M; \mathbb{R})$ for $\Psi \in \Omega^\bullet(M; E_A)$. Equivalently, $\xi = j(\hat{\xi})$ corresponds to a $G$-equivariant map $\hat{\xi} : P \to A^*$ satisfying $\hat{\xi}(p \cdot g) = \rho(g)^* \hat{\xi}(p)$. Such trace sections arise in two geometric ways:
\begin{enumerate}
    \item[(T1)] \emph{$G$-Invariant Linear Functional:} If $\tau_0 \in (A^*)^G$ is a constant $G$-invariant linear functional on $A$ (e.g., the matrix trace on $\mathfrak{gl}(k; \mathbb{R})$ or the central projection $\mathfrak{u}(k) \to \mathfrak{u}(1) \simeq \mathbb{R}$), then $\hat{\xi}(p) \equiv \tau_0$ defines a strictly $G$-invariant bundle trace morphism $\mathrm{tr}_{\tau_0} : E_A \to M \times \mathbb{R}$. Note that if $\rho = \mathrm{Ad}$ on a Lie algebra $A = \mathfrak{g}$, any $\mathrm{Ad}_G$-invariant linear functional $\tau_0 \in (\mathfrak{g}^*)^G$ vanishes identically on the derived subalgebra $[\mathfrak{g}, \mathfrak{g}]$ (though for a reductive Lie algebra $\mathfrak{g} = \mathfrak{z}(\mathfrak{g}) \oplus [\mathfrak{g}, \mathfrak{g}]$ it may be non-zero on the center $\mathfrak{z}(\mathfrak{g})$; however, any interaction term taking values strictly inside $[A, A]_*$ or $[\mathfrak{g}, \mathfrak{g}]$ is still annihilated by $\tau_0$).
    \item[(T2)] \emph{Coadjoint Background Section:} When $(A^*)^G = \{0\}$ on $[A, A]_*$, any smooth non-zero section $\xi \in \Gamma(E_A^*)$ defines a bundle linear functional $\mathrm{tr}_\xi : E_A \to M \times \mathbb{R}$. Wherever $\xi(x) \neq 0$, choosing $\xi$ acts as a symmetry-breaking background field that reduces the internal structure group $G$ at $x$ to the stabilizer subgroup $G_{\xi(x)} \subset G$.
\end{enumerate}

\begin{definition}[$A$-Connections, Field Strength, and Covariant Exterior Derivative]\label{def:A_connection}
Let $\pi : P \to M$ be a principal $G$-bundle and let $(A, *)$ be an $R$-algebra with automorphism representation $\rho : G \to \mathrm{Aut}_{\mathrm{Alg}}(A)$.
\begin{enumerate}
    \item A \emph{pure $A$-connection} on $P$ is an $A$-valued 1-form $D \in \Omega^1(P; A)$. Its \emph{field strength} (or \emph{curvature}) is the $A$-valued 2-form $F_D \in \Omega^2(P; A)$ given by
    \begin{equation}\label{eq:pure_curvature}
    F_D := dD + \frac{1}{2} D \wedge_* D = dD + \frac{1}{4} D \wedge_{[\cdot,\cdot]_*} D,
    \end{equation}
    and its \emph{commutator covariant exterior derivative} $d_D^{[\cdot,\cdot]_*} : \Omega^k(P; A) \to \Omega^{k+1}(P; A)$ is defined by
    \begin{equation}\label{eq:comm_covariant_ext_deriv}
    d_D^{[\cdot,\cdot]_*} \alpha := d\alpha + \frac{1}{2} D \wedge_{[\cdot,\cdot]_*} \alpha = d\alpha + \frac{1}{2}\big(D \wedge_* \alpha - (-1)^k \alpha \wedge_* D\big).
    \end{equation}
    For any $k$-form $\Phi \in \Omega^k(P; A)$ and 0-form $\epsilon \in \Omega^0(P; A)$, we also write $[\Phi, \epsilon]_* := \Phi \wedge_{[\cdot,\cdot]_*} \epsilon = \Phi \wedge_* \epsilon - \epsilon \wedge_* \Phi = -[\epsilon, \Phi]_* \in \Omega^k(P; A)$.
    \item Suppose $A$ is equipped with a $G$-invariant $R$-module decomposition $A \simeq A_0 \oplus A_1$. A \emph{reductive $A$-connection} on $P$ is a pair $\nabla = e + \omega \in \Omega^1(P; A_0) \oplus \Omega^1(P; A_1)$. Its \emph{full Cartan field strength} is $F_\nabla = d\nabla + \frac{1}{2} \nabla \wedge_* \nabla \in \Omega^2(P; A)$, which expands as
    \begin{equation}\label{eq:Cartan_curvature}
    F_\nabla = \Omega_\omega + \Theta_{e,\omega} + E_e,
    \end{equation}
    where $\Omega_\omega := d\omega + \frac{1}{2} \omega \wedge_* \omega$ is the \emph{pure curvature} of $\omega$, $\Theta_{e,\omega} := de + \frac{1}{2}(\omega \wedge_* e + e \wedge_* \omega) = de + \frac{1}{2} \omega \wedge_{[\cdot,\cdot]_*} e$ is the \emph{generalized torsion}, and $E_e := \frac{1}{2} e \wedge_* e$ is the \emph{quadratic soldering form} (all evaluated via the ambient multiplication $*$ of $A$).
    \item If $A = \bigoplus_{m \in \mathfrak{m}} A^m$ is an $\mathfrak{m}$-graded algebra ($A^{m_1} * A^{m_2} \subset A^{m_1+m_2}$), an $A$-connection $D \in \Omega^1(P; A)$ decomposes algebraically along the constant vector space direct sum $A = \bigoplus_{m \in \mathfrak{m}} A^m$ as $D = \sum_{m \in \mathfrak{m}} D^m$ with $D^m \in \Omega^1(P; A^m)$, and its curvature $F_D = \sum_{m \in \mathfrak{m}} (F_D)^m \in \bigoplus_{m \in \mathfrak{m}} \Omega^2(P; A^m)$ has homogeneous algebra components
    \begin{equation}\label{eq:graded_curvature_components}
    (F_D)^m = d(D^m) + \frac{1}{2} \sum_{m_1 + m_2 = m} D^{m_1} \wedge_* D^{m_2} \in \Omega^2(P; A^m).
    \end{equation}
\end{enumerate}
\end{definition}

We now establish the exact algebraic and geometric conditions under which the curvature $F_D$, the covariant exterior derivative $d_D^{[\cdot,\cdot]_*} \alpha$, and the Jacobiator forms of $D$ are horizontal and $\rho$-equivariant on $P$, so that they descend to $E_A \to M$ via the Weil isomorphism $j$.

Given any trilinear bracket $J(x,y,z)$ on $A$ (such as $\mathrm{Jac}_{[\cdot,\cdot]_*}(x,y,z)$ or $\mathrm{Jac}_\curlywedge(x,y,z) := (x \curlywedge y) \curlywedge z + (y \curlywedge z) \curlywedge x + (z \curlywedge x) \curlywedge y$), for any forms $\alpha \in \Omega^{k_1}(P; A)$, $\beta \in \Omega^{k_2}(P; A)$, and $\gamma \in \Omega^{k_3}(P; A)$ with $K = k_1 + k_2 + k_3$, we define $J(\alpha, \beta, \gamma) \in \Omega^K(P; A)$ by
\begin{equation}\label{eq:Jac_form_normalization}
J(\alpha, \beta, \gamma)_p(v_1, \dots, v_K) := \frac{1}{k_1!\, k_2!\, k_3!} \sum_{\sigma \in S_K} \mathrm{sign}(\sigma)\, J\big(\alpha_p(v_{\sigma(1)}, \dots), \beta_p(\dots), \gamma_p(\dots, v_{\sigma(K)})\big).
\end{equation}
In particular, for $D \in \Omega^1(P; A)$ and $\epsilon \in \Omega^0(P; A)$, evaluating \eqref{eq:Jac_form_normalization} on $v_1, v_2 \in T_p P$ gives $J(D, D, \epsilon)_p(v_1, v_2) = 2\, J(D_p(v_1), D_p(v_2), \epsilon(p))$, while on $v_1, v_2, v_3 \in T_p P$ it gives $J(D, D, D)_p(v_1, v_2, v_3) = 6\, J(D_p(v_1), D_p(v_2), D_p(v_3))$.

\begin{proposition}[Ehresmann Admissibility and the Lie Nucleus Constraint]\label{prop:Cartan_Ehresmann}
Let $\pi : P \to M$ be a principal $G$-bundle, let $\rho : G \to \mathrm{Aut}_{\mathrm{Alg}}(A)$ be a representation of $G$ by algebra automorphisms of $(A, *)$, and let $\eta : \mathfrak{g} \to A$ be a $G$-equivariant linear map ($\eta(\mathrm{Ad}_g X) = \rho(g)(\eta(X))$ for all $g \in G$, $X \in \mathfrak{g}$). Define the space of \emph{$\eta$-compatible Ehresmann $A$-connections} on $P$ by
\begin{equation}\label{eq:Ehresmann_conn_space}
\mathrm{Conn}_{\rho,\eta}(P; A) := \left\{ D \in \Omega^1_{\mathrm{eq},\rho}(P; A) \;\middle|\; D(\underline{X}) = \eta(X) \text{ for all } X \in \mathfrak{g} \right\}.
\end{equation}
Assume that the intertwining condition
\begin{equation}\label{eq:Ehresmann_commutator_intertwine}
d\rho(X)(y) = \frac{1}{2}[\eta(X), y]_* = \frac{1}{2}\big(\eta(X) * y - y * \eta(X)\big) \qquad (\forall X \in \mathfrak{g},\, y \in A)
\end{equation}
holds. (Notice that when $(A, *)$ is a Lie algebra $(\mathfrak{g}, [\cdot,\cdot]_{\mathfrak{g}})$ with $* = [\cdot,\cdot]_{\mathfrak{g}}$ and $\eta = \mathrm{id}_{\mathfrak{g}}$, the commutator of $*$ is $[x, y]_* = [x, y]_{\mathfrak{g}} - [y, x]_{\mathfrak{g}} = 2[x, y]_{\mathfrak{g}}$, so $\frac{1}{2}[\eta(X), y]_* = [X, y]_{\mathfrak{g}} = \mathrm{ad}_X(y)$ recovers the standard adjoint representation.) Then:
\begin{enumerate}
    \item \textbf{Lie Nucleus Constraint:} Because $d\rho(X) \in \mathrm{Der}(A)$ is a derivation of $(A, *)$ and hence of $(A, [\cdot,\cdot]_*)$, condition \eqref{eq:Ehresmann_commutator_intertwine} forces the image of $\eta$ to lie in the Lie nucleus \eqref{eq:Lie_nucleus} of $(A, [\cdot,\cdot]_*)$:
    \begin{equation}\label{eq:eta_Lie_nucleus_constraint}
    \eta(\mathfrak{g}) \subset \mathcal{N}_{\mathrm{Lie}}(A).
    \end{equation}
    More precisely, under \eqref{eq:Ehresmann_commutator_intertwine}, $\frac{1}{2}\mathrm{ad}^{[\cdot,\cdot]_*}_{\eta(X)}$ is a derivation of the commutator algebra $(A, [\cdot,\cdot]_*)$ if and only if $\eta(X) \in \mathcal{N}_{\mathrm{Lie}}(A)$ (which alone suffices for all horizontality assertions in (2) below), and is a derivation of the full algebra $(A, *)$ if and only if $\eta(X)$ also satisfies the Jordan-derivation identity $[\eta(X), y \circ_* z]_* = [\eta(X), y]_* \circ_* z + y \circ_* [\eta(X), z]_*$ for all $y, z \in A$ (which holds automatically on $\mathcal{N}_{\mathrm{Lie}}(A)$ whenever $(A, *)$ is skew-commutative or associative). In particular, if $(A, [\cdot,\cdot]_*)$ has trivial Lie nucleus $\mathcal{N}_{\mathrm{Lie}}(A) = \{0\}$ (such as the 7-dimensional octonionic Malcev algebra $\mathrm{Im}(\mathbb{O})$, where $\mathrm{Jac}_{[\cdot,\cdot]_*}(x,y,z) = 6\,\mathrm{Assoc}_{\mathbb{O}}(x,y,z)$ and $\mathcal{N}_{\mathrm{assoc}}(\mathbb{O}) = \mathbb{R}\cdot 1$, so $\mathcal{N}_{\mathrm{Lie}}(\mathrm{Im}(\mathbb{O})) = \mathrm{Im}(\mathbb{O}) \cap \mathbb{R}\cdot 1 = \{0\}$), then \eqref{eq:Ehresmann_commutator_intertwine} requires $\eta = 0$ and $d\rho = 0$.
    \item \textbf{Horizontality and Descent of Curvature, Covariant Derivative, and Jacobiator Forms:} For every $D \in \mathrm{Conn}_{\rho,\eta}(P; A)$, every $\alpha \in \Omega^k_{\mathrm{heq},\rho}(P; A)$, and every $\epsilon \in \Omega^0_{\mathrm{heq},\rho}(P; A)$:
    \begin{enumerate}
        \item[(a)] The curvature $F_D = dD + \frac{1}{2} D \wedge_* D$ lies in $\Omega^2_{\mathrm{heq},\rho}(P; A)$ and descends via $j$ to $j(F_D) \in \Omega^2(M; E_A)$;
        \item[(b)] The commutator covariant exterior derivative $d_D^{[\cdot,\cdot]_*} \alpha$ lies in $\Omega^{k+1}_{\mathrm{heq},\rho}(P; A)$ and descends via $j$ to $d_{\nabla_D}^{[\cdot,\cdot]_*} j(\alpha) := j(d_D^{[\cdot,\cdot]_*} \alpha) \in \Omega^{k+1}(M; E_A)$;
        \item[(c)] By the Lie nucleus inclusion \eqref{eq:eta_Lie_nucleus_constraint}, the Jacobiator forms $\mathrm{Jac}_{[\cdot,\cdot]_*}(D, D, D) \in \Omega^3(P; A)$ and $\mathrm{Jac}_{[\cdot,\cdot]_*}(D, D, \epsilon) \in \Omega^2(P; A)$ are horizontal and $\rho$-equivariant, i.e., lie in $\Omega^3_{\mathrm{heq},\rho}(P; A)$ and $\Omega^2_{\mathrm{heq},\rho}(P; A)$, and descend via $j$ to $\Omega^3(M; E_A)$ and $\Omega^2(M; E_A)$.
    \end{enumerate}
    \item \textbf{Affine vs.\ Linear Connection Regimes:}
    \begin{itemize}
        \item \emph{Affine Ehresmann Regime ($\eta \neq 0$):} Choosing any principal connection $\omega_0 \in \mathrm{Conn}(P; \mathfrak{g})$, $D_0 := \eta \circ \omega_0 \in \mathrm{Conn}_{\rho,\eta}(P; A)$ is a basepoint connection and $\mathrm{Conn}_{\rho,\eta}(P; A) = D_0 + \Omega^1_{\mathrm{heq},\rho}(P; A)$ is a non-empty \emph{affine space} modeled on $\Omega^1_{\mathrm{heq},\rho}(P; A) \simeq \Omega^1(M; E_A)$.
        \item \emph{Linear/Horizontal Regime ($\eta = 0$, $d\rho = 0$):} When $G$ acts trivially on $A$ (or on a trivial bundle $P = M \times G$ relative to the flat background exterior derivative $d$), $\eta = 0$ satisfies \eqref{eq:Ehresmann_commutator_intertwine} and \eqref{eq:eta_Lie_nucleus_constraint} for \emph{every} algebra $(A, *)$ without any Lie-nucleus restriction, and $\mathrm{Conn}^{\mathrm{lin}}(P; A) := \mathrm{Conn}_{\rho,0}(P; A) = \Omega^1_{\mathrm{heq},\rho}(P; A) \simeq \Omega^1(M; E_A)$ is a \emph{real vector space}.
    \end{itemize}
\end{enumerate}
\end{proposition}

\begin{proof}
For (1), since $\rho(G) \subset \mathrm{Aut}_{\mathrm{Alg}}(A)$, each $d\rho(X)$ is a derivation of $(A, *)$ and therefore a derivation of both $[y, z]_* = y * z - z * y$ and $y \circ_* z = \frac{1}{2}(y * z + z * y)$. On the other hand, expanding $\frac{1}{2}[\eta(X), [y, z]_*]_* - [\frac{1}{2}[\eta(X), y]_*, z]_* - [y, \frac{1}{2}[\eta(X), z]_*]_*$ gives $-\frac{1}{2}\mathrm{Jac}_{[\cdot,\cdot]_*}(\eta(X), y, z)$, so $\frac{1}{2}\mathrm{ad}^{[\cdot,\cdot]_*}_{\eta(X)}$ is a derivation of $[\cdot,\cdot]_*$ if and only if $\mathrm{Jac}_{[\cdot,\cdot]_*}(\eta(X), y, z) = 0$ for all $y, z \in A$, i.e., $\eta(X) \in \mathcal{N}_{\mathrm{Lie}}(A)$ (and is a derivation of the full product $*$ if and only if $\frac{1}{2}\mathrm{ad}^{[\cdot,\cdot]_*}_{\eta(X)}$ also acts as a derivation on $y \circ_* z$, which is automatic on $\mathcal{N}_{\mathrm{Lie}}(A)$ when $y \circ_* z = 0$ or when $*$ is associative).

For (2a), $\rho$-equivariance of $D$ and $\rho(g) \in \mathrm{Aut}_{\mathrm{Alg}}(A)$ immediately imply $R_g^* F_D = \rho(g^{-1}) \circ F_D$. For horizontality, differentiating $R_{\exp(tX)}^* D = \rho(\exp(-tX)) \circ D$ at $t = 0$ via Cartan's magic formula $L_{\underline{X}} D = i_{\underline{X}} dD + d(i_{\underline{X}} D)$ and using $i_{\underline{X}} D = \eta(X) = \mathrm{const}$ gives $i_{\underline{X}} dD = -d\rho(X)(D) = -\frac{1}{2}[\eta(X), D]_*$. Meanwhile, by \eqref{eq:odd_form_comm_projection}, $i_{\underline{X}}(\frac{1}{2} D \wedge_* D) = \frac{1}{4} i_{\underline{X}}(D \wedge_{[\cdot,\cdot]_*} D) = \frac{1}{2}[i_{\underline{X}} D, D]_* = \frac{1}{2}[\eta(X), D]_*$. Summing the two terms gives $i_{\underline{X}} F_D = 0$, so $F_D \in \Omega^2_{\mathrm{heq},\rho}(P; A)$.

For (2b), if $\alpha \in \Omega^k_{\mathrm{heq},\rho}(P; A)$, then $i_{\underline{X}} \alpha = 0$ and $L_{\underline{X}} \alpha = i_{\underline{X}} d\alpha = -d\rho(X)(\alpha) = -\frac{1}{2}[\eta(X), \alpha]_*$, while $i_{\underline{X}}(\frac{1}{2} D \wedge_{[\cdot,\cdot]_*} \alpha) = \frac{1}{2}[\eta(X), \alpha]_* - \frac{1}{2} D \wedge_{[\cdot,\cdot]_*} (i_{\underline{X}}\alpha) = \frac{1}{2}[\eta(X), \alpha]_*$, so $i_{\underline{X}}(d_D^{[\cdot,\cdot]_*} \alpha) = 0$.

For (2c), for any $X \in \mathfrak{g}$ and tangent vectors $v_1, v_2 \in T_p P$, \eqref{eq:Jac_form_normalization} and \eqref{eq:eta_Lie_nucleus_constraint} give
\[
\big(i_{\underline{X}}\mathrm{Jac}_{[\cdot,\cdot]_*}(D, D, D)\big)_p(v_1, v_2) = 6\, \mathrm{Jac}_{[\cdot,\cdot]_*}\big(\eta(X), D_p(v_1), D_p(v_2)\big) = 0,
\]
and $\big(i_{\underline{X}}\mathrm{Jac}_{[\cdot,\cdot]_*}(D, D, \epsilon)\big)_p(v_1) = 2\, \mathrm{Jac}_{[\cdot,\cdot]_*}(\eta(X), D_p(v_1), \epsilon(p)) = 0$ because $\eta(X) \in \mathcal{N}_{\mathrm{Lie}}(A)$. Since $\rho(G) \subset \mathrm{Aut}_{\mathrm{Alg}}(A)$ preserves $\mathrm{Jac}_{[\cdot,\cdot]_*}$, both forms are $\rho$-equivariant and horizontal. Part (3) is immediate from $i_{\underline{X}}(D_1 - D_2) = \eta(X) - \eta(X) = 0$.
\end{proof}

\subsection{Definition of \texorpdfstring{$A$}{A}-YMT Theories, Variational Equations, and the Jacobiator Anomaly}

Throughout the remainder of the paper, we work in the geometric setting of Proposition \ref{prop:Cartan_Ehresmann}: $\pi : P \to M$ is a principal $G$-bundle over a smooth $n$-manifold $M$, $\rho : G \to \mathrm{Aut}_{\mathrm{Alg}}(A)$ is an algebra automorphism representation with associated algebra bundle $E_A = P \times_\rho A \to M$ and Weil isomorphism $j : \Omega^\bullet_{\mathrm{heq},\rho}(P; A) \xrightarrow{\sim} \Omega^\bullet(M; E_A)$, and $\mathrm{Conn}_{\rho,\eta}(P; A)$ is the space of admissible $A$-connections satisfying \eqref{eq:Ehresmann_commutator_intertwine} (including the linear/horizontal case $\eta = 0$).

Let $\mathrm{Dens}(M)$ denote the $C^\infty(M)$-module of smooth 1-densities on $M$ (identified with $\Omega^n(M; \mathbb{R})$ when $M$ is oriented). For any $C^\infty(M)$-module $V$, let $V_{\mathbb{R}}$ denote the real vector space obtained by restriction of scalars along $\mathbb{R} \hookrightarrow C^\infty(M)$.

\begin{definition}[$A$-Valued Yang-Mills-Type ($A$-YMT) Theory]\label{def:A_YMT}
Let $E_A = P \times_\rho A \to M$ be an associated algebra bundle over a smooth $n$-manifold $M$.
\begin{enumerate}
    \item An \emph{$A$-valued Yang-Mills-type ($A$-YMT) pairing} on $E_A$ is an $\mathbb{R}$-bilinear map
    \begin{equation}\label{eq:YMT_pairing}
    \langle \cdot, \cdot \rangle_A : \Omega^2(M; E_A)_{\mathbb{R}} \otimes_{\mathbb{R}} \Omega^2(M; E_A)_{\mathbb{R}} \longrightarrow \mathrm{Dens}(M)_{\mathbb{R}},
    \end{equation}
    with symmetrization $\langle \Phi, \Psi \rangle_{A, \mathrm{sym}} := \frac{1}{2}(\langle \Phi, \Psi \rangle_A + \langle \Psi, \Phi \rangle_A)$. We say that $\langle \cdot, \cdot \rangle_{A, \mathrm{sym}}$ is \emph{tensorial} (or \emph{$C^\infty(M)$-bilinear}) if $\langle f\Phi, g\Psi \rangle_{A, \mathrm{sym}} = fg\langle \Phi, \Psi \rangle_{A, \mathrm{sym}}$ for all $f, g \in C^\infty(M)$ (which implies support locality $\mathrm{supp}(\langle \Phi, \Psi \rangle_{A, \mathrm{sym}}) \subset \mathrm{supp}(\Phi) \cap \mathrm{supp}(\Psi)$ and induces a smooth bundle bilinear form on $\Lambda^2 T^*M \otimes E_A$), \emph{pointwise $\mathrm{ad}_{[\cdot,\cdot]_*}$-invariant} if $\langle [\Phi, j(\epsilon)]_*, \Psi \rangle_{A, \mathrm{sym}} + \langle \Phi, [\Psi, j(\epsilon)]_* \rangle_{A, \mathrm{sym}} = 0$ in $\mathrm{Dens}(M)$ for all $\Phi, \Psi \in \Omega^2(M; E_A)$ and $\epsilon \in \Omega^0_{\mathrm{heq},\rho}(P; A)$, and \emph{non-degenerate} if $\int_M \langle \Phi, \Psi \rangle_{A, \mathrm{sym}} = 0$ for all compactly supported $\Phi \in \Omega^2_c(M; E_A)$ implies $\Psi = 0$.
    \item Assuming $M$ is compact (or restricting to connections with compactly supported curvature density), the \emph{$A$-YMT action functional} associated with $\langle \cdot, \cdot \rangle_A$ is $S^A : \mathrm{Conn}_{\rho,\eta}(P; A) \to \mathbb{R}$,
    \begin{equation}\label{eq:A_YMT_action}
    S^A[D] := \int_M \langle j(F_D), j(F_D) \rangle_A.
    \end{equation}
\end{enumerate}
We distinguish three classes of $A$-YMT theories according to the pairing $\langle \cdot, \cdot \rangle_A$:
\begin{itemize}
    \item \textbf{General $A$-YMT Theory:} $\langle \cdot, \cdot \rangle_A$ is an arbitrary $\mathbb{R}$-bilinear map \eqref{eq:YMT_pairing}.
    \item \textbf{Tensorial $A$-YMT Theory:} $\langle \cdot, \cdot \rangle_A$ is $C^\infty(M)$-bilinear, e.g., given a fiberwise symmetric bilinear form $\kappa : E_A \times_M E_A \to M \times \mathbb{R}$ and a bundle Hodge-type morphism $\star_M : \Omega^2(M; E_A) \to \Omega^{n-2}(M; E_A)$ on an oriented $n$-manifold $M$,
    \begin{equation}\label{eq:kappa_star_pairing}
    \langle \Phi, \Psi \rangle_{\kappa, \star_M} := \kappa(\Phi \wedge \star_M \Psi) \in \Omega^n(M; \mathbb{R}).
    \end{equation}
    \item \textbf{Multiplicative (or Internal-Trace) $A$-YMT Theory:} $M$ is oriented, $E_A$ is equipped with a bundle trace functional $\mathrm{tr}_\xi : E_A \to M \times \mathbb{R}$ induced by $\xi \in \Gamma(E_A^*)$, and $\hat{\star} : \Omega^2_{\mathrm{heq},\rho}(P; A) \to \Omega^{n-2}_{\mathrm{heq},\rho}(P; A)$ is an internal constitutive operator such that the pairing is formed via the \emph{internal product $\wedge_*$ of $A$} followed by $\mathrm{tr}_\xi$:
    \begin{equation}\label{eq:multiplicative_YMT}
    S^A_{\hat{\star}}[D] := \int_M \mathrm{tr}_\xi\Big( j\big(F_D \wedge_* \hat{\star} F_D\big) \Big).
    \end{equation}
\end{itemize}
\end{definition}

\begin{remark}[Internal Multiplicative Pairings vs.\ Classical Bilinear Invariant Forms]\label{rem:internal_vs_external_CW}
In classical gauge theory with $A = \mathfrak{g}$ and $\kappa \in \mathrm{inv}^2(\mathfrak{g}) = S^2(\mathfrak{g}^*)^{\mathrm{Ad}}$ (such as the Killing form $B_{\mathfrak{g}}$), both the classical Yang-Mills action $\int_M \kappa(j(F_D) \wedge *_g j(F_D))$ and the classical 4D second Chern / Pontryagin action $\int_M \kappa(j(F_D) \wedge j(F_D))$ belong to the class of \textbf{tensorial $A$-YMT theories} \eqref{eq:kappa_star_pairing}. Notice that $\kappa(j(F_D) \wedge j(F_D))$ uses the \emph{external} symmetric bilinear form $\kappa : \mathfrak{g} \times \mathfrak{g} \to \mathbb{R}$ (or equivalently, the associative matrix product $\Omega \wedge_{\mathrm{Mat}} \Omega$ in an external associative matrix envelope $\mathrm{Mat}_{k \times k}(\mathbb{R})$, which maps $\mathfrak{so}(k) \times \mathfrak{so}(k)$ into the space of symmetric matrices $\mathrm{Sym}_k(\mathbb{R})$ outside $\mathfrak{so}(k)$ before taking the matrix trace).
In contrast, in a multiplicative $A$-YMT theory \eqref{eq:multiplicative_YMT} or in the higher-degree internal Chern-Weil functional in dimension $n = 2r$ (for a parenthesization scheme $\mathcal{P} \in \{\mathrm{full}, \mathrm{bal}, \mathrm{blk}\}$),
\begin{equation}\label{eq:Chern_Weil_YMT}
S^A_{\mathrm{top}, r, \mathcal{P}}[D] := \int_M \mathrm{tr}_\xi\Big( j\big(\wedge_{*, \mathcal{P}}^r F_D\big) \Big),
\end{equation}
the curvature factors are multiplied \emph{internally within the algebra $A$ itself} using $\wedge_*$, and only afterwards is a linear functional $\mathrm{tr}_\xi : E_A \to M \times \mathbb{R}$ applied.
\end{remark}

\begin{theorem}[Variational Field Equations, Infinitesimal Gauge Covariance, and the Jacobiator Bianchi Anomaly]\label{thm:Bianchi_gauge_EL}
Let $\pi : P \to M$ be a principal $G$-bundle over a closed smooth $n$-manifold $M$, let $\rho : G \to \mathrm{Aut}_{\mathrm{Alg}}(A)$ and $\eta : \mathfrak{g} \to \mathcal{N}_{\mathrm{Lie}}(A)$ satisfy \eqref{eq:Ehresmann_commutator_intertwine}, and let $D \in \mathrm{Conn}_{\rho,\eta}(P; A)$ be an admissible $A$-connection with curvature $F_D = dD + \frac{1}{2} D \wedge_* D = dD + \frac{1}{4} D \wedge_{[\cdot,\cdot]_*} D \in \Omega^2_{\mathrm{heq},\rho}(P; A)$.
\begin{enumerate}
    \item \textbf{First Variation of Curvature and Action:} For any tangent variation $\dot{D} \in \Omega^1_{\mathrm{heq},\rho}(P; A) \simeq \Omega^1(M; E_A)$, the directional derivative of the curvature is given for \emph{every} algebra $(A, *)$ (without assuming associativity or the Jacobi identity) by
    \begin{equation}\label{eq:curvature_variation}
    \delta_{\dot{D}} F_D := \left.\frac{d}{dt}\right|_{t=0} F_{D + t\dot{D}} = d\dot{D} + \frac{1}{2}\big(D \wedge_* \dot{D} + \dot{D} \wedge_* D\big) = d_D^{[\cdot,\cdot]_*} \dot{D} \in \Omega^2_{\mathrm{heq},\rho}(P; A),
    \end{equation}
    and the first variation of the general $A$-YMT action $S^A[D]$ is
    \begin{equation}\label{eq:action_first_variation}
    \delta_{\dot{D}} S^A[D] = 2 \int_M \big\langle j(F_D),\, d_{\nabla_D}^{[\cdot,\cdot]_*} j(\dot{D}) \big\rangle_{A, \mathrm{sym}}.
    \end{equation}
    In particular, consider a tensorial $A$-YMT theory \eqref{eq:kappa_star_pairing} on a closed oriented $n$-manifold $M$, where $\kappa : E_A \times_M E_A \to M \times \mathbb{R}$ is fiberwise symmetric, non-degenerate, and $\mathrm{ad}_{[\cdot,\cdot]_*}$-invariant ($\kappa([x,y]_*, z) + \kappa(y, [x,z]_*) = 0$), and $\star_M : \Omega^2(M; E_A) \to \Omega^{n-2}(M; E_A)$ satisfies $\kappa(\Phi \wedge \star_M \Psi) = \kappa(\Psi \wedge \star_M \Phi)$. Then $D$ is a critical point of $S^A[D]$ if and only if it satisfies the \emph{$A$-Yang-Mills Euler-Lagrange equation}:
    \begin{equation}\label{eq:A_YM_EL}
    d_{\nabla_D}^{[\cdot,\cdot]_*} \big(\star_M j(F_D)\big) = j\Big( d_D^{[\cdot,\cdot]_*}\big( j^{-1}(\star_M j(F_D)) \big) \Big) = 0 \quad \text{in } \Omega^{n-1}(M; E_A).
    \end{equation}
    \item \textbf{Generalized Bianchi Identity and Jacobiator Obstruction:} For every admissible connection $D \in \mathrm{Conn}_{\rho,\eta}(P; A)$, the covariant exterior derivative of the curvature $F_D$ satisfies in $\Omega^3_{\mathrm{heq},\rho}(P; A)$:
    \begin{equation}\label{eq:Bianchi_Jacobiator}
    d_D^{[\cdot,\cdot]_*} F_D = dF_D + \frac{1}{2} D \wedge_{[\cdot,\cdot]_*} F_D = -\frac{1}{8}(D \wedge_{[\cdot,\cdot]_*} D) \wedge_{[\cdot,\cdot]_*} D = -\frac{1}{24}\mathrm{Jac}_{[\cdot,\cdot]_*}(D, D, D).
    \end{equation}
    Consequently, the classical Bianchi identity $d_{\nabla_D}^{[\cdot,\cdot]_*} j(F_D) = 0$ holds identically for all connections $D \in \mathrm{Conn}_{\rho,\eta}(P; A)$ on manifolds of dimension $n \ge 3$ \emph{if and only if} $(A, *)$ is Lie-admissible ($\mathrm{Jac}_{[\cdot,\cdot]_*} \equiv 0$).
    \item \textbf{Infinitesimal Gauge Transformations, Principal vs.\ Extended $A$-Symmetries, and Covariance Anomaly:} Given a 0-form gauge parameter $\epsilon \in \Omega^0_{\mathrm{heq}, \rho}(P; A) \simeq \Gamma(E_A)$, define the infinitesimal gauge variation of $D \in \mathrm{Conn}_{\rho,\eta}(P; A)$ by
    \begin{equation}\label{eq:inf_gauge_var_D}
    \delta_\epsilon D := d_D^{[\cdot,\cdot]_*} \epsilon = d\epsilon + \frac{1}{2}[D, \epsilon]_* \in \Omega^1_{\mathrm{heq},\rho}(P; A).
    \end{equation}
    Then $D + t\,\delta_\epsilon D \in \mathrm{Conn}_{\rho,\eta}(P; A)$, and the induced infinitesimal variation of the curvature $F_D$ is
    \begin{equation}\label{eq:inf_gauge_var_F}
    \delta_\epsilon F_D = d_D^{[\cdot,\cdot]_*} (d_D^{[\cdot,\cdot]_*} \epsilon) = \frac{1}{2}[F_D, \epsilon]_* - \frac{1}{8}\, \mathrm{Jac}_{[\cdot,\cdot]_*}(D, D, \epsilon) \in \Omega^2_{\mathrm{heq},\rho}(P; A).
    \end{equation}
    Consequently, we have a sharp dichotomy between principal $G$-gauge transformations and extended $A$-gauge transformations:
    \begin{enumerate}
        \item[(a)] \emph{Principal $G$-Gauge Transformations ($\epsilon \in \Omega^0_{\mathrm{heq},\rho}(P; \eta(\mathfrak{g})) \subset \Omega^0_{\mathrm{heq},\rho}(P; \mathcal{N}_{\mathrm{Lie}}(A))$):} For every gauge parameter $\epsilon$ valued in the Lie nucleus $\mathcal{N}_{\mathrm{Lie}}(A)$ (and in particular for every vertical principal-bundle automorphism $\epsilon = \eta(\tilde{\epsilon})$ with $\tilde{\epsilon} \in \Omega^0_{\mathrm{heq},\mathrm{Ad}}(P; \mathfrak{g})$), $\mathrm{Jac}_{[\cdot,\cdot]_*}(D, D, \epsilon) \equiv 0$ vanishes identically and $F_D$ transforms homogeneously ($\delta_\epsilon F_D = \frac{1}{2}[F_D, \epsilon]_*$) for \emph{every} algebra $(A, *)$.
        \item[(b)] \emph{Extended $A$-Gauge Transformations ($\epsilon \in \Omega^0_{\mathrm{heq},\rho}(P; A)$):} $F_D$ transforms homogeneously for all $(D, \epsilon) \in \mathrm{Conn}_{\rho,\eta}(P; A) \times \Omega^0_{\mathrm{heq},\rho}(P; A)$ on manifolds of dimension $n \ge 2$ if and only if $(A, *)$ is Lie-admissible ($\mathcal{N}_{\mathrm{Lie}}(A) = A$). Furthermore, if $\langle \cdot, \cdot \rangle_{A, \mathrm{sym}}$ is pointwise $\mathrm{ad}_{[\cdot,\cdot]_*}$-invariant, then
        \begin{equation}\label{eq:gauge_var_action_Jac}
        \delta_\epsilon S^A[D] = -\frac{1}{4}\int_M \big\langle j(F_D),\, j\big(\mathrm{Jac}_{[\cdot,\cdot]_*}(D, D, \epsilon)\big) \big\rangle_{A, \mathrm{sym}},
        \end{equation}
        which vanishes for all $(D, \epsilon)$ whenever $(A, *)$ is Lie-admissible, and conversely forces $\mathrm{Jac}_{[\cdot,\cdot]_*} \equiv 0$ on $A$ whenever $\langle \cdot, \cdot \rangle_{A, \mathrm{sym}}$ is tensorial ($C^\infty(M)$-bilinear), pointwise $\mathrm{ad}_{[\cdot,\cdot]_*}$-invariant, and non-degenerate and $n \ge 2$.
    \end{enumerate}
\end{enumerate}
\end{theorem}

\begin{proof}
By Proposition \ref{prop:Cartan_Ehresmann}(2), $F_D$, $d_D^{[\cdot,\cdot]_*}\dot{D}$, $\mathrm{Jac}_{[\cdot,\cdot]_*}(D,D,D)$, $\delta_\epsilon D$, and $\mathrm{Jac}_{[\cdot,\cdot]_*}(D,D,\epsilon)$ all lie in $\Omega^\bullet_{\mathrm{heq},\rho}(P; A)$ and descend via $j$ to $\Omega^\bullet(M; E_A)$.

For part (1), expanding $F_{D + t\dot{D}}$ to first order in $t$ and using $D \wedge_* \dot{D} + \dot{D} \wedge_* D = D \wedge_{[\cdot,\cdot]_*} \dot{D}$ gives \eqref{eq:curvature_variation} and \eqref{eq:action_first_variation}. For the tensorial pairing $\langle \Phi, \Psi \rangle_A = \kappa(\Phi \wedge \star_M \Psi)$, let $\Xi := j^{-1}(\star_M j(F_D)) \in \Omega^{n-2}_{\mathrm{heq},\rho}(P; A)$. On $P$, the graded Leibniz rule for the 1-form $\dot{D}$ gives $d(\kappa(\dot{D} \wedge \Xi)) = \kappa(d\dot{D} \wedge \Xi) - \kappa(\dot{D} \wedge d\Xi)$. On decomposable elements $D = x \otimes \alpha$, $\dot{D} = y \otimes \beta$ ($\alpha, \beta \in \Omega^1(P)$) and $\Xi = z \otimes \gamma$ ($\gamma \in \Omega^{n-2}(P)$), the Koszul sign $\beta \wedge \alpha \wedge \gamma = -\alpha \wedge \beta \wedge \gamma$ combines with $\mathrm{ad}_{[\cdot,\cdot]_*}$-invariance $\kappa([x,y]_*, z) = -\kappa(y, [x,z]_*)$ to give
\[
\kappa\big((D \wedge_{[\cdot,\cdot]_*} \dot{D}) \wedge \Xi\big) = \kappa([x,y]_*, z)\, \alpha \wedge \beta \wedge \gamma = +\kappa(y, [x,z]_*)\, \beta \wedge \alpha \wedge \gamma = +\kappa\big(\dot{D} \wedge (D \wedge_{[\cdot,\cdot]_*} \Xi)\big).
\]
Applying Stokes' theorem on the closed manifold $M$ yields $\delta_{\dot{D}} S^A[D] = 2 \int_M \kappa(j(\dot{D}) \wedge d_{\nabla_D}^{[\cdot,\cdot]_*} (\star_M j(F_D)))$, which vanishes for all $\dot{D} \in \Omega^1_{\mathrm{heq},\rho}(P; A)$ if and only if \eqref{eq:A_YM_EL} holds.

For part (2), since $d^2 D = 0$ and $d(D \wedge_{[\cdot,\cdot]_*} D) = 2\, dD \wedge_{[\cdot,\cdot]_*} D = -2\, D \wedge_{[\cdot,\cdot]_*} dD$, we compute:
\begin{align*}
d_D^{[\cdot,\cdot]_*} F_D &= d\Big(dD + \frac{1}{4} D \wedge_{[\cdot,\cdot]_*} D\Big) + \frac{1}{2} D \wedge_{[\cdot,\cdot]_*} \Big(dD + \frac{1}{4} D \wedge_{[\cdot,\cdot]_*} D\Big) \\
&= -\frac{1}{2} D \wedge_{[\cdot,\cdot]_*} dD + \frac{1}{2} D \wedge_{[\cdot,\cdot]_*} dD + \frac{1}{8} D \wedge_{[\cdot,\cdot]_*} (D \wedge_{[\cdot,\cdot]_*} D) = -\frac{1}{8}\, (D \wedge_{[\cdot,\cdot]_*} D) \wedge_{[\cdot,\cdot]_*} D.
\end{align*}
Since $(D \wedge_{[\cdot,\cdot]_*} D)_p(u, w) = 2[D_p(u), D_p(w)]_*$, evaluating the 3-form $((D \wedge_{[\cdot,\cdot]_*} D) \wedge_{[\cdot,\cdot]_*} D)(v_1, v_2, v_3)$ on tangent vectors $v_1, v_2, v_3 \in T_p P$ via \eqref{eq:wedge_star_pointwise} sums the terms $2[[D(v_{\sigma(1)}), D(v_{\sigma(2)})]_*, D(v_{\sigma(3)})]_*$ with prefactor $\frac{1}{2!\, 1!}$ over $\sigma \in S_3$. Grouping the $6$ permutations into the $3$ cyclic pairs $(123), (231), (312)$ gives $2\,\mathrm{Jac}_{[\cdot,\cdot]_*}(D(v_1), D(v_2), D(v_3)) = \frac{1}{3}\mathrm{Jac}_{[\cdot,\cdot]_*}(D, D, D)(v_1, v_2, v_3)$ by \eqref{eq:Jac_form_normalization}, proving \eqref{eq:Bianchi_Jacobiator}. When $n \ge 3$, choosing a local trivialization of $P$ over an open chart $U \subset M$ around $x \in M$ and evaluating \eqref{eq:Bianchi_Jacobiator} on horizontal tangent vectors $v_1, v_2, v_3 \in T_p P$ projecting to linearly independent vectors in $T_x M$ with $D_p(v_i) = y_i \in A$ arbitrary shows that $\mathrm{Jac}_{[\cdot,\cdot]_*}(D, D, D) = 0$ for all $D$ if and only if $\mathrm{Jac}_{[\cdot,\cdot]_*}(y_1, y_2, y_3) = 0$ for all $y_1, y_2, y_3 \in A$.

For part (3), substituting $\dot{D} = \delta_\epsilon D = d\epsilon + \frac{1}{2}[D, \epsilon]_*$ into \eqref{eq:curvature_variation} and using the graded Leibniz identity $d(\frac{1}{2}[D, \epsilon]_*) = \frac{1}{2}[dD, \epsilon]_* - \frac{1}{2} D \wedge_{[\cdot,\cdot]_*} d\epsilon$ (since $D$ has degree $1$ and $\epsilon$ has degree $0$) gives
\begin{align*}
\delta_\epsilon F_D &= d\Big(d\epsilon + \frac{1}{2}[D, \epsilon]_*\Big) + \frac{1}{2} D \wedge_{[\cdot,\cdot]_*} \Big(d\epsilon + \frac{1}{2}[D, \epsilon]_*\Big) \\
&= \frac{1}{2}[dD, \epsilon]_* - \frac{1}{2} D \wedge_{[\cdot,\cdot]_*} d\epsilon + \frac{1}{2} D \wedge_{[\cdot,\cdot]_*} d\epsilon + \frac{1}{4} D \wedge_{[\cdot,\cdot]_*} [D, \epsilon]_* \\
&= \frac{1}{2}\Big[F_D - \frac{1}{4} D \wedge_{[\cdot,\cdot]_*} D,\, \epsilon\Big]_* + \frac{1}{4} D \wedge_{[\cdot,\cdot]_*} [D, \epsilon]_*.
\end{align*}
Evaluating $\frac{1}{4} D \wedge_{[\cdot,\cdot]_*} [D, \epsilon]_* - \frac{1}{8}[D \wedge_{[\cdot,\cdot]_*} D, \epsilon]_*$ on $(v_1, v_2)$ gives $-\frac{1}{4}\, \mathrm{Jac}_{[\cdot,\cdot]_*}(D(v_1), D(v_2), \epsilon) = -\frac{1}{8}\, \mathrm{Jac}_{[\cdot,\cdot]_*}(D, D, \epsilon)(v_1, v_2)$ by \eqref{eq:Jac_form_normalization}, proving \eqref{eq:inf_gauge_var_F}. Whenever $\epsilon \in \Omega^0_{\mathrm{heq},\rho}(P; \mathcal{N}_{\mathrm{Lie}}(A))$, equation \eqref{eq:Lie_nucleus} yields $\mathrm{Jac}_{[\cdot,\cdot]_*}(D, D, \epsilon) = 0$, proving (3a). In general, substituting \eqref{eq:inf_gauge_var_F} into $\delta_\epsilon S^A[D] = 2\int_M \langle j(F_D), j(\delta_\epsilon F_D)\rangle_{A,\mathrm{sym}}$ and using pointwise $\mathrm{ad}_{[\cdot,\cdot]_*}$-invariance $\langle [j(F_D), j(\epsilon)]_*, j(F_D)\rangle_{A,\mathrm{sym}} = 0$ yields \eqref{eq:gauge_var_action_Jac}.

Finally, to prove the converse in (3b) rigorously when $\langle \cdot, \cdot \rangle_{A, \mathrm{sym}}$ is tensorial ($C^\infty(M)$-bilinear), pointwise $\mathrm{ad}_{[\cdot,\cdot]_*}$-invariant, and non-degenerate and $n \ge 2$, suppose \eqref{eq:gauge_var_action_Jac} vanishes identically for all $D \in \mathrm{Conn}_{\rho,\eta}(P; A)$ and $\epsilon \in \Omega^0_{\mathrm{heq},\rho}(P; A)$. Because $\langle \cdot, \cdot \rangle_{A, \mathrm{sym}}$ is $C^\infty(M)$-bilinear and non-degenerate, it is given by a fiberwise non-degenerate symmetric bilinear bundle form $(\cdot, \cdot)_{A, x} : (\Lambda^2 T^*_x M \otimes (E_A)_x) \times (\Lambda^2 T^*_x M \otimes (E_A)_x) \to \mathrm{Dens}_x(M)$ at every point $x \in M$. Replacing $\epsilon$ by $f \epsilon$ for arbitrary $f \in C^\infty_c(M)$ and using $C^\infty(M)$-linearity $\mathrm{Jac}_{[\cdot,\cdot]_*}(D, D, f\epsilon) = f\,\mathrm{Jac}_{[\cdot,\cdot]_*}(D, D, \epsilon)$ implies that the pointwise pairing $\big(j(F_D)_{x_0},\, j(\mathrm{Jac}_{[\cdot,\cdot]_*}(D, D, \epsilon))_{x_0}\big)_{A, x_0} = 0$ vanishes identically at every point $x_0 \in M$ for all $D \in \mathrm{Conn}_{\rho,\eta}(P; A)$ and all $\epsilon \in \Omega^0_{\mathrm{heq},\rho}(P; A)$. Now fix $x_0 \in M$, a coordinate chart $(U, x^1, \dots, x^n)$ centered at $x_0$ ($x^\mu(x_0) = 0$), a basepoint connection $D_0 = \eta \circ \omega_0 \in \mathrm{Conn}_{\rho,\eta}(P; A)$, and a local section $s_0 : U \to P$ chosen in synchronous gauge at $x_0$ so that $(s_0^* D_0)_{x_0} = 0$. Writing $D = D_0 + \mathcal{A}$ with $\mathcal{A} \in \Omega^1_{\mathrm{heq},\rho}(P; A)$, given any $y_1, y_2, z \in A$ and any 2-form $\Psi_0 = \frac{1}{2}\sum_{\mu,\nu=1}^n \Psi_{\mu\nu}\, dx^\mu \wedge dx^\nu \in \Lambda^2 T^*_{x_0} M \otimes A$ ($\Psi_{\mu\nu} = -\Psi_{\nu\mu} \in A$), choose $\mathcal{A} \in \Omega^1_{\mathrm{heq},\rho}(P; A)$ such that on $U$,
\[
s_0^* \mathcal{A} = y_1\, dx^1 + y_2\, dx^2 + \frac{1}{2}\sum_{\mu,\nu=1}^n x^\mu \Psi_{\mu\nu}\, dx^\nu,
\]
and choose $\epsilon \in \Omega^0_{\mathrm{heq},\rho}(P; A)$ with $(s_0^*\epsilon)(x_0) = z$. Evaluating at $x_0$ (where $x^\mu(x_0) = 0$ and $(s_0^* D_0)_{x_0} = 0$) gives $(s_0^* D)_{x_0} = (s_0^* \mathcal{A})_{x_0} = y_1\, dx^1 + y_2\, dx^2$, while $(s_0^* d\mathcal{A})_{x_0} = \Psi_0$ can be chosen arbitrarily so that $(s_0^* F_D)_{x_0} = (s_0^* dD_0)_{x_0} + \Psi_0 + \frac{1}{2}[y_1, y_2]_*\, dx^1 \wedge dx^2$ equals any prescribed element of $\Lambda^2 T^*_{x_0} M \otimes A$. By pointwise non-degeneracy of $(\cdot, \cdot)_{A, x_0}$, it follows that $j(\mathrm{Jac}_{[\cdot,\cdot]_*}(D, D, \epsilon))_{x_0} = 2\,\mathrm{Jac}_{[\cdot,\cdot]_*}(y_1, y_2, z)\, dx^1 \wedge dx^2 = 0$, so $\mathrm{Jac}_{[\cdot,\cdot]_*}(y_1, y_2, z) = 0$ for all $y_1, y_2, z \in A$.
\end{proof}

\section{Algebraic and Polynomial Obstructions on \texorpdfstring{$A$}{A}-YMT Theories}\label{sec:algebraic_obstructions}

Substituting $F_D = dD + \frac{1}{2} D \wedge_* D$ into the $A$-YMT action functional \eqref{eq:A_YMT_action} yields the expansion
\begin{equation}\label{eq:YMT_expansion}
S^A[D] = S^A_{\mathrm{free}}[D] + S^A_{\mathrm{cubic}}[D] + S^A_{\mathrm{quartic}}[D],
\end{equation}
where $S^A_{\mathrm{free}}[D] := \int_M \langle j(dD), j(dD) \rangle_A$, $S^A_{\mathrm{cubic}}[D] := \int_M \langle j(dD), j(D \wedge_* D) \rangle_{A, \mathrm{sym}}$, and $S^A_{\mathrm{quartic}}[D] := \frac{1}{4} \int_M \langle j(D \wedge_* D), j(D \wedge_* D) \rangle_A$ (in the linear/horizontal setting $\eta = 0$, or relative to a basepoint connection $D_0$ in the affine setting).

\subsection{Commutator Projection, Abelianization, and Subbundle Localization}

By \eqref{eq:odd_form_comm_projection}, for any 1-form $D \in \Omega^1(P; A)$, the binary product $D \wedge_* D = \frac{1}{2} D \wedge_{[\cdot,\cdot]_*} D$ depends solely on the commutator bracket $[x,y]_* = x*y - y*x$. We record the immediate consequences of this projection in a single proposition.

\begin{proposition}[Abelianization, Derived Filtration Shift, and Subbundle Localization]\label{prop:abelianization}
Let $\pi : P \to M$ be a principal $G$-bundle over an $n$-manifold $M$ ($n \ge 2$) and let $\rho : G \to \mathrm{Aut}_{\mathrm{Alg}}(A)$ be an algebra automorphism representation on an $\mathfrak{m}$-graded algebra $A = \bigoplus_{m \in \mathfrak{m}} A^m$.
\begin{enumerate}
    \item \textbf{Abelianization:} If $A$ is a weak $(1,1)$-solvable $\mathfrak{m}$-graded algebra (i.e., $A = \bigoplus_{i=1}^r V_i$ decomposes into mutually orthogonal commutative graded submodules), or if $A = \bigoplus_{i=1}^r A_i$ is a vector space direct sum of commutative subalgebras and $D \in \mathcal{C}^1_{\mathrm{split}}(P; A)$ is algebraically split-adapted, then $D \wedge_* D = 0$ identically, reducing $F_D = dD$ and $S^A[D] = S^A_{\mathrm{free}}[D]$ to a free abelian theory in all dimensions $n \ge 2$.
    \item \textbf{Derived Filtration Shift:} If $A$ is a solvable algebra of derived length $s$ ($A^{(s+1)} = 0$), then for every $D \in \Omega^1(P; A)$, $D \wedge_* D \in \Omega^2(P; A^{(1)})$, and the iterated squares $(D \wedge_* D)^{[m]} \in \Omega^{2^{m+1}}(P; A^{(m+1)})$ vanish identically at step $m = s$.
    \item \textbf{Commutator Subbundle Localization:} Because $[A, A]_* \subset A^{(1)}$ is $\mathrm{Aut}_{\mathrm{Alg}}(A)$-invariant, $E_{[A,A]_*} = P \times_\rho [A, A]_* \subset E_{A^{(1)}} \subset E_A$ is a canonical vector subbundle. For every $D \in \mathrm{Conn}^{\mathrm{lin}}(P; A)$, $j(D \wedge_* D) \in \Omega^2(M; E_{[A,A]_*})$, so $S^A_{\mathrm{quartic}}[D]$ vanishes identically whenever $\Omega^2(M; E_{[A,A]_*})$ lies in the radical of $\langle \cdot, \cdot \rangle_A$.
\end{enumerate}
\end{proposition}

\begin{proof}
In part (1), writing $D = \sum_{i=1}^r \sum_{m \in \mathfrak{m}} D_i^m$ with $D_i^m \in \Omega^1(P; V_i^m)$, orthogonality $V_i * V_j = 0$ ($i \neq j$) and weak $(1,1)$-nilpotency $D_i^{m_1} \wedge_* D_i^{m_2} = 0$ give $D \wedge_* D = 0$ (and similarly for $D \in \mathcal{C}^1_{\mathrm{split}}(P; A)$). Part (2) follows by induction from $A^{(m)} * A^{(m)} = A^{(m+1)}$. For part (3), $(D \wedge_* D)_p(v_1, v_2) = [D_p(v_1), D_p(v_2)]_* \in [A, A]_* \subset A^{(1)}$, which descends via $j$ to $\Omega^2(M; E_{[A,A]_*})$.
\end{proof}

\subsection{Obstructions on Internal Multiplicative, Self-Dual, and Topological \texorpdfstring{$A$}{A}-YMT Theories}

We now examine internal multiplicative $A$-YMT theories \eqref{eq:multiplicative_YMT} and higher-degree internal topological (Chern-Weil) $A$-YMT theories \eqref{eq:Chern_Weil_YMT}, in which the pairing on $F_D$ is constructed from the internal product $\wedge_*$ of $\Omega(P; A)$ and a bundle trace functional $\mathrm{tr}_\xi : E_A \to M \times \mathbb{R}$ (Remark \ref{rem:internal_vs_external_CW}). By \eqref{eq:even_form_jordan_projection}, the exterior square of the 2-form $F_D \wedge_* F_D = F_D \wedge_{\circ_*} F_D$ projects onto the symmetric Jordan product $x \circ_* y = \frac{1}{2}(x*y + y*x)$, so $(2,s)$-solvability of $(A, *)$ yields direct dimensional and self-duality obstructions.

\begin{theorem}[Obstruction to Internal-Trace Topological and Self-Dual $A$-YMT Theories]\label{thm:topological_obstruction}
Let $\pi : P \to M$ be a principal $G$-bundle over an oriented $n$-manifold $M$, let $\rho : G \to \mathrm{Aut}_{\mathrm{Alg}}(A)$ be an automorphism representation on a weak $\mathcal{P}$-$(2, s)$-solvable $\mathfrak{m}$-graded algebra $A = \bigoplus_{m \in \mathfrak{m}} A^m$ (for $\mathcal{P} \in \{\mathrm{full}, \mathrm{bal}, \mathrm{blk}\}$), and let $\mathrm{tr}_\xi : E_A \to M \times \mathbb{R}$ be any bundle trace functional induced by $\xi \in \Gamma(E_A^*)$.
\begin{enumerate}
    \item \textbf{Higher-Dimensional Internal Chern-Weil Obstruction:} Suppose $n = 2r$ is even. If $n \ge 2(s + 1)$ (i.e., $r \ge s + 1$), then for every admissible connection $D \in \mathrm{Conn}_{\rho,\eta}(P; A)$ (including $\mathfrak{m}$-graded connections of Definition \ref{def:A_connection}(3); or for every homogeneous connection $D \in \Omega^1(P; A^m)$ with $m + m = m$ when $A$ is only homogeneous weak $\mathcal{P}$-$(2,s)$-solvable), $\wedge_{*, \mathcal{P}}^r F_D = 0$ identically in $\Omega^{2r}(P; A)$, and hence the $r$-th internal topological $A$-YMT functional
    \[
    S^A_{\mathrm{top}, r, \mathcal{P}}[D] = \int_M \mathrm{tr}_\xi\Big( j\big(\wedge_{*, \mathcal{P}}^r F_D\big) \Big)
    \]
    vanishes identically on $\mathrm{Conn}_{\rho,\eta}(P; A)$ for every trace section $\xi \in \Gamma(E_A^*)$.
    \item \textbf{Four-Dimensional Internal Topological and Self-Dual Obstruction ($s = 1$):} Suppose $s = 1$ (so $A$ is weak $(2,1)$-solvable, which by \eqref{eq:even_form_jordan_projection} includes every skew-commutative algebra since $F_D \wedge_* F_D = F_D \wedge_{\circ_*} F_D = 0$). Then for every $D \in \Omega^1(P; A)$, $F_D \wedge_* F_D = 0$ in $\Omega^4(P; A)$. Consequently, in dimension $n = 4$:
    \begin{enumerate}
        \item[(a)] The 4D internal topological Yang-Mills functional $S^A_{\mathrm{top}, 2}[D] = \int_M \mathrm{tr}_\xi(j(F_D \wedge_* F_D))$ is identically zero;
        \item[(b)] Any internal multiplicative 4D $A$-YMT functional $S^A_{\hat{\star}}[D] = \int_M \mathrm{tr}_\xi(j(F_D \wedge_* \hat{\star} F_D))$ vanishes identically on all $\hat{\star}$-self-dual and $\hat{\star}$-anti-self-dual connections ($\hat{\star} F_D = \pm F_D$).
    \end{enumerate}
\end{enumerate}
\end{theorem}

\begin{proof}
Decomposing $A = \bigoplus_{i=1}^k W_i$ into mutually orthogonal weak $\mathcal{P}$-$(2, s)$-nilpotent graded submodules $W_i = \bigoplus_{m \in \mathfrak{m}} W_i^m$ ($W_i * W_j = 0$ for $i \neq j$), we write $F_D = \sum_{i=1}^k F_{D, i}$ with $F_{D, i} = \sum_{m \in \mathfrak{m}} F_{D, i}^m \in \Omega^2(P; W_i)$ (where $F_{D,i}^m$ is given by \eqref{eq:graded_curvature_components}). Expanding $\wedge_{*, \mathcal{P}}^r F_D$ and using $W_i * W_j = 0$ ($i \neq j$) yields $\sum_{i=1}^k \sum_{m_1, \dots, m_r \in \mathfrak{m}} p_r(F_{D, i}^{m_1}, \dots, F_{D, i}^{m_r})$, where each $\mathcal{P}$-parenthesized monomial of degree $r \ge s + 1$ vanishes by Definition \ref{def:ks_solv}(4). Thus $\wedge_{*, \mathcal{P}}^r F_D = 0$ for all $r \ge s + 1$, proving (1). Setting $s = 1$ and $r = 2$ gives $F_D \wedge_* F_D = 0$, which immediately implies (2a) and (2b).
\end{proof}

Finally, given a surjective $R$-linear projection $Q : E \twoheadrightarrow A$ and an $R$-linear section $S : A \hookrightarrow E$ with $Q \circ S = \mathrm{id}_A$ (for instance, a split short exact sequence $0 \to H \to E \stackrel{Q}{\to} A \to 0$), endowing $E$ with the $(Q, S)$-pulled-back product $x *_{Q,S} y := S(Q(x) * Q(y))$ yields the following invariance property.

\begin{proposition}[Invariance under $(Q,S)$-Splitting Extensions]\label{prop:splitting_invariance}
For $x *' y := S(Q(x) * Q(y))$ on $E$ with $Q \circ S = \mathrm{id}_A$, any non-associative monomial $p_r$ satisfies the relation $p_r^{*'}(\alpha_1, \dots, \alpha_r) = (\Omega S)\big(p_r^*((\Omega Q)(\alpha_1), \dots, (\Omega Q)(\alpha_r))\big)$ on $\Omega(P; E)$. Since $\Omega S$ is injective and $\Omega Q$ is surjective, $(E, *')$ is $\mathcal{P}$-$(k,s)$-nil if and only if $(A, *)$ is $\mathcal{P}$-$(k,s)$-nil.
\end{proposition}

\begin{proof}
Follows by induction on $r \ge 2$ from $\alpha \wedge_{*'} \beta = (\Omega S)((\Omega Q)(\alpha) \wedge_* (\Omega Q)(\beta))$ and $(\Omega Q) \circ (\Omega S) = \mathrm{id}_{\Omega(P; A)}$.
\end{proof}

\section{Parenthesization Cutoffs and Holonomy Reductions in Reductive Cartan-Yang-Mills Theories}\label{sec:holonomy_obstructions}

In approaches that unify gravity and Yang-Mills gauge theory---such as MacDowell-Mansouri gravity \cite{WiseMacDowell, Wise} and Cartan-Yang-Mills theories---one works with a reductive $A$-connection $\nabla = e + \omega \in \Omega^1(P; A_0) \oplus \Omega^1(P; A_1)$ and uses the soldering 1-form $e$ to construct an internal constitutive operator on the Cartan curvature $F_\nabla$. We now analyze the dimensional cutoffs of reductive Cartan-Yang-Mills theories under both quartic-block parenthesization \eqref{eq:quartic_block_power} and recursive binary-tree parenthesization \eqref{eq:balanced_power}, establish the 4D first-Bianchi contraction identity and proportionality between left-grouped and right-nested EHP terms, pair the dimensional cutoffs with Berger's holonomy classification in dimensions $4 \le n \le 7$, and establish the representation-theoretic dichotomy at the critical $G_2$ dimension $n = k = 7$.

\subsection{\texorpdfstring{$H$}{H}-Equivariant Reductive Soldering Products and \texorpdfstring{$k$}{k}-Proper Bundles}

Let us establish the algebraic and geometric setup:
\begin{itemize}
    \item Let $\mathfrak{h} \subset \mathfrak{so}(k_1) \oplus \dots \oplus \mathfrak{so}(k_r) \subset \mathfrak{so}(k)$ ($k_1 + \dots + k_r = k$) be a matrix Lie subalgebra equipped with the skew-commutative product $X \curlywedge Y := \frac{1}{2}(XY - YX) = \frac{1}{2}[X, Y]$ in $\mathfrak{h}$ (a \emph{$k$-algebra}, with connected \emph{$k$-group} $H \subset SO(k_1) \times \dots \times SO(k_r)$) \cite{MartinsBiezunerEHP}.
    \item On the reductive direct sum $A = A_0 \oplus A_1 = \mathbb{R}^k \oplus \mathfrak{h}$ (where $H$ acts on $A_0 = \mathbb{R}^k$ via the defining vector representation and on $A_1 = \mathfrak{h}$ via $\mathrm{Ad}_H$), we identify $\mathbb{R}^k \wedge \mathbb{R}^k \simeq \mathfrak{so}(k)$ via $(u \wedge v)_{ab} := u_a v_b - u_b v_a$ for $u, v \in \mathbb{R}^k$, and let $\sigma : \mathbb{R}^k \wedge \mathbb{R}^k \to \mathfrak{h}$ be the canonical $H$-equivariant skew-symmetric soldering projection defined by the orthogonal projection $\mathfrak{so}(k) \twoheadrightarrow \mathfrak{h}$ with respect to the positive-definite inner product $\langle X, Y \rangle_{\mathfrak{so}(k)} := -\frac{1}{2}\mathrm{Tr}(XY) = \frac{1}{2}\sum_{a,b=1}^k X_{ab} Y_{ab}$. Explicitly, for any $\langle \cdot, \cdot \rangle_{\mathfrak{so}(k)}$-orthonormal basis $\{T_A\}$ of $\mathfrak{h}$,
    \begin{equation}\label{eq:sigma_basis_def}
    \sigma(u \wedge v) = \sum_A \langle T_A,\, u \wedge v \rangle_{\mathfrak{so}(k)} T_A = \sum_A \Big(\sum_{a,b=1}^k (T_A)_{ab} u_a v_b\Big) T_A,
    \end{equation}
    so that for $\mathbb{R}^k$-valued forms $\alpha \in \Omega^p(P; \mathbb{R}^k)$ and $\beta \in \Omega^q(P; \mathbb{R}^k)$, the $\mathfrak{h}$-valued $(p+q)$-form $\sigma(\alpha \wedge \beta) = \sum_A (\sum_{a,b=1}^k (T_A)_{ab}\, \alpha^a \wedge \beta^b) T_A$ is the orthogonal projection of the skew-symmetric matrix $(\alpha \wedge \beta)_{ab} = \alpha_a \wedge \beta_b - \alpha_b \wedge \beta_a = \alpha_a \wedge \beta_b - (-1)^{pq}\beta_a \wedge \alpha_b \in \Omega^{p+q}(P; \mathfrak{so}(k))$. Let $\times_H : \mathbb{R}^k \wedge \mathbb{R}^k \to \mathbb{R}^k$ be an $H$-equivariant skew-symmetric bilinear vector product on $\mathbb{R}^k$ (for irreducible $H \subset SO(k)$, $\times_H = 0$ except when $k = 7$ and $H = G_2 \subset SO(7)$, where $(u \times_{G_2} v)^a = \varphi^a_{bc} u^b v^c$ is the octonionic cross product induced by the associative $G_2$-invariant 3-form $\varphi_0 \in (\Lambda^3(\mathbb{R}^7)^*)^{G_2}$ \cite{BaezOctonions, Salamon}). We define the \emph{$H$-equivariant reductive Cartan-MacDowell-Mansouri product} $\curlywedge_\rtimes$ on $A = \mathbb{R}^k \oplus \mathfrak{h}$ by
    \begin{equation}\label{eq:MM_product}
    (v_1, X_1) \curlywedge_\rtimes (v_2, X_2) := \Big( \frac{1}{2}(X_1 v_2 - X_2 v_1) + v_1 \times_H v_2,\; X_1 \curlywedge X_2 + \sigma(v_1 \wedge v_2) \Big),
    \end{equation}
    where $(X v)^a := X^a{}_b v^b = \sum_{b=1}^k X_{ab} v^b$ denotes the standard matrix action of $X \in \mathfrak{h} \subset \mathfrak{so}(k)$ on $v \in \mathbb{R}^k$.
    The product \eqref{eq:MM_product} satisfies three key properties:
    \begin{enumerate}
        \item[(P1)] \emph{$H$-Equivariance:} $H$ acts on $(A, \curlywedge_\rtimes)$ by algebra automorphisms, inducing the associated algebra bundle $E_A = P \times_H (\mathbb{R}^k \oplus \mathfrak{h}) \to M$.
        \item[(P2)] \emph{Skew-Commutativity, Parenthesization, and Rank Identities:} Because \eqref{eq:MM_product} is skew-commutative on $A = \mathbb{R}^k \oplus \mathfrak{h}$, $(A, \curlywedge_\rtimes)$ is $(2,1)$-nil ($\Phi \curlywedge_\rtimes \Phi = 0$ for all $\Phi \in \Omega^{\mathrm{even}}(P; A)$) and block-balanced $(1,3)$-nil ($\curlywedge_{\rtimes, \mathrm{blk}}^m \alpha = 0$ for all $\alpha \in \Omega^1(P; A)$ and $m \ge 4$ by Lemma \ref{lem:skew_comm_powers}(1)). Furthermore, because any $A_0$-valued soldering 1-form $e = (e, 0) \in \Omega^1(P; \mathbb{R}^k)$ takes values in the $k$-dimensional subspace $A_0 = \mathbb{R}^k$, Lemma \ref{lem:skew_comm_powers}(2c) implies that $\curlywedge_{\rtimes, \mathcal{P}}^m e = 0$ identically on $P$ for all $m > k$. Moreover, when $\times_H = 0$, $A = A_0 \oplus A_1$ is $\mathbb{Z}_2$-graded ($A_0 \curlywedge_\rtimes A_0 \subset A_1$, $A_1 \curlywedge_\rtimes A_0 \subset A_0$, $A_1 \curlywedge_\rtimes A_1 \subset A_1$), so $e \curlywedge_\rtimes e = (0, \sigma(e \wedge e)) \in \Omega^2(P; \mathfrak{h})$ lies in $A_1 = \mathfrak{h}$ and $\curlywedge_{\rtimes, \mathrm{bal}}^3 e = (e \curlywedge_\rtimes e) \curlywedge_\rtimes e \in \Omega^3(P; \mathbb{R}^k)$ lies in $A_0 = \mathbb{R}^k$.
        \item[(P3)] \emph{Classical Cartan Torsion on $A_0 = \mathbb{R}^k$:} For $\omega \in \Omega^1(P; \mathfrak{h})$ and $e \in \Omega^1(P; \mathbb{R}^k)$, the generalized torsion in \eqref{eq:Cartan_curvature} is $\Theta_{e,\omega} = (de + \frac{1}{2}\omega \wedge e, 0)$ (since by \eqref{eq:wedge_star_pointwise} and \eqref{eq:MM_product}, $(\omega \curlywedge_\rtimes e + e \curlywedge_\rtimes \omega)_p(v_1, v_2) = 2\,\omega_p(v_1)\cdot \frac{1}{2}e_p(v_2) - 2\,\omega_p(v_2)\cdot \frac{1}{2}e_p(v_1) = (\omega \wedge e)_p(v_1, v_2)$), which vanishes if and only if $\tilde{\omega} := \frac{1}{2}\omega$ satisfies the classical Cartan torsion-free equation $de + \tilde{\omega} \wedge e = 0$.
    \end{enumerate}
    \item Let $N$ be a simply connected smooth $k$-manifold ($k \le n$). Following \cite{MartinsBiezunerEHP}, a principal $H$-bundle $\pi : P \to M$ over an $n$-manifold $M$ is called \emph{$k$-proper} if there exists an immersed simply connected $k$-manifold $i : N \hookrightarrow M$ whose orthonormal frame bundle $FN$ reduces to an $H$-subbundle $P_N \subset FN$ isomorphic as a principal $H$-bundle over $N$ to $i^* P$ via $\kappa : P_N \xrightarrow{\sim} i^* P$, and we define the \emph{partially torsionless irreducible configuration space} $\mathrm{Conn}^{\Theta_N = 0, \mathrm{irr}}_A(P)$ to be the space of pairs of horizontal $H$-equivariant soldering 1-forms $e \in \Omega^1_{\mathrm{heq}}(P; \mathbb{R}^k)$ and Ehresmann connections $\tilde{\omega} = \frac{1}{2}\omega \in \mathrm{Conn}(P; \mathfrak{h})$ whose pullback along $i \circ \kappa : P_N \to P$ satisfies the Cartan torsion-free equation $(i \circ \kappa)^*(de + \tilde{\omega} \wedge e) = 0$ on $P_N$ (and $de + \tilde{\omega} \wedge e = 0$ globally on $P$ whenever $k = n$ and $M = N$, in which case we write $\mathrm{Conn}^{\Theta = 0, \mathrm{irr}}_A(P)$) such that $(i \circ \kappa)^* e \in \Omega^1(P_N; \mathbb{R}^k)$ is the canonical orthonormal soldering 1-form on $P_N \subset FN$ inducing a locally irreducible, locally non-symmetric Riemannian metric $g_N$ on $N$ (so that $(N, g_N)$ is a \emph{Berger $k$-manifold} \cite{Berger, Besse, Olmos} with Levi-Civita connection $\tilde{\omega}_N := \frac{1}{2}(i \circ \kappa)^*\omega$ and curvature $\Omega_{\tilde{\omega}_N} = \frac{1}{2}(i \circ \kappa)^*\Omega_\omega$).
\end{itemize}

\begin{remark}[Geometric Realizations of $k$-Proper Bundles and the Role of Mixed Torsion when $k < n$]\label{rem:k_proper_foliated}
When $k = n$ and $M = N$ is simply connected, $P \subset FM$ is an $H$-structure on $M$ with globally torsion-free Levi-Civita connection $\tilde{\omega} = \frac{1}{2}\omega$ ($\Theta_{e,\omega} = de + \tilde{\omega} \wedge e = 0$ globally on $P$) and orthonormal coframe $e$. When $k < n$, $e \in \Omega^1_{\mathrm{heq}}(P; \mathbb{R}^k)$ has a horizontal kernel distribution $\mathcal{V}^\perp = \ker(e) \subset TM$ of rank $n - k$. Suppose $\pi_M : M \to N$ is a smooth submersion (for instance, a Riemannian product $M = N^k \times K^{n-k}$) with $i : N \hookrightarrow M$ a section, $P = \pi_M^* P_N$, and $e = \pi_M^* e_N$. Any connection $\tilde{\omega} = \pi_M^* \tilde{\omega}_N + \beta$ with $\beta \in \Omega^1_{\mathrm{heq}}(P; \mathfrak{h})$ vanishing along $TN$ ($i_{\underline{v}}\beta = 0$ for $v \in TN$) satisfies the pulled-back torsion-free condition $(i \circ \kappa)^*(de + \tilde{\omega} \wedge e) = de_N + \tilde{\omega}_N \wedge e_N = 0$ on $P_N$.
Notice that on $M^n = N^k \times K^{n-k}$ with $e = \pi_M^* e_N$, $de = \pi_M^* de_N = -\pi_M^*(\tilde{\omega}_N \wedge e_N)$, so the full torsion on $M^n$ is $\Theta_{e,\omega} = de + \tilde{\omega}\wedge e = \beta \wedge \pi_M^* e_N$. Since $\beta$ is along $T^*K$, $e_N$ is an orthonormal coframe along $T^*N$, and $H \subset SO(k)$ acts faithfully on $\mathbb{R}^k$, imposing \emph{global} torsion-freeness $\Theta_{e,\omega} = 0$ on all of $M^n$ would force $\beta(w) \cdot e_N(v) = 0$ for all $w \in TK, v \in TN$ and hence $\beta = 0$, which would make both $e$ and $\Omega_\omega$ pullbacks from $N^k$ and force every $n$-form built from $(e, \Omega_\omega)$ to vanish identically for $n > k$ (since $\Lambda^n(T^* N^k) = 0$)!
Conversely, because $\mathrm{Conn}^{\Theta_N=0,\mathrm{irr}}_A(P)$ imposes torsion-freeness $(i \circ \kappa)^*\Theta_{e,\omega} = 0$ only along $N^k$ while allowing non-zero mixed torsion $\Theta_{e,\omega}(w, v) = \beta(w)\cdot e_N(v) \neq 0$ on $TK \times TN$, a non-zero horizontal 1-form $\beta$ along $TK$ contributes non-zero transverse curvature $\Omega_\omega$ along $T^*K \wedge T^*K$ (or $T^*K \wedge T^*N$). Since the two factors of $\Omega_\omega$ in the pure Yang-Mills $n$-form $\Omega_\omega \curlywedge_\rtimes ((\curlywedge_{\rtimes,\mathcal{P}}^{n-4} e)\curlywedge_\rtimes \Omega_\omega)$ have total degree $4$ and can saturate at most $4$ cotangent directions along $T^*K$ while $\curlywedge_{\rtimes,\mathcal{P}}^{n-4} e$ saturates $n - 4 \le k$ directions along $T^*N$, the pure Yang-Mills $n$-form is non-vanishing on $\mathrm{Conn}^{\Theta_N=0,\mathrm{irr}}_A(P)$ whenever $n - 4 \le k < n$.
\end{remark}

Let us clarify the relationship between the quadratic constitutive expansion of the full Cartan curvature $F_\nabla = \Omega_\omega + \Theta_{e,\omega} + E_e$ ($E_e = \frac{1}{2} e \curlywedge_\rtimes e$) in \eqref{eq:multiplicative_YMT} and the canonical left-grouped Cartan-Yang-Mills sectors. In a non-associative algebra $(A, \curlywedge_\rtimes)$, if one applies the internal constitutive operator $\hat{\star}_{\mathcal{P}}(\Phi) := (\curlywedge_{\rtimes, \mathcal{P}}^{n-4} e) \curlywedge_\rtimes \Phi$ directly to the torsionless Cartan curvature $F_\nabla^{\Theta=0} := \Omega_\omega + \frac{1}{2} e \curlywedge_\rtimes e$, expanding $F_\nabla^{\Theta=0} \curlywedge_\rtimes \hat{\star}_{\mathcal{P}}(F_\nabla^{\Theta=0})$ produces the quadratic Yang-Mills term $\Omega_\omega \curlywedge_\rtimes ((\curlywedge_{\rtimes, \mathcal{P}}^{n-4} e) \curlywedge_\rtimes \Omega_\omega)$, together with the cross-terms $\frac{1}{2}(e \curlywedge_\rtimes e) \curlywedge_\rtimes ((\curlywedge_{\rtimes, \mathcal{P}}^{n-4} e) \curlywedge_\rtimes \Omega_\omega) + \frac{1}{2}\Omega_\omega \curlywedge_\rtimes ((\curlywedge_{\rtimes, \mathcal{P}}^{n-4} e) \curlywedge_\rtimes (e \curlywedge_\rtimes e))$ and the cosmological term $\frac{1}{4}(e \curlywedge_\rtimes e) \curlywedge_\rtimes ((\curlywedge_{\rtimes, \mathcal{P}}^{n-4} e) \curlywedge_\rtimes (e \curlywedge_\rtimes e))$. Because $(A, \curlywedge_\rtimes)$ is non-associative, in order to place all three gravitational and gauge sectors on a uniform footing where the constitutive soldering factors of $e$ are grouped into a single monomial $\curlywedge_{\rtimes, \mathcal{P}}^m e$ prior to multiplying with $\Omega_\omega$ (matching the $A$-EHP convention of \cite{MartinsBiezunerEHP}), we define for $\mathcal{P} \in \{\mathrm{blk}, \mathrm{bal}\}$ in \eqref{eq:balanced_power}--\eqref{eq:quartic_block_power} the \emph{left-grouped Cartan-Yang-Mills Lagrangian $n$-form} $\mathcal{L}^{\mathrm{CYM}}_{n, \mathcal{P}} \in \Omega^n_{\mathrm{heq}}(P; A)$ by
\begin{equation}\label{eq:CYM_expansion}
\mathcal{L}^{\mathrm{CYM}}_{n, \mathcal{P}} := \Omega_\omega \curlywedge_\rtimes \Big(\big(\curlywedge_{\rtimes, \mathcal{P}}^{n-4} e\big) \curlywedge_\rtimes \Omega_\omega\Big) + \big(\curlywedge_{\rtimes, \mathcal{P}}^{n-2} e\big) \curlywedge_\rtimes \Omega_\omega + \frac{1}{4} \curlywedge_{\rtimes, \mathcal{P}}^n e,
\end{equation}
where the three terms are respectively the \emph{quadratic Yang-Mills term}, the \emph{left-grouped Einstein-Hilbert-Palatini (EHP) cross-term}, and the \emph{cosmological term}. (Notice that at $n = 6$, the ungrouped cross-term $\frac{1}{2}(e \curlywedge_\rtimes e) \curlywedge_\rtimes ((e \curlywedge_\rtimes e) \curlywedge_\rtimes \Omega_\omega)$ does not factor through $\curlywedge_{\rtimes, \mathrm{bal}}^4 e = 0$, whereas in \eqref{eq:CYM_expansion} each sector depends directly on $\curlywedge_{\rtimes, \mathcal{P}}^m e$ for $m \in \{n-4, n-2, n\}$.) Recall from Section \ref{sec:foundations} that when $n = 4$ ($n - 4 = 0$), our $m = 0$ convention $\big(\curlywedge_{\rtimes, \mathcal{P}}^0 e\big) \curlywedge_\rtimes \Omega_\omega := \Omega_\omega$ omits the $e$-factor, so the quadratic Yang-Mills term at $n = 4$ is simply $\Omega_\omega \curlywedge_\rtimes \Omega_\omega$, which vanishes identically by $(2,1)$-nilpotency of $(A, \curlywedge_\rtimes)$!

To compare the two natural parenthesizations of the EHP cross-term at $n = 4$, we also consider the \emph{right-nested EHP cross-term}
\begin{equation}\label{eq:right_nested_EHP}
\mathcal{L}^{\mathrm{EHP, right}}_4 := e \curlywedge_\rtimes (e \curlywedge_\rtimes \Omega_\omega) \in \Omega^4_{\mathrm{heq}}(P; A),
\end{equation}
as opposed to the left-grouped EHP cross-term $\mathcal{L}^{\mathrm{EHP, left}}_4 := (e \curlywedge_\rtimes e) \curlywedge_\rtimes \Omega_\omega$ in \eqref{eq:CYM_expansion}.

\begin{definition}[Cartan-Yang-Mills Functional]\label{def:CYM}
For $n \ge 4$, $\mathcal{P} \in \{\mathrm{blk}, \mathrm{bal}\}$, and any bundle trace functional $\mathrm{tr}_\xi : E_A \to M \times \mathbb{R}$ induced by $\xi \in \Gamma(E_A^*)$, the \emph{$\mathcal{P}$-balanced Cartan-Yang-Mills ($A$-CYM) functional} on $\mathrm{Conn}^{\Theta_N = 0, \mathrm{irr}}_A(P)$ is
\begin{equation}\label{eq:CYM_action}
S^{\mathrm{CYM}}_{n, \xi, \mathcal{P}}[e, \omega] := \int_M \mathrm{tr}_\xi\big( j(\mathcal{L}^{\mathrm{CYM}}_{n, \mathcal{P}}) \big).
\end{equation}
\end{definition}

\subsection{Dimensional Cutoffs, the 4D First-Bianchi Contraction Identity, and Berger Holonomy Reductions}

Before stating our main theorem on $S^{\mathrm{CYM}}_{n, \xi, \mathcal{P}}[e, \omega]$, we prove a fundamental geometric lemma establishing the exact proportionality between the left-grouped EHP 4-form $(e \curlywedge_\rtimes e) \curlywedge_\rtimes \Omega_\omega$ and the right-nested EHP 4-form $e \curlywedge_\rtimes (e \curlywedge_\rtimes \Omega_\omega)$, and showing that on 4-manifolds with $U(2)$ or $SU(2)$ holonomy both parenthesizations vanish identically on-shell (when torsion $\Theta_{e,\omega} = 0$) while remaining non-zero off-shell (when $\Theta_{e,\omega} \neq 0$).

\begin{lemma}[4D First-Bianchi Contraction Identity and Proportionality of Left-Grouped and Right-Nested EHP Terms]\label{lem:4D_Bianchi_dichotomy}
Let $n = k = 4$ and let $H \subset SO(4)$ be either $SU(2)_-$ (hyperk\"ahler holonomy) or $U(2) = (U(1)_+ \times SU(2)_-)/\mathbb{Z}_2$ (K\"ahler 4-manifold holonomy), under the self-dual/anti-self-dual Lie algebra decomposition $\mathfrak{so}(4) \simeq \mathfrak{su}(2)_+ \oplus \mathfrak{su}(2)_-$. Let $(e, \omega) \in \mathrm{Conn}^{\Theta = 0}_A(P)$ be any torsion-free pair ($de + \tilde{\omega}\wedge e = 0$, $\tilde{\omega} = \frac{1}{2}\omega$). Then:
\begin{enumerate}
    \item \textbf{On-Shell Vanishing of the Left-Grouped EHP Term $(e \curlywedge_\rtimes e) \curlywedge_\rtimes \Omega_\omega$:} Under \eqref{eq:MM_product}, $(e \curlywedge_\rtimes e) \curlywedge_\rtimes \Omega_\omega = (0, \frac{1}{2}[\sigma(e\wedge e), \Omega_\omega]) \in \Omega^4(P; [\mathfrak{h}, \mathfrak{h}])$. Because $de + \tilde{\omega} \wedge e = 0$ implies the algebraic first Bianchi identity $\Omega_\omega \wedge e = 0$, the curvature 2-form $\Omega_\omega$ has symmetric Riemann components $R_{ab} = R_{ba}$ in the standard basis $\{\omega_1^-, \omega_2^-, \omega_3^-\}$ of $\Lambda^2_- \mathbb{R}^4 \simeq \mathfrak{su}(2)_-$. Since $\omega_a^- \wedge \omega_b^- = -2\delta_{ab}\,\mathrm{vol}_4$ is also symmetric in $(a, b)$ while $[T_a, T_b] = \epsilon_{abc} T_c$ is skew-symmetric, we have
    \begin{equation}\label{eq:left_grouped_4D_vanishing}
    (e \curlywedge_\rtimes e) \curlywedge_\rtimes \Omega_\omega = \big(0,\, \frac{1}{2}[\sigma(e\wedge e), \Omega_\omega]\big) = 0 \quad \text{identically on } \mathrm{Conn}^{\Theta=0}_A(P)
    \end{equation}
    for both $H = SU(2)_-$ and $H = U(2)$. (Off-shell, i.e., for connections $\omega$ with non-zero torsion $\Theta_{e,\omega} \neq 0$ whose curvature matrix $R_{ab} \neq R_{ba}$ has a non-zero skew-symmetric part, $[\sigma(e\wedge e), \Omega_\omega] = -2\sum_{a,b} R_{ab}\epsilon_{abc} T_c\,\mathrm{vol}_4 \neq 0$ does not vanish.)
    \item \textbf{Exact Proportional Relation with the Right-Nested EHP Term $e \curlywedge_\rtimes (e \curlywedge_\rtimes \Omega_\omega)$:} For any subalgebra $\mathfrak{h} \subset \mathfrak{so}(k)$ with $\times_H = 0$, the right-nested parenthesization \eqref{eq:right_nested_EHP} satisfies
    \begin{equation}\label{eq:right_nested_4D_formula}
    e \curlywedge_\rtimes (e \curlywedge_\rtimes \Omega_\omega) = \Big(0,\; -\frac{1}{2}\sigma\big(e \wedge (\Omega_\omega \cdot e)\big)\Big) = +\frac{1}{2}\,(e \curlywedge_\rtimes e) \curlywedge_\rtimes \Omega_\omega = \Big(0,\; +\frac{1}{4}\big[\sigma(e\wedge e),\, \Omega_\omega\big]\Big).
    \end{equation}
    Consequently, at $n = k = 4$ with $H = SU(2)_-$ or $U(2)$, both $(e \curlywedge_\rtimes e) \curlywedge_\rtimes \Omega_\omega$ and $e \curlywedge_\rtimes (e \curlywedge_\rtimes \Omega_\omega)$ vanish identically on the torsion-free space $\mathrm{Conn}^{\Theta=0}_A(P)$ and are non-vanishing off-shell when $\Theta_{e,\omega} \neq 0$.
\end{enumerate}
\end{lemma}

\begin{proof}
Let us verify both parts directly. In any subalgebra $\mathfrak{h} \subset \mathfrak{so}(k)$ with $\times_H = 0$, for $e \in \Omega^1(P; \mathbb{R}^k)$ and $\Omega_\omega \in \Omega^2(P; \mathfrak{h})$, \eqref{eq:MM_product} gives $e \curlywedge_\rtimes \Omega_\omega = (-\frac{1}{2}\Omega_\omega \cdot e, 0) \in \Omega^3(P; \mathbb{R}^k \oplus \mathfrak{h})$, and multiplying on the left by $e = (e, 0)$ gives $e \curlywedge_\rtimes (e \curlywedge_\rtimes \Omega_\omega) = (0, -\frac{1}{2}\sigma(e \wedge (\Omega_\omega \cdot e)))$. For any basis $\{T_A\}$ of $\mathfrak{h} \subset \mathfrak{so}(k)$ orthonormal with respect to $\langle T_A, T_B \rangle_{\mathfrak{so}(k)} = -\frac{1}{2}\mathrm{Tr}(T_A T_B) = \delta_{AB}$, note that since $e^i$ is a 1-form and $(\Omega_\omega \cdot e)^j = (\Omega_\omega)_{jk} \wedge e^k$ is a 3-form, the $(i,j)$-entry of the skew-symmetric matrix $e \wedge (\Omega_\omega \cdot e) \in \Omega^4(P; \mathfrak{so}(k))$ is $e^i \wedge (\Omega_\omega \cdot e)^j - e^j \wedge (\Omega_\omega \cdot e)^i = e^i \wedge (\Omega_\omega \cdot e)^j + (\Omega_\omega \cdot e)^i \wedge e^j$. By \eqref{eq:sigma_basis_def}, the $T_A$-component of $-\frac{1}{2}\sigma(e \wedge (\Omega_\omega \cdot e))$ is therefore
\[
\frac{1}{4}\mathrm{Tr}\big(T_A \cdot (e \wedge (\Omega_\omega \cdot e)^T + (\Omega_\omega \cdot e) \wedge e^T)\big) = -\frac{1}{2} (T_A)_{ij} (\Omega_\omega)_{jk} \wedge e^i \wedge e^k = -\frac{1}{4} [T_A, \Omega_\omega]_{ik} \wedge e^i \wedge e^k,
\]
because the symmetric part $\frac{1}{2}(T_A \Omega_\omega + \Omega_\omega T_A)_{ik}$ vanishes when contracted with the skew-symmetric 2-form $e^i \wedge e^k$. On the other hand, since $\langle X, [Y, Z] \rangle_{\mathfrak{so}(k)} = -\langle [X, Z], Y \rangle_{\mathfrak{so}(k)}$ by $\mathrm{ad}$-invariance of $\langle \cdot, \cdot \rangle_{\mathfrak{so}(k)}$, \eqref{eq:sigma_basis_def} shows that the $T_A$-component of $+\frac{1}{4}[\sigma(e\wedge e), \Omega_\omega]$ is
\[
\frac{1}{4}\big\langle T_A,\, [\sigma(e\wedge e), \Omega_\omega] \big\rangle_{\mathfrak{so}(k)} = -\frac{1}{4}\big\langle [T_A, \Omega_\omega],\, \sigma(e\wedge e) \big\rangle_{\mathfrak{so}(k)} = -\frac{1}{4} [T_A, \Omega_\omega]_{ik} \wedge e^i \wedge e^k.
\]
Consequently, for \emph{every} subalgebra $\mathfrak{h} \subset \mathfrak{so}(k)$ with $\times_H = 0$, the associative matrix identity $(T_A)_{ij}(T_B)_{jk} e^i \wedge e^k = \frac{1}{2}[T_A, T_B]_{ik} e^i \wedge e^k$ implies \eqref{eq:right_nested_4D_formula}.

Now suppose $n = k = 4$ and $\mathfrak{h} \subset \mathfrak{u}(2) = \mathfrak{u}(1)_+ \oplus \mathfrak{su}(2)_-$. Since $\mathfrak{u}(1)_+$ is central in $\mathfrak{u}(2)$, the commutator $[\sigma(e\wedge e), \Omega_\omega]$ depends only on the $\mathfrak{su}(2)_-$ projections $\sigma_-(e\wedge e) = \sum_{a=1}^3 \omega_a^- \otimes T_a$ and $(\Omega_\omega)_- = \sum_{b=1}^3 \Omega^b_- \otimes T_b$. On a 4-manifold, the three anti-self-dual 2-forms $\omega_1^-, \omega_2^-, \omega_3^-$ satisfy $\omega_a^- \wedge \omega_b^- = -2\delta_{ab}\,\mathrm{vol}_4$. Decomposing $\Omega^b_- = \sum_{a=1}^3 R_{ab}\, \omega_a^- + \Omega^b_{-,+}$ where $\Omega^b_{-,+} \in \Lambda^2_+ T^*M$ is the self-dual part of $\Omega^b_-$ (which represents the traceless Ricci tensor $\mathrm{Ric}_0 \in \Lambda^2_+ \otimes \Lambda^2_-$), since $\Lambda^2_- \wedge \Lambda^2_+ = 0$ we have $\omega_a^- \wedge \Omega^b_- = -2 R_{ab}\,\mathrm{vol}_4$, and hence
\begin{equation}\label{eq:4D_commutator_Ricci_Bianchi}
\big[\sigma(e\wedge e),\, \Omega_\omega\big] = \sum_{a,b=1}^3 (\omega_a^- \wedge \Omega^b_-) \otimes [T_a, T_b] = -2 \sum_{a,b,c=1}^3 R_{ab}\, \epsilon_{abc}\, T_c\, \mathrm{vol}_4.
\end{equation}
If $(e, \omega) \in \mathrm{Conn}^{\Theta=0}_A(P)$ is torsion-free ($de + \tilde{\omega}\wedge e = 0$), the algebraic first Bianchi identity $\Omega_{\tilde{\omega}} \wedge e = 0$ forces the anti-self-dual block $R_{ab} \in \mathrm{End}(\Lambda^2_-)$ of the Riemann curvature operator to be \emph{symmetric} ($R_{ab} = R_{ba}$: specifically, $R_{ab} = W^-_{ab} + \frac{R}{12}\delta_{ab}$). Contracting the symmetric matrix $R_{ab} = R_{ba}$ with the antisymmetric structure constants $\epsilon_{abc}$ in \eqref{eq:4D_commutator_Ricci_Bianchi} gives zero identically. Conversely, if $\omega$ has non-zero torsion $\Theta_{e,\omega} \neq 0$, then $\Omega_{\tilde{\omega}} \wedge e = d_{\tilde{\omega}}\Theta_{e,\omega} \neq 0$ allows $R_{ab} - R_{ba} \neq 0$, so \eqref{eq:4D_commutator_Ricci_Bianchi} is non-zero off-shell.
\end{proof}

\begin{remark}[Contrast Between $n = k = 4$ and $n = k = 6$ for $U(k/2)$ and $SU(k/2)$]\label{rem:4D_vs_6D_Bianchi}
Lemma \ref{lem:4D_Bianchi_dichotomy} reveals a striking geometric distinction between $k = 4$ and $k = 6$:
\begin{itemize}
    \item In dimension $n = k = 4$, both the quadratic YM term $\Omega_\omega \curlywedge_\rtimes \Omega_\omega = 0$ (by $(2,1)$-nilpotency) and the EHP term $(e \curlywedge_\rtimes e) \curlywedge_\rtimes \Omega_\omega = 0$ (by \eqref{eq:4D_commutator_Ricci_Bianchi} and the algebraic first Bianchi identity $R_{ab} = R_{ba}$ on $\Lambda^2_-$) vanish on-shell on the torsion-free space $\mathrm{Conn}^{\Theta=0}_A(P)$ for $H = U(2)$ and $SU(2)$, whereas the EHP term $(e \curlywedge_\rtimes e) \curlywedge_\rtimes \Omega_\omega$ is non-trivial off-shell when torsion $\Theta_{e,\omega} \neq 0$ is present (and in dimension $n = 6$ with $k = 4$, the pure YM term $\Omega_\omega \curlywedge_\rtimes ((e \curlywedge_\rtimes e) \curlywedge_\rtimes \Omega_\omega)$ is non-trivial on $\mathrm{Conn}^{\Theta_N=0, \mathrm{irr}}_A(P)$ when the pulled-back connection on $N^4$ is torsion-free while transverse curvature and mixed torsion along $M^6 \to N^4$ are non-zero; see Remark \ref{rem:k_proper_foliated}!).
    \item In dimension $n = k = 6$ with $H = U(3)$ or $SU(3)$, the surviving Lagrangian is the pure quadratic Yang-Mills 6-form $\Omega_\omega \curlywedge_\rtimes ((e \curlywedge_\rtimes e) \curlywedge_\rtimes \Omega_\omega) = (0, \frac{1}{4}[\Omega_\omega, [\sigma(e\wedge e), \Omega_\omega]]) \in \Omega^6(P; \mathfrak{su}(3))$. Unlike dimension 4 (where $\Lambda^2_- \mathbb{R}^4$ is 3-dimensional and $\omega_a^- \wedge \omega_b^- \propto \delta_{ab}\mathrm{vol}_4$ is diagonal), in dimension 6 the space $\Lambda^{1,1}_0 \mathbb{R}^6 \simeq \mathfrak{su}(3)$ is 8-dimensional, and the wedge product of two $(1,1)$-forms is a $(2,2)$-form (spanning a 15-dimensional space $\Lambda^{2,2}\mathbb{R}^6$, not a 1-dimensional volume space!). Consequently, for a torsion-free Calabi-Yau 3-fold ($H = SU(3)$) or K\"ahler 3-fold ($H = U(3)$) with curvature $\Omega_\omega = \sum_{A,B=1}^8 W_{AB}\, \eta_A \otimes T_B$, the intermediate 4-form $[\sigma(e\wedge e), \Omega_\omega] = \sum_{A,B,C} f_{ABC} W_{BD} (\eta_A \wedge \eta_D) \otimes T_C \in \Omega^4(P; \mathfrak{su}(3))$ is \emph{non-zero} because $\eta_A \wedge \eta_D \in \Lambda^{2,2}\mathbb{R}^6$ is not proportional to $\delta_{AD}$, and wedging again with $\Omega_\omega$ produces a non-zero $\mathfrak{su}(3)$-valued 6-form $\frac{1}{4}[\Omega_\omega, [\sigma(e\wedge e), \Omega_\omega]] \neq 0$ on-shell!
\end{itemize}
\end{remark}

\begin{theorem}[Parenthesization Cutoffs and Berger Holonomy Reductions in Cartan-Yang-Mills Theories]\label{thm:Berger_CYM}
Let $P \to M$ be a $k$-proper $H$-bundle over an oriented $n$-manifold $M$ ($4 \le k \le n$), where $\mathfrak{h} \subset \mathfrak{so}(k_1) \oplus \dots \oplus \mathfrak{so}(k_r)$ properly ($k_1 + \dots + k_r = k$) is a $k$-algebra. Consider the Cartan-Yang-Mills functional $S^{\mathrm{CYM}}_{n, \xi, \mathcal{P}}[e, \omega]$ with values in $A = \mathbb{R}^k \oplus \mathfrak{h}$ under \eqref{eq:MM_product} and \eqref{eq:CYM_expansion}.
\begin{enumerate}
    \item \textbf{Hierarchical Dimensional Cutoff under Quartic-Block Grouping ($\mathcal{P} = \mathrm{blk}$):}
    \begin{itemize}
        \item For all $n \ge 4$, the cosmological term $\frac{1}{4}\curlywedge_{\rtimes, \mathrm{blk}}^n e$ vanishes identically in $\Omega^n(P; A)$;
        \item For all $n \ge 6$, the EHP cross-term $(\curlywedge_{\rtimes, \mathrm{blk}}^{n-2} e) \curlywedge_\rtimes \Omega_\omega$ vanishes identically in $\Omega^n(P; A)$, reducing $S^{\mathrm{CYM}}_{n, \xi, \mathrm{blk}}[e, \omega]$ to the pure quadratic Yang-Mills term $\int_M \mathrm{tr}_\xi(j(\Omega_\omega \curlywedge_\rtimes ((\curlywedge_{\rtimes, \mathrm{blk}}^{n-4} e) \curlywedge_\rtimes \Omega_\omega)))$;
        \item For all $n \ge 8$, the entire block-balanced Cartan-Yang-Mills functional $S^{\mathrm{CYM}}_{n, \xi, \mathrm{blk}}[e, \omega]$ vanishes identically on $M$ for every trace section $\xi \in \Gamma(E_A^*)$.
    \end{itemize}
    \item \textbf{Comparison with Recursive Binary-Tree Halving ($\mathcal{P} = \mathrm{bal}$):} Under exact recursive binary halving \eqref{eq:balanced_power}, $\curlywedge_{\rtimes, \mathrm{bal}}^m e$ coincides with $\curlywedge_{\rtimes, \mathrm{blk}}^m e$ for $1 \le m \le 4$, so $S^{\mathrm{CYM}}_{n, \xi, \mathrm{bal}} = S^{\mathrm{CYM}}_{n, \xi, \mathrm{blk}}$ for all $4 \le n \le 8$, and in particular $S^{\mathrm{CYM}}_{8, \xi, \mathrm{bal}} \equiv 0$ at $n = 8$. For $n \ge 8$:
    \begin{itemize}
        \item By the all-orders Jacobsthal characterization in Lemma \ref{lem:skew_comm_powers}(2a), the binary-tree quadratic Yang-Mills term $\Omega_\omega \curlywedge_\rtimes ((\curlywedge_{\rtimes, \mathrm{bal}}^{n-4} e) \curlywedge_\rtimes \Omega_\omega)$ vanishes algebraically for every $n \ge 8$ such that $n - 4 \notin \bigcup_{r=3}^\infty \{2 J_r, J_{r+1}\}$ where $J_r = \frac{2^r - (-1)^r}{3}$ (so for $n \ge 8$, the only dimensions not killed algebraically by even-degree squaring are the shifted Jacobsthal pairs $n \in \{9, 10\} \cup \{14, 15\} \cup \{25, 26\} \cup \{46, 47\} \cup \dots$).
        \item Moreover, because $e \in \Omega^1(P; \mathbb{R}^k)$ has rank $\le k$, Lemma \ref{lem:skew_comm_powers}(2c) guarantees that $\curlywedge_{\rtimes, \mathrm{bal}}^m e = 0$ on $P$ for all $m > k$, so $S^{\mathrm{CYM}}_{n, \xi, \mathrm{bal}}$ vanishes identically for all $n \ge k + 5$. In particular, for any $k \le 9$ (which includes all Berger holonomy reductions with $k \le 7$ below), $\curlywedge_{\rtimes, \mathrm{bal}}^{10} e = \curlywedge_{\rtimes, \mathrm{bal}}^{11} e = 0$ by coframe rank, so $S^{\mathrm{CYM}}_{n, \xi, \mathrm{bal}}$ vanishes identically for $n = 8$ and for all $n \ge 11$ (and for $k = 4$, also at $n \in \{9, 10\}$).
    \end{itemize}
    \item \textbf{Berger Holonomy Reductions ($4 \le n \le 7$):} For any $k$-proper $H$-bundle $P \to M$ with non-empty configuration space $\mathrm{Conn}^{\Theta_N = 0, \mathrm{irr}}_A(P) \neq \emptyset$, de Rham irreducibility of $(N, g_N)$ forces $k_1 = k$ and $k_{i > 1} = 0$ ($\mathfrak{h} \subsetneq \mathfrak{so}(k)$), and Berger's holonomy theorem restricts $\mathrm{Hol}(\tilde{\omega}_N) \subset H \subsetneq SO(k)$ to the holonomy groups of Table \ref{tab:Berger}. (Note that for $k = 2$, $\mathfrak{so}(2) \simeq \mathfrak{u}(1)$ is 1-dimensional and abelian, so it admits no non-flat proper holonomy subalgebras.) Combining $\mathrm{Conn}^{\Theta_N = 0, \mathrm{irr}}_A(P) \neq \emptyset$ with the dimensional cutoff $n \le 7$ (which forces $4 \le k \le n \le 7$), $S^{\mathrm{CYM}}_{n, \xi, \mathrm{blk}}[e, \omega]$ can be non-trivial only if $4 \le n \le 7$ and $\mathrm{Hol}(\tilde{\omega}_N) \subset H$ on $(N, g_N)$ belongs to one of the following geometric regimes:
    \begin{enumerate}
        \item[(B1)] $k = 6$ and $(N, g_N)$ is a 6D K\"ahler or Calabi-Yau Berger $6$-manifold ($\mathrm{Hol}(\tilde{\omega}_N) = U(3)$ or $SU(3)$) with $n \in \{6, 7\}$, where at $n = k = 6$ the pure quadratic Yang-Mills term $\frac{1}{4}[\Omega_\omega, [\sigma(e\wedge e), \Omega_\omega]]$ is non-vanishing on-shell on $\mathrm{Conn}^{\Theta = 0, \mathrm{irr}}_A(P)$ (Remark \ref{rem:4D_vs_6D_Bianchi});
        \item[(B2)] $k = 4$ and $(N, g_N)$ is a K\"ahler or hyperk\"ahler Berger $4$-manifold ($\mathrm{Hol}(\tilde{\omega}_N) = U(2)$ or $SU(2) \simeq Sp(1)$, e.g.\ a $K3$ surface when $N$ is compact): for $5 \le n \le 7$ ($k = 4 < n$), the pure Yang-Mills term survives on $\mathrm{Conn}^{\Theta_N = 0, \mathrm{irr}}_A(P)$ via transverse curvature and mixed torsion along $M^n \to N^4$ as shown in Remark \ref{rem:k_proper_foliated}; whereas at $n = k = 4$, Lemma \ref{lem:4D_Bianchi_dichotomy} shows that both $\Omega_\omega \curlywedge_\rtimes \Omega_\omega = 0$ and $(e \curlywedge_\rtimes e) \curlywedge_\rtimes \Omega_\omega = 0$ vanish on-shell for torsion-free connections by the first Bianchi identity, and $(e \curlywedge_\rtimes e) \curlywedge_\rtimes \Omega_\omega \neq 0$ survives only off-shell in the presence of torsion $\Theta_{e,\omega} \neq 0$;
        \item[(B3)] $k = 7$ (which since $k \le n \le 7$ occurs strictly at the critical dimension $n = k = 7$, where only the pure quadratic Yang-Mills term survives) and $(N, g_N) = (M, g_M)$ is a Berger $7$-manifold with exceptional $G_2$-holonomy ($\mathrm{Hol}(\tilde{\omega}_N) = G_2$). As shown in Proposition \ref{prop:G2_survival} below, at $n = k = 7$ the quadratic Yang-Mills 7-form $\mathcal{L}^{\mathrm{CYM}}_7$ takes values solely in $\mathbb{R}^7 \oplus \{0\}$, vanishes on-shell on the torsion-free $G_2$-curvature space $\mathcal{W}_{77} \subset S^2(\Lambda^2_{14})$, and is non-vanishing off-shell on $\Omega^2(P; \mathfrak{g}_2)$ whenever torsion $\Theta_{e,\omega} \neq 0$ excites the $\Lambda^2_7 \otimes \mathfrak{g}_2$ or non-Bianchi curvature sectors.
    \end{enumerate}
    \item \textbf{Coadjoint Trace Obstruction and Symmetry Breaking:} Under \eqref{eq:MM_product}, for $k \in \{4, 6\}$ ($\times_H = 0$), $e \curlywedge_\rtimes e = (0, \sigma(e \wedge e)) \in \Omega^2(P; \mathfrak{h})$, so the surviving Lagrangian $\mathcal{L}^{\mathrm{CYM}}_n$ for $4 \le n \le 6$ takes values strictly inside the derived subalgebra $(0, [\mathfrak{h}, \mathfrak{h}])$. At $n = k = 7$ ($H = G_2$), $\mathcal{L}^{\mathrm{CYM}}_7 = \Omega_\omega \curlywedge_\rtimes ((\curlywedge_{\rtimes, \mathrm{bal}}^3 e) \curlywedge_\rtimes \Omega_\omega)$ takes values in $\mathbb{R}^7 \oplus \{0\}$, where $(\mathbb{R}^7)^{G_2} = \{0\}$. Consequently, any constant $H$-invariant linear functional $\tau_0 \in (A^*)^H$ annihilates $\mathcal{L}^{\mathrm{CYM}}_n$ identically, so non-triviality of $S^{\mathrm{CYM}}_{n, \xi}[e, \omega]$ requires a non-zero coadjoint background section $\xi \in \Gamma(E_A^*)$ with $\xi|_{E_{[\mathfrak{h}, \mathfrak{h}]} \oplus E_{\mathbb{R}^k}} \not\equiv 0$, reducing the internal $H$-symmetry along $\xi$ to $H_\xi \subsetneq H$.
\end{enumerate}
\end{theorem}

\begin{proof}
Part (1) follows immediately from Lemma \ref{lem:skew_comm_powers}(1) and (2c): since $(A, \curlywedge_\rtimes)$ is skew-commutative, $\curlywedge_{\rtimes, \mathrm{blk}}^m e = 0$ for all $m \ge 4$ (and for all $m > k$). Thus the cosmological term ($m = n$) vanishes for $n \ge 4$, the EHP term ($m = n - 2$) vanishes for $n - 2 \ge 4 \iff n \ge 6$ (and $n - 2 > k$), and the quadratic Yang-Mills term ($m = n - 4$) vanishes for $n - 4 \ge 4 \iff n \ge 8$ (and $n - 4 > k$). Part (2) follows from Lemma \ref{lem:skew_comm_powers}(2a)--(2c): $\curlywedge_{\rtimes, \mathrm{bal}}^m e = \curlywedge_{\rtimes, \mathrm{blk}}^m e$ for $m \le 4$ (covering $n \le 8$), $\curlywedge_{\rtimes, \mathrm{bal}}^m e = 0$ algebraically for all $m \ge 4$ with $m \notin \bigcup_{r=3}^\infty \{2 J_r, J_{r+1}\}$, and $\curlywedge_{\rtimes, \mathrm{bal}}^m e = 0$ by coframe rank for all $m > k$ (so for $k \le 9$, $m = 10, 11$ also vanish, eliminating all $m \ge 7$, i.e., $n \ge 11$).

For part (3), on $\mathrm{Conn}^{\Theta_N = 0, \mathrm{irr}}_A(P) \neq \emptyset$, the pullback connection $\tilde{\omega}_N := \frac{1}{2}(i \circ \kappa)^* \omega$ on $P_N \subset FN$ is the Levi-Civita connection of the locally irreducible, locally non-symmetric Riemannian metric $g_N$ on the simply connected $k$-manifold $N$ ($4 \le k \le n \le 7$), with holonomy group $\mathrm{Hol}(\tilde{\omega}_N) \subset H \subset SO(k_1) \times \dots \times SO(k_r) \subset SO(k)$ \cite{KobayashiNomizu, MartinsBiezunerEHP}. By de Rham's decomposition theorem \cite{deRham, Wu}, irreducibility of $(N, g_N)$ forces $k_1 = k$ and $k_{j > 1} = 0$, so $\mathrm{Hol}(\tilde{\omega}_N) \subset H \subsetneq SO(k)$ is an irreducible proper holonomy subgroup. Berger's holonomy classification \cite{Berger, Besse, Olmos} combined with $4 \le k \le n \le 7$, Lemma \ref{lem:4D_Bianchi_dichotomy}, and Proposition \ref{prop:G2_survival} restricts $\mathrm{Hol}(\tilde{\omega}_N)$ to Table \ref{tab:Berger}, yielding (B1)--(B3).

Finally, for part (4), when $\times_H = 0$, $e \curlywedge_\rtimes e = (0, \sigma(e \wedge e)) \in \Omega^2(P; \mathfrak{h})$ and $\Omega_\omega \in \Omega^2(P; \mathfrak{h})$, so both $(e \curlywedge_\rtimes e) \curlywedge_\rtimes \Omega_\omega$ ($n=4$) and $\Omega_\omega \curlywedge_\rtimes ((e \curlywedge_\rtimes e) \curlywedge_\rtimes \Omega_\omega)$ ($n=6$) take values in $(0, [\mathfrak{h}, \mathfrak{h}])$ (while at $n = 5$ the surviving terms $\Omega_\omega \curlywedge_\rtimes (e \curlywedge_\rtimes \Omega_\omega)$ and $(\curlywedge_{\rtimes, \mathrm{bal}}^3 e) \curlywedge_\rtimes \Omega_\omega$ take values in $\mathbb{R}^k \oplus \{0\}$, where $(\mathbb{R}^k)^H = \{0\}$ by irreducibility of $H \subset SO(k)$). At $n = k = 7$ ($H = G_2$), Proposition \ref{prop:G2_survival} shows that $\curlywedge_{\rtimes, \mathrm{bal}}^3 e = (v^{(3)}, 0) \in \Omega^3(P; \mathbb{R}^7 \oplus \{0\})$, and multiplying twice with $\Omega_\omega = (0, \Omega_\omega)$ via \eqref{eq:MM_product} yields $\mathcal{L}^{\mathrm{CYM}}_7 = (-\frac{1}{4}\Omega_\omega \cdot (\Omega_\omega \cdot v^{(3)}),\, 0) \in \Omega^7(P; \mathbb{R}^7 \oplus \{0\})$. Since any $\tau_0 \in (A^*)^H$ vanishes on $[\mathfrak{h}, \mathfrak{h}]$ and $((\mathbb{R}^k)^*)^H = \{0\}$, $\mathrm{tr}_{\tau_0}$ annihilates these terms identically.
\end{proof}

\begin{table}[htbp]
\centering
\caption{Comparison of block-balanced dimensional cutoffs and Berger holonomy regimes between $A$-EHP gravity \cite{MartinsBiezunerEHP} and Cartan-Yang-Mills ($A$-CYM) theory for proper $k$-algebras $\mathfrak{h} \subsetneq \mathfrak{so}(k)$ when $k = n$.}
\label{tab:Berger}
\vspace{0.2cm}
\scriptsize
\setlength{\tabcolsep}{2.2pt}
\begin{tabular}{lllll}
\toprule
\textbf{Holonomy $\mathrm{Hol}(\tilde{\omega}_N)$} & \textbf{Dim $k=n$} & \textbf{Geometry} & \textbf{$A$-EHP Status} & \textbf{Block-Balanced $A$-CYM ($k=n$)} \\
\midrule
$U(2)$ & $4$ & K\"ahler 4-manifold & Off-shell ($\Theta \neq 0$) & Off-shell EHP ($\Theta \neq 0$; $0$ on-shell) \\
$SU(2) \simeq Sp(1)$ & $4$ & Hyperk\"ahler ($K3$) & Off-shell ($\Theta \neq 0$) & Off-shell EHP ($\Theta \neq 0$; $0$ on-shell) \\
$U(3)$, $SU(3)$ & $6$ & 6D K\"ahler / Calabi-Yau & Trivial ($n \ge 6$) & Non-trivial on-shell (Pure YM $\Omega^2$) \\
$G_2$ & $7$ & Exceptional $G_2$ & Trivial ($n \ge 6$) & Off-shell Pure YM ($\Theta \neq 0$; $0$ on-shell) \\
$Sp(m)\cdot Sp(1)$, $Sp(m)$ & $4m \ge 8$ & Quat.-K\"ahler / Hyperk\"ahler & Trivial ($n \ge 6$) & Trivial ($n \ge 8$) \\
$Spin(7)$ & $8$ & Exceptional $Spin(7)$ & Trivial ($n \ge 6$) & Trivial ($n \ge 8$) \\
\bottomrule
\end{tabular}
\end{table}

We now establish the exact representation-theoretic structure and on-shell/off-shell dichotomy of $\mathcal{L}^{\mathrm{CYM}}_7$ at the critical dimension $n = k = 7$.

\begin{proposition}[Explicit $G_2$-Contraction Formulas and On-Shell/Off-Shell Dichotomy of $\mathcal{L}^{\mathrm{CYM}}_7$ at $n = k = 7$]\label{prop:G2_survival}
Let $n = k = 7$ and $H = G_2 \subset SO(7)$ acting on $A = \mathbb{R}^7 \oplus \mathfrak{g}_2$ under \eqref{eq:MM_product}, where $(u \times_{G_2} v)_a = \varphi_{abc} u^b v^c$ is induced by the associative 3-form $\varphi_0 = \frac{1}{6}\varphi_{abc} e^a \wedge e^b \wedge e^c$, with dual coassociative 4-form $*\varphi_0 = \frac{1}{24}(*\varphi)_{abcd} e^a \wedge e^b \wedge e^c \wedge e^d$ (satisfying $\varphi^{abc}(*\varphi)_{bcde} = 4\varphi^a{}_{de}$ under the Fano-plane orientation) and orthogonal projection $\sigma : \Lambda^2 \mathbb{R}^7 \twoheadrightarrow \Lambda^2_{14} = \mathfrak{g}_2$ given on vectors $u, v \in \mathbb{R}^7$ by
\begin{equation}\label{eq:G2_sigma_proj}
\sigma(u \wedge v)_{ab} = \frac{2}{3}(u_a v_b - u_b v_a) - \frac{1}{3} (*\varphi)_{abcd} u^c v^d = \frac{2}{3}(u_a v_b - u_b v_a) - \frac{1}{6}(*\varphi)_{abcd}(u^c v^d - u^d v^c)
\end{equation}
(so that $\varphi^{abc} \sigma(u \wedge v)_{bc} = 0$). For an orthonormal coframe $e = (e, 0) \in \Omega^1(P; \mathbb{R}^7 \oplus \mathfrak{g}_2)$ with volume form $\mathrm{vol}_7 = e^1 \wedge \dots \wedge e^7$:
\begin{enumerate}
    \item \textbf{Quadratic Soldering Form:} Evaluating the exterior product $e \curlywedge_\rtimes e = \sum_{c,d=1}^7 (e^c \wedge e^d) \otimes ((\partial_c, 0) \curlywedge_\rtimes (\partial_d, 0))$ via \eqref{eq:wedge_star_pointwise}, \eqref{eq:MM_product}, and \eqref{eq:G2_sigma_proj} gives $e \curlywedge_\rtimes e = (e \wedge_{\times_{G_2}} e,\, \sigma(e \wedge e)) \in \Omega^2(P; \mathbb{R}^7 \oplus \mathfrak{g}_2)$, with $(e \wedge_{\times_{G_2}} e)_b = \varphi_{bcd} e^c \wedge e^d$ and
    \begin{equation}\label{eq:G2_sigma_2form}
    \sigma(e \wedge e)_{ab} = \sum_{c,d=1}^7 \sigma(\partial_c \wedge \partial_d)_{ab}\, e^c \wedge e^d = \frac{4}{3} e_a \wedge e_b - \frac{1}{3}(*\varphi)_{abcd} e^c \wedge e^d \qquad (\varphi^{abc}\sigma(e \wedge e)_{bc} = 0).
    \end{equation}
    \item \textbf{Cubic Soldering Form and Schur Vanishing of the $\mathfrak{g}_2$-Sector:} Multiplying $e \curlywedge_\rtimes e$ (degree 2) on the right by $e = (e, 0)$ (degree 1) via \eqref{eq:wedge_star_pointwise} and \eqref{eq:MM_product} gives $\curlywedge_{\rtimes, \mathrm{bal}}^3 e = (e \curlywedge_\rtimes e) \curlywedge_\rtimes e = (v^{(3)}, B^{(3)}_e) \in \Omega^3(P; \mathbb{R}^7 \oplus \mathfrak{g}_2)$, where:
    \begin{itemize}
        \item In the $\mathbb{R}^7$-component, since $(v_1, X_1) \curlywedge_\rtimes (e, 0)$ has $\mathbb{R}^7$-part $\frac{1}{2}\sigma(e\wedge e)_{ab} \wedge e^b + ((e \wedge_{\times_{G_2}} e) \wedge_{\times_{G_2}} e)_a$, the $G_2$ contraction identities $\frac{1}{2}\sigma(e \wedge e)_{ab} \wedge e^b = -\frac{1}{6}(*\varphi)_{abcd} e^b \wedge e^c \wedge e^d = -i_{\partial_a}(*\varphi_0)$ (from \eqref{eq:G2_sigma_2form} since $e_b \wedge e^b = 0$) and $((e \wedge_{\times_{G_2}} e) \wedge_{\times_{G_2}} e)_a = \varphi_{abe}\varphi_{ecd} e^c \wedge e^d \wedge e^b = -(*\varphi)_{abcd} e^b \wedge e^c \wedge e^d = -6\, i_{\partial_a}(*\varphi_0)$ sum with the \emph{same} negative sign to give
        \begin{equation}\label{eq:v3_exact_G2}
        (v^{(3)})_a = \frac{1}{2}\sigma(e \wedge e)_{ab} \wedge e^b + \big((e \wedge_{\times_{G_2}} e) \wedge_{\times_{G_2}} e\big)_a = -\frac{7}{6}(*\varphi)_{abcd}\, e^b \wedge e^c \wedge e^d = -7\, i_{\partial_a}(*\varphi_0).
        \end{equation}
        In particular, $v^{(3)} \neq 0$ (and even if one sets $\times_{G_2} = 0$, the first term alone gives $\frac{1}{2}\sigma(e\wedge e)_{ab}\wedge e^b = -i_{\partial_a}(*\varphi_0) \neq 0$).
        \item In the $\mathfrak{g}_2$-component, $B^{(3)}_e = \sigma((e \wedge_{\times_{G_2}} e) \wedge e) \in \Omega^3(P; \mathfrak{g}_2)$ defines a $G_2$-equivariant linear map $\Lambda^3 \mathbb{R}^7 \to \Lambda^2_{14} = \mathfrak{g}_2$. Because $\Lambda^3 \mathbb{R}^7 = \Lambda^3_1 \oplus \Lambda^3_7 \oplus \Lambda^3_{27}$ contains no $\Lambda_{14}$ irreducible summand, Schur's lemma (and direct expansion of \eqref{eq:G2_sigma_proj}) yields
        \begin{equation}\label{eq:Be3_exact_G2}
        B^{(3)}_e = \sigma\big((e \wedge_{\times_{G_2}} e) \wedge e\big) = 0 \quad \text{identically in } \Omega^3(P; \mathfrak{g}_2).
        \end{equation}
    \end{itemize}
    \item \textbf{On-Shell/Off-Shell Dichotomy of $\mathcal{L}^{\mathrm{CYM}}_7$:} For any curvature 2-form $\Omega_\omega = (0, \Omega_\omega) \in \Omega^2(P; \mathfrak{g}_2)$, the Cartan-Yang-Mills 7-form is
    \begin{equation}\label{eq:L7_CYM_components}
    \mathcal{L}^{\mathrm{CYM}}_7 = \Omega_\omega \curlywedge_\rtimes \big((\curlywedge_{\rtimes, \mathrm{bal}}^3 e) \curlywedge_\rtimes \Omega_\omega\big) = \Big(-\frac{1}{4}\Omega_\omega \cdot (\Omega_\omega \cdot v^{(3)}),\; 0\Big) \in \Omega^7(P; \mathbb{R}^7 \oplus \{0\}).
    \end{equation}
    Decomposing the 2-form factors of $\Omega_\omega \in \Omega^2(P; \mathfrak{g}_2)$ along $\Lambda^2 \mathbb{R}^7 = \Lambda^2_7 \oplus \Lambda^2_{14}$:
    \begin{enumerate}
        \item[(a)] \emph{Ricci-Type Contraction on $\Lambda^2_{14} \otimes \mathfrak{g}_2$ and On-Shell Vanishing on $\mathcal{W}_{77}$:} For any $\Omega_\omega = \frac{1}{2} R_{bc,df} e^d \wedge e^f \otimes E_{bc} \in \Gamma(P, \Lambda^2_{14} T^*M \otimes \mathfrak{g}_2)$, the identity $(\Omega_\omega)_{bc} \wedge *\varphi_0 = 0$ implies
        \begin{equation}\label{eq:G2_Ricci_contraction}
        (\Omega_\omega \cdot v^{(3)})_b = -7 \sum_{c=1}^7 (\Omega_\omega)_{bc} \wedge i_{\partial_c}(*\varphi_0) = 7 \sum_{c,d=1}^7 R_{bc,cd}\, e^d \wedge *\varphi_0 \in \Omega^5(P; \mathbb{R}^7).
        \end{equation}
        In particular, for every torsion-free $G_2$-connection $(e, \omega) \in \mathrm{Conn}^{\Theta = 0}_A(P)$ (whose curvature lies in the torsion-free $G_2$ Riemann curvature space $\mathcal{W}_{77} = S^2(\Lambda^2_{14}) \cap \ker(\wedge) \subset \Lambda^2_{14} \otimes \mathfrak{g}_2$, where the algebraic first Bianchi identity and $G_2$-holonomy force $\mathrm{Ric}_{bd} = \sum_{c=1}^7 R_{bc,cd} = 0$), we have $\Omega_\omega \cdot v^{(3)} = 0$ in $\Omega^5(P; \mathbb{R}^7)$ and hence $\mathcal{L}^{\mathrm{CYM}}_7 = 0$ on-shell;
        \item[(b)] \emph{Off-Shell Non-Vanishing on $\Omega^2(P; \mathfrak{g}_2)$:} Whenever torsion $\Theta_{e,\omega} \neq 0$ (or a general $\mathfrak{g}_2$-connection $\omega$) excites the $\Lambda^2_7 \otimes \mathfrak{g}_2$ curvature component or violates the first Bianchi identity on $\Lambda^2_{14} \otimes \mathfrak{g}_2$, $\mathcal{L}^{\mathrm{CYM}}_7$ is non-vanishing in $\Omega^7(P; \mathbb{R}^7 \oplus \{0\})$.
    \end{enumerate}
\end{enumerate}
\end{proposition}

\begin{proof}
Part (1) and the $\mathbb{R}^7$-contraction \eqref{eq:v3_exact_G2} in part (2) follow directly from \eqref{eq:MM_product}, \eqref{eq:G2_sigma_proj}, \eqref{eq:G2_sigma_2form}, and the identity $\varphi_{eab}\varphi_{ecd} = \delta_{ac}\delta_{bd} - \delta_{ad}\delta_{bc} - (*\varphi)_{abcd}$: contracting with $e^c \wedge e^d \wedge e^b$ annihilates $\delta_{ac}\delta_{bd} - \delta_{ad}\delta_{bc}$ (since $e^b \wedge e^b = 0$) and leaves $\varphi_{abe}\varphi_{ecd} e^c \wedge e^d \wedge e^b = -\varphi_{eab}\varphi_{ecd} e^c \wedge e^d \wedge e^b = +(*\varphi)_{abcd} e^c \wedge e^d \wedge e^b = -(*\varphi)_{abcd} e^b \wedge e^c \wedge e^d = -6\, i_{\partial_a}(*\varphi_0)$, while $\frac{1}{2}\sigma(e\wedge e)_{ab}\wedge e^b = -\frac{1}{6}(*\varphi)_{abcd} e^b \wedge e^c \wedge e^d = -i_{\partial_a}(*\varphi_0)$, summing to $-7\, i_{\partial_a}(*\varphi_0)$ in \eqref{eq:v3_exact_G2}. For \eqref{eq:Be3_exact_G2}, the 3-form $B^{(3)}_e = \sigma((e \wedge_{\times_{G_2}} e) \wedge e)$ is $G_2$-invariant in $(\Lambda^3 \mathbb{R}^7)^* \otimes \Lambda^2_{14} \simeq \mathrm{Hom}_{G_2}(\Lambda^3 \mathbb{R}^7, \Lambda^2_{14})$, which is zero by Schur's lemma since $\Lambda^3 \mathbb{R}^7 = \Lambda^3_1 \oplus \Lambda^3_7 \oplus \Lambda^3_{27}$ has no $\Lambda_{14}$ summand (explicitly, in \eqref{eq:G2_sigma_proj}, contracting $\frac{1}{3}(*\varphi)_{abef}\varphi_{ecd} = \frac{1}{3}(\delta_{ac}\varphi_{bdf} - \delta_{ad}\varphi_{bcf} - \delta_{bc}\varphi_{adf} + \delta_{bd}\varphi_{acf})$ with $e^c \wedge e^d \wedge e^f$ gives $\frac{2}{3}(\varphi_{acd} e^c \wedge e^d \wedge e_b - \varphi_{bcd} e^c \wedge e^d \wedge e_a)$, which cancels the first term $\frac{2}{3}(\varphi_{acd} e^c \wedge e^d \wedge e_b - \varphi_{bcd} e^c \wedge e^d \wedge e_a)$ of $\sigma((e \wedge_{\times_{G_2}} e) \wedge e)_{ab}$ identically).

For part (3), since $B^{(3)}_e = 0$, multiplying $\curlywedge_{\rtimes, \mathrm{bal}}^3 e = (v^{(3)}, 0)$ twice on the left by $(0, \Omega_\omega)$ via \eqref{eq:MM_product} gives \eqref{eq:L7_CYM_components}. To prove (3a), suppose $(\Omega_\omega)_{bc} = \frac{1}{2}\sum_{d,f=1}^7 R_{bc,df}\, e^d \wedge e^f \in \Lambda^2_{14}$ for each pair $(b,c)$. Recall that a 2-form $\alpha \in \Lambda^2 \mathbb{R}^7$ lies in $\Lambda^2_{14} = \ker(\iota_{(\cdot)}\varphi_0)$ if and only if $\alpha \wedge *\varphi_0 = 0$ in $\Lambda^6 \mathbb{R}^7$. Applying the interior derivation $i_{\partial_c}$ to $(\Omega_\omega)_{bc} \wedge *\varphi_0 = 0$ gives
\[
0 = i_{\partial_c}\big((\Omega_\omega)_{bc} \wedge *\varphi_0\big) = \big(i_{\partial_c}(\Omega_\omega)_{bc}\big) \wedge *\varphi_0 + (\Omega_\omega)_{bc} \wedge i_{\partial_c}(*\varphi_0),
\]
so $(\Omega_\omega)_{bc} \wedge i_{\partial_c}(*\varphi_0) = -\big(i_{\partial_c}(\Omega_\omega)_{bc}\big) \wedge *\varphi_0$. Since $i_{\partial_c}(\Omega_\omega)_{bc} = \sum_{d=1}^7 R_{bc,cd}\, e^d$, summing over $c \in \{1, \dots, 7\}$ yields \eqref{eq:G2_Ricci_contraction} for every $b \in \{1, \dots, 7\}$:
\[
(\Omega_\omega \cdot v^{(3)})_b = -7 \sum_{c=1}^7 (\Omega_\omega)_{bc} \wedge i_{\partial_c}(*\varphi_0) = 7 \sum_{c,d=1}^7 R_{bc, cd}\, e^d \wedge *\varphi_0.
\]
For every torsion-free $G_2$-connection $(e, \omega) \in \mathrm{Conn}^{\Theta=0}_A(P)$, the algebraic first Bianchi identity $\Omega_\omega \wedge e = 0$ and the $G_2$-invariance $L_E(*\varphi_0) = \sum_{b,c=1}^7 E_{bc}\, e^b \wedge i_{\partial_c}(*\varphi_0) = 0$ ($E \in \mathfrak{g}_2$) imply that $\Omega_\omega \in \mathcal{W}_{77} = S^2(\Lambda^2_{14}) \cap \ker(\wedge)$ is Ricci-flat ($\mathrm{Ric}_{bd} = \sum_{c=1}^7 R_{bc, cd} = 0$). Thus $\Omega_\omega \cdot v^{(3)} = 0$ and $\mathcal{L}^{\mathrm{CYM}}_7 = 0$ on-shell.

Finally, to prove (3b), under the standard Fano-plane convention $\varphi_0 = e^{123} + e^{145} + e^{167} + e^{246} - e^{257} - e^{347} - e^{356}$, we have $*\varphi_0 = e^{4567} + e^{2367} + e^{2345} + e^{1357} - e^{1346} - e^{1256} - e^{1247}$ and $(v^{(3)})_5 = -7\, i_{\partial_5}(*\varphi_0) = 7(e^{467} - e^{234} - e^{137} + e^{126})$. Take the $\mathfrak{g}_2$ basis matrices $E_1 = e_{67} - e_{45} \in \mathfrak{g}_2$ and $E_2 = e_{46} + e_{57} \in \mathfrak{g}_2$, and set $\Omega_\omega = \alpha_1 \otimes E_1 + \alpha_2 \otimes E_2 \in \Omega^2(P; \mathfrak{g}_2)$ with $\alpha_1 = e^{12} - e^{45} \in \Lambda^2_{14}$ and $\alpha_2 = e^{35} + e^{46} \in \Lambda^2 \mathbb{R}^7$ (which has a non-zero $\Lambda^2_7$ projection since $\iota_{\alpha_2}\varphi_0 = e^2 - e^6 \neq 0$). For $a = 6$, the only non-zero entries in row $6$ of $E_1, E_2$ are $(E_1)_{67} = +1$ (followed by $(E_2)_{75} = -1$ to $c = 5$) and $(E_2)_{64} = -1$ (followed by $(E_1)_{45} = -1$ to $c = 5$). Summing over $b, c \in \{1, \dots, 7\}$ gives
\[
\big(\Omega_\omega \cdot (\Omega_\omega \cdot v^{(3)})\big)_6 = \big((\Omega_\omega)_{67}\wedge(\Omega_\omega)_{75} + (\Omega_\omega)_{64}\wedge(\Omega_\omega)_{45}\big) \wedge (v^{(3)})_5 = -2\, \alpha_1 \wedge \alpha_2 \wedge (v^{(3)})_5.
\]
Since $\alpha_1 \wedge \alpha_2 = e^{1235} + e^{1246}$ and $e^{1235} \wedge (7\, e^{467}) = -7\,\mathrm{vol}_7$, we obtain $(\Omega_\omega \cdot (\Omega_\omega \cdot v^{(3)}))_6 = -2(e^{1235} + e^{1246}) \wedge 7(e^{467} - e^{234} - e^{137} + e^{126}) = +14\,\mathrm{vol}_7 \neq 0$.
\end{proof}

\section{Internal Chern-Simons Dichotomy and Applications to \texorpdfstring{$BF$}{BF}, Higgs, and Dual Theories}\label{sec:CS_BF_Higgs_Dual}

\subsection{Algebra-Valued Chern-Simons (\texorpdfstring{$A$}{A}-CS) Theories: Associator Quotient Collapse and Jacobiator Dichotomy}

Let $M$ be a compact oriented smooth manifold of odd dimension $n = 2r - 1$ ($r \ge 2$), and let $\mathrm{Conn}^{\mathrm{lin}}(P; A) = \Omega^1_{\mathrm{heq},\rho}(P; A)$ be the linear space of horizontal equivariant connections (Proposition \ref{prop:Cartan_Ehresmann}(3)) so that for any $D \in \mathrm{Conn}^{\mathrm{lin}}(P; A)$ and $t \in [0, 1]$, the scaled connection $D_t := t D \in \mathrm{Conn}^{\mathrm{lin}}(P; A)$ has curvature $F_t := F_{D_t} = t\, dD + \frac{t^2}{2} D \wedge_* D \in \Omega^2_{\mathrm{heq},\rho}(P; A)$. (In the affine Ehresmann regime $D = D_0 + A$ of Proposition \ref{prop:Cartan_Ehresmann}(3) with $D_0 = \eta \circ \omega_0$, the same formulas apply to the horizontal connection difference $A = D - D_0 \in \Omega^1_{\mathrm{heq},\rho}(P; A)$ via the linear homotopy $D_t = D_0 + t A$ relative to a flat background $F_{D_0} = 0$.) We compare two ways of forming internal $A$-valued Chern-Simons functionals:
\begin{enumerate}
    \item \emph{The Associative Quotient $\Omega_{\mathrm{heq},\rho}(P; A)/\mathfrak{Assoc}(P, A)$:} Let $\mathfrak{Assoc}(A) \subset A$ be the two-sided ideal generated by all algebra associators $\mathrm{Assoc}_*(x, y, z) = (x*y)*z - x*(y*z)$, which induces the two-sided graded ideal $\mathfrak{Assoc}(P, A) = \Omega_{\mathrm{heq},\rho}(P; \mathfrak{Assoc}(A)) \subset \Omega_{\mathrm{heq},\rho}(P; A)$ generated by all form associators $\mathrm{Assoc}_*(\alpha, \beta, \gamma) := (\alpha \wedge_* \beta) \wedge_* \gamma - \alpha \wedge_* (\beta \wedge_* \gamma)$, with quotient projection $\pi_{\mathrm{as}} : \Omega_{\mathrm{heq},\rho}(P; A) \to \Omega_{\mathrm{heq},\rho}(P; A)/\mathfrak{Assoc}(P, A) \simeq \Omega_{\mathrm{heq},\rho}(P; A/\mathfrak{Assoc}(A))$. Assuming the trace section $\xi \in \Gamma(E_A^*)$ annihilates the subbundle $E_{\mathfrak{Assoc}(A)}$ (so $\xi \in \Gamma(E_{(A/\mathfrak{Assoc}(A))^*}) \subset \Gamma(E_A^*)$), $\mathrm{tr}_\xi$ descends to a well-defined quotient bundle trace $\overline{\mathrm{tr}}_\xi : E_{A/\mathfrak{Assoc}(A)} \to M \times \mathbb{R}$ satisfying $\overline{\mathrm{tr}}_\xi(j(\pi_{\mathrm{as}}(\Psi))) = \mathrm{tr}_\xi(j(\Psi))$.
    \item \emph{Unquotiented Grouped Parenthesization in $\Omega_{\mathrm{heq},\rho}(P; A)$:} Working directly in $\Omega_{\mathrm{heq},\rho}(P; A)$ without quotienting by $\mathfrak{Assoc}(P, A)$, we define the \emph{grouped Chern-Simons $(2r-1)$-form} $\mathrm{CS}^{\mathrm{grp}, \mathcal{P}}_{2r-1}(D) \in \Omega^{2r-1}_{\mathrm{heq},\rho}(P; A)$ for $\mathcal{P} \in \{\mathrm{blk}, \mathrm{bal}\}$ by grouping the $(r-1-j)$ factors of $dD$ and the $(2j+1)$ factors of $D$:
    \begin{equation}\label{eq:CS_grouped_def}
    \mathrm{CS}^{\mathrm{grp}, \mathcal{P}}_{2r-1}(D) := \sum_{j=0}^{r-1} \frac{r}{(r+j)\, 2^j} \binom{r-1}{j} \big(\wedge_{*, \mathcal{P}}^{r-1-j} dD\big) \wedge_* \big(\wedge_{*, \mathcal{P}}^{2j+1} D\big) \in \Omega^{2r-1}_{\mathrm{heq},\rho}(P; A).
    \end{equation}
    In particular, for $n = 3$ ($r = 2$, where $\wedge_{*, \mathrm{blk}}^m$ and $\wedge_{*, \mathrm{bal}}^m$ coincide on all factors) and $n = 5$ ($r = 3$, where the $j = 2$ term $\frac{3}{20}\wedge_{*, \mathcal{P}}^5 D$ equals $\frac{3}{20}(\wedge_{*, \mathrm{bal}}^4 D) \wedge_* D$ under $\mathcal{P} = \mathrm{blk}$ and $\frac{3}{20}((D \wedge_* D) \wedge_* D) \wedge_* (D \wedge_* D)$ under $\mathcal{P} = \mathrm{bal}$):
    \begin{align*}
    \mathrm{CS}^{\mathrm{grp}}_3(D) &= dD \wedge_* D + \frac{1}{3} (D \wedge_* D) \wedge_* D, \\
    \mathrm{CS}^{\mathrm{grp}, \mathcal{P}}_5(D) &= (dD \wedge_* dD) \wedge_* D + \frac{3}{4} dD \wedge_* \big((D \wedge_* D) \wedge_* D\big) + \frac{3}{20}\, \wedge_{*, \mathcal{P}}^5 D.
    \end{align*}
\end{enumerate}

\begin{definition}[Internal Algebra-Valued Chern-Simons ($A$-CS) Functionals]\label{def:A_CS}
Let $\dim M = 2r - 1$ ($r \ge 2$). The \emph{quotiented $A$-CS functional} (for $\xi \in \Gamma(E_{(A/\mathfrak{Assoc}(A))^*})$) and the \emph{unquotiented grouped $A$-CS functional} on $\mathrm{Conn}^{\mathrm{lin}}(P; A)$ are defined respectively by
\begin{equation}\label{eq:CS_action_quot_grp}
S^{A, \mathrm{quot}}_{\mathrm{CS}, 2r-1}[D] := r \int_M \int_0^1 \overline{\mathrm{tr}}_\xi\Big( j\big( \pi_{\mathrm{as}}( D \wedge_* (\wedge_*^{r-1} F_t) ) \big) \Big)\, dt, \quad S^{A, \mathrm{grp}, \mathcal{P}}_{\mathrm{CS}, 2r-1}[D] := \int_M \mathrm{tr}_\xi\Big( j\big( \mathrm{CS}^{\mathrm{grp}, \mathcal{P}}_{2r-1}(D) \big) \Big).
\end{equation}
\end{definition}

\begin{remark}[Internal $A$-CS vs.\ Classical Bilinear Chern-Simons Theory]\label{rem:internal_vs_classical_CS}
Just as noted for Chern-Weil theory in Remark \ref{rem:internal_vs_external_CW}, classical 3D Chern-Simons theory on a Lie algebra $\mathfrak{g}$ uses an \emph{external symmetric} $\mathrm{Ad}$-invariant bilinear form $\kappa \in S^2(\mathfrak{g}^*)^{\mathrm{Ad}}$ to form $\int_M (\kappa(D \wedge dD) + \frac{1}{3}\kappa(D \wedge [D \wedge D]_{\mathfrak{g}}))$ (or uses the associative matrix product in $\mathrm{Mat}_{k \times k}(\mathbb{R})$). Because $\kappa$ is symmetric while $[\cdot, \cdot]_{\mathfrak{g}}$ is skew-symmetric, $\kappa(x, [y, z]_{\mathfrak{g}})$ is totally skew-symmetric in $(x, y, z)$, so the classical cubic term is non-zero for every non-abelian semi-simple Lie algebra. In contrast, Definition \ref{def:A_CS} multiplies all factors \emph{internally} within $(A, *)$ before applying a linear functional $\mathrm{tr}_\xi : E_A \to M \times \mathbb{R}$. When $(A, \curlywedge)$ is skew-commutative, the outer multiplication in $(D \curlywedge D) \curlywedge D$ is also skew-symmetric, which forces $(D \curlywedge D) \curlywedge D$ to project onto the Jacobiator $\mathrm{Jac}_\curlywedge(D, D, D)$! Moreover, for a skew-commutative product $\curlywedge$, because $dD$ has degree $2$ and $D$ has degree $1$, the quadratic term $dD \curlywedge D$ takes values in the commutator subspace $[A, A]_\curlywedge$: thus $dD \curlywedge D$ vanishes identically if $(A, \curlywedge)$ is abelian, is annihilated by any $G$-invariant trace $\tau_0 \in (A^*)^G$, and is globally exact on $P$ since $d(D \curlywedge D) = dD \curlywedge D - D \curlywedge dD = dD \curlywedge D - (-1)^{1 \cdot 2} dD \curlywedge D = 2\, dD \curlywedge D$ (so $dD \curlywedge D = \frac{1}{2} d(D \curlywedge D)$ and $d(dD \curlywedge D) = -dD \curlywedge dD = 0$).
\end{remark}

\begin{theorem}[Associator Quotient Collapse, Jacobiator Dichotomy, and Transgression Anomaly in Internal-Trace $A$-CS Theories]\label{thm:CS_obstructions}
Let $M$ be a compact oriented manifold of odd dimension $n = 2r - 1$ ($r \ge 2$), and let $E_A = P \times_\rho A \to M$ be equipped with a trace functional $\mathrm{tr}_\xi$.
\begin{enumerate}
    \item \textbf{Cubic Annihilation of Skew-Commutative Algebras in the Associative Quotient:} Suppose $(A, *)$ is any skew-commutative algebra ($x * y = -y * x$). Then in the associative quotient algebra $A/\mathfrak{Assoc}(A)$, every 3-fold product vanishes identically ($x * y * z \in \mathfrak{Assoc}(A)$ for all $x, y, z \in A$). Consequently, in $\Omega_{\mathrm{heq},\rho}(P; A)/\mathfrak{Assoc}(P, A)$:
    \begin{itemize}
        \item For $n = 3$ ($r = 2$), the cubic term $\frac{1}{3}\pi_{\mathrm{as}}(D \wedge_* D \wedge_* D) = 0$ vanishes identically, leaving only $\int_M \overline{\mathrm{tr}}_\xi(j(\pi_{\mathrm{as}}(D \wedge_* dD))) = -\int_M \overline{\mathrm{tr}}_\xi(j(\pi_{\mathrm{as}}(dD \wedge_* D)))$ (where $D \wedge_* dD = -(-1)^{1\cdot 2} dD \wedge_* D = -dD \wedge_* D$ by \eqref{eq:graded_comm_symmetry});
        \item For all odd dimensions $n = 2r - 1 \ge 5$ ($r \ge 3$), $S^{A, \mathrm{quot}}_{\mathrm{CS}, 2r-1}[D] \equiv 0$ vanishes identically.
    \end{itemize}
    \item \textbf{Unquotiented Vertex Truncation under $\mathcal{P}$-$(1, s)$-Solvability:} If $A = \bigoplus_{m \in \mathfrak{m}} A^m$ is a weak $\mathcal{P}$-$(1, s)$-solvable $\mathfrak{m}$-graded algebra ($\mathcal{P} \in \{\mathrm{blk}, \mathrm{bal}\}$), then for every $j$ satisfying $2j + 1 \ge s + 1$ (i.e., $j \ge \lceil s/2 \rceil$), the $(r + j)$-point vertex $(\wedge_{*, \mathcal{P}}^{r-1-j} dD) \wedge_* (\wedge_{*, \mathcal{P}}^{2j+1} D)$ in \eqref{eq:CS_grouped_def} vanishes identically in $\Omega^{2r-1}(P; A)$.
    \item \textbf{Unquotiented Skew-Commutative Dichotomy (Lie vs.\ Non-Lie Algebras):} Suppose $(A, \curlywedge)$ is any skew-commutative algebra ($x \curlywedge y = -y \curlywedge x$), with Jacobiator $\mathrm{Jac}_\curlywedge(x, y, z) := (x \curlywedge y) \curlywedge z + (y \curlywedge z) \curlywedge x + (z \curlywedge x) \curlywedge y$. For every 1-form $D \in \Omega^1(P; A)$, direct evaluation via \eqref{eq:wedge_star_pointwise} and \eqref{eq:Jac_form_normalization} gives the exact identity
    \begin{equation}\label{eq:cubic_D_Jacobiator}
    (D \curlywedge D) \curlywedge D = \frac{1}{3}\, \mathrm{Jac}_{\curlywedge}(D, D, D) \in \Omega^3(P; A), \quad \text{and} \quad dD \curlywedge D = \frac{1}{2}\, d(D \curlywedge D) \in \Omega^3(P; A).
    \end{equation}
    Consequently, for the unquotiented grouped Chern-Simons functional $S^{A, \mathrm{grp}, \mathcal{P}}_{\mathrm{CS}, 2r-1}[D]$:
    \begin{enumerate}
        \item[(a)] \emph{Lie Case ($\mathrm{Jac}_\curlywedge \equiv 0$, e.g., any Lie subalgebra $\mathfrak{h} \subset \mathfrak{so}(k)$ under $X \curlywedge Y = \frac{1}{2}[X,Y]$):} By \eqref{eq:cubic_D_Jacobiator}, $\curlywedge^3_{\mathrm{bal}} D = (D \curlywedge D) \curlywedge D = 0$ vanishes identically in $\Omega^3(P; \mathfrak{h})$, and hence by Lemma \ref{lem:skew_comm_powers}(2a), $\curlywedge^m_{\mathrm{bal}} D = \curlywedge^m_{\mathrm{blk}} D = 0$ vanishes for \emph{all} $m \ge 3$ under both $\mathrm{blk}$ and $\mathrm{bal}$ parenthesizations! Therefore:
        \begin{itemize}
            \item In dimension $n = 3$ ($r = 2$), the cubic vertex $\frac{1}{3}(D \curlywedge D) \curlywedge D = 0$ vanishes identically, reducing $S^{\mathfrak{h}, \mathrm{grp}}_{\mathrm{CS}, 3}[D]$ to the quadratic commutator functional $\frac{1}{2}\int_M \mathrm{tr}_\xi(j(d(D \curlywedge D)))$ (which vanishes identically by Stokes' theorem on the closed manifold $M$ whenever the equivariant lift $\hat{\xi} : P \to \mathfrak{h}^*$ is constant, $d\hat{\xi} = 0$, in the linear/horizontal regime $d\rho = 0$);
            \item In all odd dimensions $n = 2r - 1 \ge 5$ ($r \ge 3$), $S^{\mathfrak{h}, \mathrm{grp}, \mathcal{P}}_{\mathrm{CS}, 2r-1}[D] \equiv 0$ vanishes identically in $\Omega^{2r-1}(P; \mathfrak{h})$ under \emph{both} $\mathcal{P} = \mathrm{blk}$ and $\mathcal{P} = \mathrm{bal}$.
        \end{itemize}
        \item[(b)] \emph{Non-Lie Skew-Commutative Case ($\mathrm{Jac}_\curlywedge \not\equiv 0$, e.g., the octonionic Malcev algebra $\mathrm{Im}(\mathbb{O})$, reductive extensions $A = \mathbb{R}^k \oplus \mathfrak{h}$ under $\curlywedge_\rtimes$, or Cayley-Dickson algebras $\mathfrak{u}(k; \mathrm{CD}^l(\mathbb{R}))$ for $l \ge 3$):}
        \begin{itemize}
            \item In dimension $n = 3$ ($r = 2$), the cubic vertex $\frac{1}{3}(D \curlywedge D) \curlywedge D = \frac{1}{9}\mathrm{Jac}_\curlywedge(D, D, D)$ is non-vanishing in $\Omega^3(P; A)$;
            \item In dimension $n = 5$ ($r = 3$), under quartic-block grouping ($\mathcal{P} = \mathrm{blk}$), both $(dD \curlywedge dD) \curlywedge D = 0$ and $\frac{3}{20}\curlywedge_{\mathrm{blk}}^5 D = \frac{3}{20} (\curlywedge_{\mathrm{bal}}^4 D) \curlywedge D = 0$ vanish by $(2,1)$-nilpotency, collapsing the unquotiented 5D block-grouped Chern-Simons functional to the pure quartic Jacobiator vertex:
            \begin{equation}\label{eq:5D_CS_quartic}
            S^{A, \mathrm{grp}, \mathrm{blk}}_{\mathrm{CS}, 5}[D] = \frac{3}{4}\! \int_M \!\mathrm{tr}_\xi\big(j(dD \curlywedge ((D \curlywedge D) \curlywedge D))\big) = \frac{1}{4}\! \int_M \!\mathrm{tr}_\xi\big(j(dD \curlywedge \mathrm{Jac}_\curlywedge(D, D, D))\big)
            \end{equation}
            (whereas under recursive binary halving $\mathcal{P} = \mathrm{bal}$, $S^{A, \mathrm{grp}, \mathrm{bal}}_{\mathrm{CS}, 5}[D]$ additionally retains the $J_4 = 5$ quintic Jacobsthal vertex $\frac{3}{20}\int_M \mathrm{tr}_\xi(j(\curlywedge_{\mathrm{bal}}^5 D)) = \frac{1}{20}\int_M \mathrm{tr}_\xi(j(\mathrm{Jac}_\curlywedge(D,D,D) \curlywedge (D \curlywedge D)))$);
            \item In all odd dimensions $n = 2r - 1 \ge 7$ ($r \ge 4$), the block-grouped $A$-valued Chern-Simons functional $S^{A, \mathrm{grp}, \mathrm{blk}}_{\mathrm{CS}, 2r-1}[D]$ vanishes identically in $\Omega^{2r-1}(P; A)$, while under recursive binary halving ($\mathcal{P} = \mathrm{bal}$), $S^{A, \mathrm{grp}, \mathrm{bal}}_{\mathrm{CS}, n}[D]$ vanishes algebraically for every odd dimension $n \ge 7$ with $n \notin \bigcup_{m=4}^\infty \{J_m, J_m + 2\} \cap [7, \infty) = \{7\} \cup \{11, 13\} \cup \{21, 23\} \cup \{43, 45\} \cup \dots$ (where $J_m = \frac{2^m - (-1)^m}{3}$ is the $m$-th odd Jacobsthal number) and vanishes by rank for all $n \ge \dim_{\mathbb{R}} A + 3$.
        \end{itemize}
    \end{enumerate}
    \item \textbf{Exterior Derivative in 3D and 5D:} For any skew-commutative algebra $(A, \curlywedge)$, since $d(dD \curlywedge D) = -dD \curlywedge dD = 0$, the exterior derivatives of $\mathrm{CS}^{\mathrm{grp}}_3(D)$ and $\mathrm{CS}^{\mathrm{grp}, \mathrm{blk}}_5(D)$ are given purely by their Jacobiator terms:
    \begin{equation}\label{eq:3D_5D_CS_transgression_anomaly}
    d\,\mathrm{CS}^{\mathrm{grp}}_3(D) = \frac{1}{9}\, d\big(\mathrm{Jac}_\curlywedge(D, D, D)\big), \qquad d\,\mathrm{CS}^{\mathrm{grp}, \mathrm{blk}}_5(D) = -\frac{1}{4}\, dD \curlywedge d\big(\mathrm{Jac}_\curlywedge(D, D, D)\big),
    \end{equation}
    where $\frac{1}{3}d(\mathrm{Jac}_\curlywedge(D, D, D)) = d((D \curlywedge D) \curlywedge D) = 2(dD \curlywedge D) \curlywedge D - dD \curlywedge (D \curlywedge D)$.
\end{enumerate}
\end{theorem}

\begin{proof}
To prove (1), in the associative quotient $A/\mathfrak{Assoc}(A)$, alternating use of skew-commutativity and associativity gives $(x * y) * z = -(y * x) * z = -y * (x * z) = +(x * z) * y = x * (z * y) = -x * (y * z) = -(x * y) * z$, so $2(x * y * z) = 0 \implies x * y * z = 0$ in $A/\mathfrak{Assoc}(A)$. Directly on differential forms in $\Omega_{\mathrm{heq},\rho}(P; A)/\mathfrak{Assoc}(P, A)$, since $D \wedge_* D$ is a 2-form and $D$ is a 1-form while $\alpha \wedge_* \beta = -(-1)^{kl}\beta \wedge_* \alpha$ by \eqref{eq:graded_comm_symmetry}, we have $(D \wedge_* D) \wedge_* D = \allowbreak -(-1)^{2 \cdot 1} D \wedge_* (D \wedge_* D) = -D \wedge_* (D \wedge_* D) \equiv -(D \wedge_* D) \wedge_* D \pmod{\mathfrak{Assoc}(P, A)}$, so $\pi_{\mathrm{as}}((D \wedge_* D) \wedge_* D) = 0$ (and similarly any $\ge 3$-fold product of $D$ and $F_t$ vanishes in the quotient).

Part (2) is immediate since $\wedge_{*, \mathcal{P}}^{2j+1} D = 0$ whenever $2j + 1 \ge s + 1$.

For (3), since $(D \curlywedge D)_p(u, w) = 2\, D_p(u) \curlywedge D_p(w)$, evaluating $((D \curlywedge D) \curlywedge D)_p(v_1, v_2, v_3)$ directly from \eqref{eq:wedge_star_pointwise} sums $2(D_p(v_{\sigma(1)}) \curlywedge D_p(v_{\sigma(2)})) \curlywedge D_p(v_{\sigma(3)})$ with prefactor $\frac{1}{2!\, 1!}$ over $\sigma \in S_3$; grouping into the $3$ cyclic pairs $(123), (231), (312)$ yields $2\,\mathrm{Jac}_\curlywedge(D_p(v_1), D_p(v_2), D_p(v_3)) = \frac{1}{3}\mathrm{Jac}_\curlywedge(D, D, D)_p(v_1, v_2, v_3)$ by \eqref{eq:Jac_form_normalization}. Furthermore, since $\alpha \curlywedge \beta = -(-1)^{kl} \beta \curlywedge \alpha$, we have $d(D \curlywedge D) = dD \curlywedge D - D \curlywedge dD = dD \curlywedge D - (-1)^{1 \cdot 2} dD \curlywedge D = 2\, dD \curlywedge D$, proving \eqref{eq:cubic_D_Jacobiator}. In the $j$-th vertex $(\curlywedge_{\mathcal{P}}^{r-1-j} dD) \curlywedge (\curlywedge_{\mathcal{P}}^{2j+1} D)$ of \eqref{eq:CS_grouped_def}, $dD \curlywedge dD = 0$ kills all terms with $r - 1 - j \ge 2 \iff j \le r - 3$. If $\mathrm{Jac}_\curlywedge \equiv 0$, then $\curlywedge^3_{\mathrm{bal}} D = 0$ and $\curlywedge^4_{\mathrm{bal}} D = 0$, and by Lemma \ref{lem:skew_comm_powers}(2a), $\curlywedge^m_{\mathrm{bal}} D = \curlywedge^m_{\mathrm{blk}} D = 0$ for all $m \ge 3$ ($2j+1 \ge 3 \iff j \ge 1$), leaving only $j = 0$ at $r = 2$ ($n = 3$) and killing all terms for $r \ge 3$ ($n \ge 5$). If $\mathrm{Jac}_\curlywedge \not\equiv 0$, under $\mathcal{P} = \mathrm{blk}$ we have $\curlywedge_{\mathrm{blk}}^{2j+1} D = 0$ for $2j + 1 \ge 4 \iff j \ge 2$, so only $r - 2 \le j \le 1$ can survive: $j \in \{0, 1\}$ survive at $n = 3$, $j = 1$ survives at $n = 5$ ($r = 3$), and $r - 2 \le j \le 1$ is empty for $r \ge 4$ ($n \ge 7$). Under $\mathcal{P} = \mathrm{bal}$, since $dD \curlywedge dD = 0$ restricts $j$ to $j \in \{r-2, r-1\}$ where $2j+1 \in \{n-2, n\}$ is odd, Lemma \ref{lem:skew_comm_powers}(2a),(2c) implies that $\curlywedge_{\mathrm{bal}}^{2j+1} D = 0$ unless $2j+1 = J_m$ for an odd Jacobsthal number $J_m$ ($m \ge 3$) with $J_m \le \dim_{\mathbb{R}} A$.

Finally, for (4), since $d(dD \curlywedge D) = \frac{1}{2} d^2(D \curlywedge D) = 0$, differentiating $\mathrm{CS}^{\mathrm{grp}}_3(D) = dD \curlywedge D + \frac{1}{9}\mathrm{Jac}_\curlywedge(D, D, D)$ and $\mathrm{CS}^{\mathrm{grp}, \mathrm{blk}}_5(D) = \frac{1}{4} dD \curlywedge \mathrm{Jac}_\curlywedge(D, D, D)$ immediately gives \eqref{eq:3D_5D_CS_transgression_anomaly}.
\end{proof}

\subsection{Applications to \texorpdfstring{$A$-$BF$}{A-BF} Theories, the Yang-Mills-Higgs Mechanism, and Reductive Dualities}

We conclude this section by applying our framework to $A$-valued $BF$ theories \cite{BaezBF, CattaneoBF}, the $A$-valued Yang-Mills-Higgs mechanism \cite{Higgs1964, EnglertBrout, Westenholz, Sardanashvily} (including an explicit octonionic Malcev Proca mass spectrum computation), and reductive dualities.

\begin{proposition}[Cattaneo-Plebanski $A$-$BF$ Relations]\label{prop:BF_obstructions}
Let $M$ be a compact oriented $n$-manifold ($n \ge 2$) and $E_A = P \times_\rho A \to M$.
\begin{enumerate}
    \item Given pairings $\langle \cdot, \cdot \rangle_{\mathrm{BF}}$ and $\langle \cdot, \cdot \rangle_B$ on $\Omega^2(M; E_A)$ and $\Omega^{n-2}(M; E_A)$ related via a linear isomorphism $\star_M : \Omega^2(M; E_A) \xrightarrow{\sim} \Omega^{n-2}(M; E_A)$ by $\langle \Phi, \mathcal{B} \rangle_{\mathrm{BF}} = \langle \star_M \Phi, \mathcal{B} \rangle_B$ with $\langle \cdot, \cdot \rangle_B$ symmetric and non-degenerate, the deformed $A$-$BF$ functional
    \begin{equation}\label{eq:general_BF}
    S^A_{\mathrm{BF}, \mu}[D, B] := \int_M \Big( \langle j(F_D), j(B) \rangle_{\mathrm{BF}} - \frac{\mu}{2} \langle j(B), j(B) \rangle_B \Big) \qquad (\mu \neq 0)
    \end{equation}
    has the unique auxiliary Euler-Lagrange solution $j(B_{\mathrm{on\text{-}shell}}(D)) = \frac{1}{\mu}\star_M j(F_D)$, whose substitution into \eqref{eq:general_BF} recovers $\frac{1}{2\mu} S^A[D]$ with $\langle \Phi, \Psi \rangle_A := \langle \star_M \Phi, \star_M \Psi \rangle_B$.
    \item In the multiplicative $A$-$BF$ functional $S^A_{\mathrm{BF}, *}[D, B] := \int_M \mathrm{tr}_\xi(j(F_D \wedge_* B))$, if $A \simeq A_0 \oplus A_1$ has a weak $\mathcal{P}$-$(k,s)$-solvable submodule $A_0$ and $B_e = \wedge_{*, \mathcal{P}}^{n-2} e$ for $e \in \Omega^1_{\mathrm{heq},\rho}(P; A_0)$, then $B_e = 0$ for $n \ge k + s + 3$ \cite{MartinsBiezunerEHP}; if $n = 2r$ is even, $A$ is weak $\mathcal{P}$-$(2,s)$-solvable, and $B = \wedge_{*, \mathcal{P}}^{r-1} \beta$ for $\beta \in \Omega^2_{\mathrm{heq},\rho}(P; A)$, then $B = 0$ for $n \ge 2(s+2)$.
\end{enumerate}
\end{proposition}

\begin{proof}
In (1), varying \eqref{eq:general_BF} with respect to $\dot{B}$ gives $\int_M \langle \star_M j(F_D) - \mu j(B), j(\dot{B}) \rangle_B = 0 \iff j(B) = \frac{1}{\mu}\star_M j(F_D)$. Part (2) follows directly from $\wedge_{*, \mathcal{P}}^{k+s+1} e = 0$ and $\wedge_{*, \mathcal{P}}^{s+1} \beta = 0$.
\end{proof}

Next, let $\iota : G \hookrightarrow \hat{G}$ be a closed Lie subgroup inclusion inducing the extended principal $\hat{G}$-bundle $\hat{P} = P \times_G \hat{G} \to M$ and canonical reduction $\iota_P : P \hookrightarrow \hat{P}$ \cite{KobayashiNomizu, MichorGauge}, which defines the canonical global Higgs section $\phi_P \in \mathrm{Higgs}(\hat{P}; G) := \Gamma(\hat{P} \times_{\hat{G}} (\hat{G}/G))$. Suppose $(\hat{A}, *)$ is a finite-dimensional real algebra with representation $\rho : \hat{G} \to \mathrm{Aut}_{\mathrm{Alg}}(\hat{A})$ admitting a $\hat{G}$-equivariant map $\theta : \hat{G}/G \to \hat{A}$, $\theta(agG) = \rho(a)(\theta(gG))$, and let $\Phi_\phi := j^{-1}(\theta_*(\phi)) \in \Omega^0_{\mathrm{heq}, \rho}(\hat{P}; \hat{A})$ be the lifted Higgs 0-form (where $\theta_*$ denotes the induced bundle map $\hat{P} \times_{\hat{G}} (\hat{G}/G) \to E_{\hat{A}}$). Along the unitary-gauge reduction $\iota_P : P \hookrightarrow \hat{P}$, the canonical section $\phi_P$ takes the constant value $\Phi_0 := \theta(eG) \in \hat{A}^G$, so $d(\iota_P^* \Phi_{\phi_P}) = 0$. Consequently:
\begin{itemize}
    \item In the Ehresmann-compatible setting of Proposition \ref{prop:Cartan_Ehresmann} ($d\rho(X)(y) = \frac{1}{2}[\eta(X), y]_*$), the commutator covariant derivative $\nabla^{\hat{A}, [\cdot,\cdot]_*}_{\hat{D}} \Phi_\phi := d\Phi_\phi + \frac{1}{2}\hat{D} \wedge_{[\cdot,\cdot]_*} \Phi_\phi \in \Omega^1_{\mathrm{heq}, \rho}(\hat{P}; \hat{A})$ pulls back along $\iota_P : P \hookrightarrow \hat{P}$ at $\phi = \phi_P$ to $\mathcal{R}_{\Phi_0}(\iota_P^* \hat{D})$, where $\mathcal{R}_{\Phi_0} : \hat{A} \to \hat{A}$ is the linear mass operator $\mathcal{R}_{\Phi_0}(X) = \frac{1}{2}[X, \Phi_0]_*$.
    \item In the horizontal/flat setting ($\eta = 0, d = d_0$) with one-sided coupling $\nabla^{\hat{A}}_{\hat{D}} \Phi_0 = \hat{D} \wedge_* \Phi_0$, the linear mass operator is $\mathcal{R}_{\Phi_0}(X) = X * \Phi_0$.
\end{itemize}
Given a positive-definite fiberwise inner product $\langle \cdot, \cdot \rangle_{\hat{A}}$ on $\hat{A}$ and a Riemannian metric $g$ on $M$ inducing the kinetic pairing $\langle \alpha, \beta \rangle_{\mathrm{kin}} = \langle \alpha \wedge *_g \beta \rangle_{\hat{A}}$ on $\Omega^1(M; E_{\hat{A}})$, let $N_{\mathrm{mass}}(\Phi_0, x) = \mathrm{rk}(\mathcal{R}_{\Phi_0})$ denote the rank of the Proca mass quadratic form $\langle j(\mathcal{R}_{\Phi_0}(\hat{D})), j(\mathcal{R}_{\Phi_0}(\hat{D})) \rangle_{\mathrm{kin}, x}$ at $x \in M$, and let $\mathfrak{g}_{\Phi_0} := \ker(\mathcal{R}_{\Phi_0}) \subset \hat{A}$ be the algebraic stabilizer of $\Phi_0$.

\begin{proposition}[Unitary-Gauge Proca Mass Bounds and Octonionic/Reductive Spectra in $A$-YMH Theories]\label{prop:Higgs_obstructions}
Let $\Phi_0 = \theta(eG) \in \hat{A}^G$ be the unitary-gauge Higgs vacuum on $P \hookrightarrow \hat{P}$.
\begin{enumerate}
    \item \textbf{Splitting Kernel Decoupling:} If $\hat{A}$ arises from a split extension $0 \to H \to \hat{A} \stackrel{q}{\to} A \to 0$ with pulled-back product $x *' y = s(q(x) * q(y))$ and $\Phi_0 \in H = \ker(q)$, then $\mathcal{R}_{\Phi_0} \equiv 0$ and $N_{\mathrm{mass}}(\Phi_0, x) = 0$.
    \item \textbf{Derived and Commutator Rank Bounds:} Because $\mathrm{im}(\mathcal{R}_{\Phi_0}) \subset [\hat{A}, \hat{A}]_*$ for the commutator coupling (resp.\ $\mathrm{im}(\mathcal{R}_{\Phi_0}) \subset \hat{A}^{(1)} = \hat{A} * \hat{A}$ for the one-sided coupling), the Proca mass rank and the algebraic stabilizer dimension $\dim_{\mathbb{R}} \mathfrak{g}_{\Phi_0} = \dim_{\mathbb{R}} \hat{A} - \mathrm{rk}(\mathcal{R}_{\Phi_0})$ satisfy
    \begin{align}
    N_{\mathrm{mass}}(\Phi_0, x) &\le \mathrm{rk}(\mathcal{R}_{\Phi_0}) \le \dim_{\mathbb{R}} [\hat{A}, \hat{A}]_*, & \dim_{\mathbb{R}} \mathfrak{g}_{\Phi_0} &\ge \dim_{\mathbb{R}} \hat{A} - \dim_{\mathbb{R}} [\hat{A}, \hat{A}]_* \; (\text{comm.}), \label{eq:Proca_comm_bound} \\
    N_{\mathrm{mass}}(\Phi_0, x) &\le \mathrm{rk}(\mathcal{R}_{\Phi_0}) \le \dim_{\mathbb{R}} \hat{A}^{(1)}, & \dim_{\mathbb{R}} \mathfrak{g}_{\Phi_0} &\ge \dim_{\mathbb{R}} \hat{A} - \dim_{\mathbb{R}} \hat{A}^{(1)} \;\; (\text{one-sided}). \label{eq:Proca_mass_bound}
    \end{align}
    If $\Phi_0 \in \hat{A}^{(r)}$ (resp.\ $\Phi_0 \in \hat{A}^r$), then $N_{\mathrm{mass}}(\Phi_0, x) \le \dim_{\mathbb{R}}(\hat{A} * \hat{A}^{(r)} + \hat{A}^{(r)} * \hat{A})$ (resp.\ $\le \dim_{\mathbb{R}} \hat{A}^{r+1}$).
    \item \textbf{Explicit Octonionic Malcev and Reductive Cartan Mass Spectra:}
    \begin{enumerate}
        \item[(a)] \emph{Octonionic Gauge Theory ($\hat{A} = \mathbb{O} = \mathbb{R} \cdot 1 \oplus \mathrm{Im}(\mathbb{O})$):} For any vacuum $\Phi_0 = c_0 \cdot 1 + \vec{\phi}_0 \in \mathbb{O}$ with $|\vec{\phi}_0| = v > 0$ under the commutator coupling $\mathcal{R}_{\Phi_0}(X) = \frac{1}{2}[X, \Phi_0]_* = \vec{X} \times_{G_2} \vec{\phi}_0$, the central unit $\mathbb{R}\cdot 1$ and the parallel imaginary line $\mathbb{R}\cdot \vec{\phi}_0$ span a 2-dimensional massless stabilizer $\mathfrak{g}_{\Phi_0} = \mathbb{R}\cdot 1 \oplus \mathbb{R}\cdot \vec{\phi}_0 \simeq \mathbb{C}$, while on the orthogonal 6-plane $\vec{\phi}_0^\perp \subset \mathrm{Im}(\mathbb{O})$ the operator $\mathcal{R}_{\Phi_0}^2 = -v^2\,\mathrm{id}_{\vec{\phi}_0^\perp}$ gives $N_{\mathrm{mass}}(\Phi_0) = 6$ degenerate massive Proca gauge bosons of mass $m = v$ (breaking $G_2 \to SU(3)$, the stabilizer of $\vec{\phi}_0$ in $G_2$).
        \item[(b)] \emph{Reductive Cartan Algebra ($\hat{A} = \mathbb{R}^k \oplus \mathfrak{so}(k)$ under $\curlywedge_\rtimes$ with $\times_H = 0$):} For a pure translational vacuum $\Phi_0 = (v_0, 0) \in \mathbb{R}^k \oplus \{0\}$ with $|v_0| = v > 0$, $\mathcal{R}_{\Phi_0}(u, X) = (u, X) \curlywedge_\rtimes (v_0, 0) = (\frac{1}{2}X v_0,\, \sigma(u \wedge v_0))$. Here $\ker(\mathcal{R}_{\Phi_0}) = \mathbb{R} v_0 \oplus \mathfrak{so}(k-1)_{v_0}$ has dimension $1 + \binom{k-1}{2}$, while $N_{\mathrm{mass}}(\Phi_0) = 2(k-1)$ gauge modes ($k-1$ boost/coset generators in $\mathfrak{so}(k)/\mathfrak{so}(k-1)$ with mass $m_{\mathrm{rot}} = \frac{1}{2}v$ and $k-1$ orthogonal soldering modes in $v_0^\perp \subset \mathbb{R}^k$ with mass $m_{\mathrm{sold}} = v$) acquire Proca masses.
    \end{enumerate}
\end{enumerate}
\end{proposition}

\begin{proof}
Part (1) follows from $q(\Phi_0) = 0 \implies \mathcal{R}_{\Phi_0}(X) = 0$, and part (2) follows from $\mathrm{rk}(\mathcal{R}_{\Phi_0}) = \dim_{\mathbb{R}} \mathrm{im}(\mathcal{R}_{\Phi_0})$ together with $\frac{1}{2}[X, \Phi_0]_* \in [\hat{A}, \hat{A}]_*$ and $X * \Phi_0 \in \hat{A}^{(1)}$. For (3a), in $\mathbb{O}$, $\frac{1}{2}[x_0 + \vec{x}, c_0 + \vec{\phi}_0]_* = \vec{x} \times_{G_2} \vec{\phi}_0$, and by the octonionic cross-product identity $(\vec{x} \times_{G_2} \vec{\phi}_0) \times_{G_2} \vec{\phi}_0 = \langle \vec{x}, \vec{\phi}_0 \rangle \vec{\phi}_0 - |\vec{\phi}_0|^2 \vec{x}$, the operator $\mathcal{R}_{\Phi_0}^*\mathcal{R}_{\Phi_0}$ has eigenvalue $0$ on $\mathrm{span}\{1, \vec{\phi}_0\}$ (multiplicity 2) and eigenvalue $v^2$ on $\vec{\phi}_0^\perp \subset \mathrm{Im}(\mathbb{O})$ (multiplicity 6). For (3b), $X v_0 = 0 \iff X \in \mathfrak{so}(k-1)_{v_0}$ and $u \wedge v_0 = 0 \iff u \in \mathbb{R} v_0$, giving $\dim\ker(\mathcal{R}_{\Phi_0}) = 1 + \binom{k-1}{2}$ and $\mathrm{rk}(\mathcal{R}_{\Phi_0}) = 2(k-1)$.
\end{proof}

Finally, given a reductive decomposition $A \simeq A_0 \oplus A_1$ where all products are evaluated via the ambient multiplication $*$ of $A$, we define the \emph{geometric dual} ${}^* S^A[e, \omega] := \int_M \langle j({}^* F_{(e,\omega)}), j({}^* F_{(e,\omega)}) \rangle_A$ on $\Omega^1_{\mathrm{heq},\rho}(P; A_0) \oplus \Omega^1_{\mathrm{heq},\rho}(P; A_1)$ by interchanging the roles of $e$ and $\omega$ in the differential and quadratic sectors (${}^* F_{(e,\omega)} := (de + \frac{1}{2} e \wedge_* e) + (d\omega + \frac{1}{2}\omega \wedge_{[\cdot,\cdot]_*} e) + \frac{1}{2}\omega \wedge_* \omega$), and the \emph{algebraic dual} $(S^A)^*[e', \omega'] := \int_M \langle j(F_{(e',\omega')}), j(F_{(e',\omega')}) \rangle_A$ on the swapped configuration space $(e', \omega') \in \Omega^1_{\mathrm{heq},\rho}(P; A_1) \oplus \Omega^1_{\mathrm{heq},\rho}(P; A_0)$ with $F_{(e',\omega')} := (d\omega' + \frac{1}{2}\omega' \wedge_* \omega') + (de' + \frac{1}{2}\omega' \wedge_{[\cdot,\cdot]_*} e') + \frac{1}{2} e' \wedge_* e'$. Applying Proposition \ref{prop:abelianization} and Theorem \ref{thm:Berger_CYM}(1) directly to the swapped 1-forms yields:

\begin{proposition}[Duality Transfer of Obstructions]\label{prop:dual_obstructions}
Let $A \simeq A_0 \oplus A_1$ be an $\mathfrak{m}$-graded algebra. If $A_0$ is weak $(1,1)$-solvable under $*$, then $e \wedge_* e = 0$ and $\omega' \wedge_* \omega' = 0$, abelianizing the differential curvature sector of both ${}^* S^A$ and $(S^A)^*$ in all dimensions $n \ge 2$. If $A_1 = \mathfrak{h} \subset \mathfrak{so}(k_1) \oplus \dots \oplus \mathfrak{so}(k_r)$ is a $k$-algebra under $\curlywedge$, then the block-balanced dual Cartan-Yang-Mills functionals ${}^* S^{\mathrm{CYM}}_{n, \xi, \mathrm{blk}}[e, \omega]$ and $(S^{\mathrm{CYM}}_{n, \xi, \mathrm{blk}})^*[e', \omega']$ (formed with constitutive powers $\curlywedge_{\mathrm{blk}}^m \omega$ and $\curlywedge_{\mathrm{blk}}^m e'$) have vanishing cosmological terms for $n \ge 4$, vanishing EHP terms for $n \ge 6$, and vanish identically for $n \ge 8$.
\end{proposition}

\section{Realizations and Concrete/Abstract Examples}\label{sec:examples}

We conclude by applying our theorems to concrete matrix geometries, Cayley-Dickson geometries, and abstract non-associative and graded algebras.

\subsection{Classical, Generalized Complex, and Exceptional Geometries}

By Theorem \ref{thm:topological_obstruction}(2) and Theorem \ref{thm:Berger_CYM}(1), any algebra $(A, *)$ that is $(2,1)$-nil and block-balanced $(1,3)$-nil obstructs 4D internal-product topological/self-dual multiplicative $A$-YMT theories and cuts off block-balanced Cartan-Yang-Mills theories for $n \ge 8$. Since every subalgebra $\mathfrak{h} \subset \mathfrak{so}(k_1) \oplus \dots \oplus \mathfrak{so}(k_r)$ (as well as $\mathfrak{h} \otimes_{\mathbb{R}} \mathbb{C} \subset \mathfrak{so}(k; \mathbb{C})$ and $\mathbb{R}^k \oplus \mathfrak{h}$ under $\curlywedge_\rtimes$) is skew-commutative, Lemma \ref{lem:skew_comm_powers}(1) shows that this encompasses:
\begin{enumerate}
    \item \textbf{Classical $G$-Structures in Berger's List and Spinor Embeddings:} $\mathfrak{so}(k)$, $\mathfrak{u}(k)$, $\mathfrak{su}(k)$, $\mathfrak{sp}(k)$, $\mathfrak{sp}(k) \oplus \mathfrak{sp}(1)$, $\mathfrak{g}_2 \subset \mathfrak{so}(7)$, $\mathfrak{spin}(7) \subset \mathfrak{so}(8)$, and $\mathfrak{spin}(9) \subset \mathfrak{so}(16)$, as well as their covering groups $\mathrm{Spin}(k)$ and product distributions \cite{Berger, Besse, MartinsBiezunerEHP}.
    \item \textbf{Type II String and Hitchin's Generalized Geometries:} The Narain orthogonal reduction $O(k) \times O(k) \hookrightarrow O(k, k)$ \cite{Narain, Coimbra}, Hitchin and Gualtieri's generalized complex geometry $U(k, k) \hookrightarrow O(2k, 2k)$ \cite{HitchinGCY, Gualtieri}, generalized Calabi-Yau geometry $SU(k, k) \hookrightarrow O(2k, 2k)$, and generalized K\"ahler geometry $U(k) \times U(k) \hookrightarrow O(2k, 2k)$ \cite{Gualtieri}.
    \item \textbf{Exotic Orthogonal Subgroups and Exceptional Lie Algebras:} Obata's 15 infinite families of subgroups $H \subset O(k)$ \cite{Obata}, maximal tori $T(q; \mathbb{K}) \hookrightarrow O(q; \mathbb{K})$ ($\mathrm{ch}(\mathbb{K}) \neq 2$ \cite{BayerFluckiger}), and the exceptional orthogonal embeddings $O(3) \hookrightarrow G_2$, $O(9) \hookrightarrow F_4$, $O(10) \hookrightarrow E_6$, $O(12) \hookrightarrow E_7$, and $O(16) \hookrightarrow E_8$.
\end{enumerate}

\subsection{Infinite Families of Real and Generalized Cayley-Dickson Geometries}

Let $(A, *)$ be a finite-dimensional real algebra with involution $\overline{(-)} : A \to A$, and let $\mathrm{CD}(A) = A \oplus A$ be its Cayley-Dickson double with $(x, y) * (z, w) = (x * z - \overline{w} * y, w * x + y * \overline{z})$ and $\overline{(x, y)} = (\overline{x}, -y)$ \cite{SchaferCD, BaezOctonions}. For $k \ge 1$, let $U(k; A) \subset \mathrm{Mat}_{k \times k}(A)$ denote the $A$-unitary matrices preserving $s(x, y) = \sum_{i=1}^k \overline{x}_i * y_i$ \cite{MartinsBiezunerEHP}. Iterating $U(k; \mathrm{CD}(A)) \hookrightarrow U(2k; A)$ yields for every $l \ge 1$ \cite{MartinsBiezunerEHP}:
\begin{align}
U\big(k; \mathrm{CD}^l(\mathbb{R})\big) \hookrightarrow \dots \hookrightarrow U(2^{l-2}k; \mathbb{H}) \hookrightarrow U(2^{l-1}k; \mathbb{C}) &\hookrightarrow O(2^l k; \mathbb{R}), \label{eq:CD_real} \\
U\big(k, k; \mathrm{CD}^l(A)\big) \hookrightarrow O\big(2^l k, 2^l k; \mathrm{CD}^l(A)\big) &\simeq O\big(2^{l+1} k; \mathrm{CD}^l(A)\big), \label{eq:CD_gen}
\end{align}
recovering Hitchin's generalized complex geometry for $l=1$ \cite{Gualtieri} and Pantilie's generalized quaternionic geometry for $l=2$ \cite{Pantilie, Deschamps}. Because the algebra $\mathfrak{o}(m; B)$ of $B$-skew-Hermitian matrices under $X \curlywedge Y = \frac{1}{2}(XY - YX)$ is skew-commutative for any involutive algebra $B$, $\mathfrak{u}(k; \mathrm{CD}^l(\mathbb{R}))$ is $(2,1)$-nil and block-balanced $(1,3)$-nil for all $l \ge 1$. Moreover, $\mathfrak{u}(k; \mathrm{CD}^l(\mathbb{R}))$ is a Lie algebra ($\mathrm{Jac}_\curlywedge = 0$) for $l \in \{1, 2\}$ ($\mathbb{C}, \mathbb{H}$), whereas for $l \ge 3$ ($\mathbb{O}$, sedenions, etc.) it is a non-Lie skew-commutative algebra with $\mathrm{Jac}_\curlywedge \neq 0$.

\begin{corollary}[Obstructions in Cayley-Dickson Geometries of Arbitrary Order]\label{cor:universal_triviality}
For every $k \ge 1$ and $l \ge 1$, the algebras $\mathfrak{u}(k; \mathrm{CD}^l(\mathbb{R}))$ and $\mathfrak{u}(k, k; \mathrm{CD}^l(A))$ (and their reductive extensions by $\mathbb{R}^{2^l k}$) are $(2,1)$-nil and block-balanced $(1,3)$-nil under $\curlywedge$ and $\curlywedge_\rtimes$. Therefore:
\begin{enumerate}
    \item Every 4D internal-product topological $A$-YMT theory $\int_M \mathrm{tr}_\xi(j(F_D \curlywedge F_D))$ and every 4D self-dual internal multiplicative $A$-YMT theory vanishes identically;
    \item Every block-balanced Cartan-Yang-Mills theory \eqref{eq:CYM_action} (and its geometric and algebraic duals) has a vanishing cosmological term for $n \ge 4$, a vanishing EHP term for $n \ge 6$, and vanishes identically for $n \ge 8$;
    \item For the unquotiented grouped $A$-valued Chern-Simons functional $S^{A, \mathrm{grp}, \mathcal{P}}_{\mathrm{CS}, 2r-1}[D]$ valued in the algebra $\mathfrak{u}(k; \mathrm{CD}^l(\mathbb{R}))$: when $l \in \{1, 2\}$ ($\mathrm{Jac}_\curlywedge = 0$), the 3D cubic vertex vanishes and all $\ge 5$D internal $A$-CS theories vanish identically under both $\mathrm{blk}$ and $\mathrm{bal}$; when $l \ge 3$ ($\mathrm{Jac}_\curlywedge \neq 0$), the 3D cubic and 5D quartic Jacobiator vertices \eqref{eq:5D_CS_quartic} survive under $\mathrm{blk}$ (together with the 5D quintic Jacobsthal vertex under $\mathrm{bal}$), while for odd dimensions $n = 2r - 1 \ge 7$ all block-grouped $A$-CS theories vanish identically (and binary-balanced $A$-CS theories vanish outside $n \in \bigcup_{m \ge 4}\{J_m, J_m + 2\} \cap [7, \infty) = \{7\} \cup \{11, 13\} \cup \dots$ and for $n \ge \dim_{\mathbb{R}} \mathfrak{u}(k; \mathrm{CD}^l(\mathbb{R})) + 3$).
\end{enumerate}
\end{corollary}

\subsection{Abstract Examples: Lie Algebras, Octonionic Malcev Algebras, Graded Lie Superalgebras, and Nil Algebras}

\begin{enumerate}
    \item \textbf{Lie Algebras and Graded Lie Superalgebras under $[\wedge]_{\mathfrak{g}}$:} Let $(\mathfrak{g}, [\cdot, \cdot])$ be a Lie algebra. Because the Lie bracket is skew-commutative and satisfies the Jacobi identity, $\mathrm{Jac}_{[\cdot,\cdot]} \equiv 0$ (so $d_D^{[\cdot,\cdot]} F_D = 0$ and infinitesimal gauge covariance hold universally by Theorem \ref{thm:Bianchi_gauge_EL}), while $\mathfrak{g}$ is $(2,1)$-nil under $[\wedge]_{\mathfrak{g}}$, meaning $F_D [\wedge]_{\mathfrak{g}} F_D = 0$ identically. More generally, if $\mathfrak{g} = \mathfrak{g}^0 \oplus \mathfrak{g}^1$ is an $\mathfrak{m}$-graded Lie superalgebra, its even part $\mathfrak{g}^0$ is $(2,1)$-nil. By Theorem \ref{thm:topological_obstruction} and Remark \ref{rem:internal_vs_external_CW}, any topological or self-dual multiplicative $A$-YMT functional built solely from the internal Lie bracket $[\wedge]_{\mathfrak{g}}$ and a linear trace functional $\mathrm{tr}_\xi : E_{\mathfrak{g}} \to M \times \mathbb{R}$ vanishes identically, underscoring why classical 4D Chern-Weil theory requires an external symmetric bilinear invariant polynomial $\kappa \in \mathrm{inv}^2(\mathfrak{g})$.
    \item \textbf{Non-Lie Octonionic and Malcev Gauge Theories:} Let $A = \mathbb{O}$ be the octonions (or its 7-dimensional imaginary subspace $\mathrm{Im}(\mathbb{O}) \simeq \mathbb{R}^7$ equipped with $[x,y]_* = xy - yx = 2\, x \times_{G_2} y$, which is a non-Lie Malcev algebra \cite{BaezOctonions, SchaferBook}). For any $\mathbb{O}$-valued connection 1-form $D = D_{\mathrm{re}} \cdot 1 + \vec{D} \in \Omega^1(P; \mathbb{R} \oplus \mathrm{Im}(\mathbb{O}))$, the real unit part $D_{\mathrm{re}} \cdot 1$ lies in the center ($[\mathbb{O}, 1]_* = 0$) and the real inner-product part $-\langle \vec{D} \wedge \vec{D} \rangle = 0$ vanishes by symmetry of $\langle \cdot, \cdot \rangle$, so $D_{\mathrm{re}} \cdot 1$ decouples from $D \wedge_* D = \frac{1}{2} D \wedge_{[\cdot,\cdot]_*} D = \vec{D} \wedge_{\times_{G_2}} \vec{D} \in \Omega^2(P; \mathrm{Im}(\mathbb{O}))$ by Proposition \ref{prop:abelianization}(3). Because $\mathrm{Jac}_{[\cdot,\cdot]_*}(x,y,z) = 6\, \mathrm{Assoc}_*(x,y,z) \neq 0$ on non-associative octonionic triples, the Lie nucleus of $\mathrm{Im}(\mathbb{O})$ is trivial ($\mathcal{N}_{\mathrm{Lie}}(\mathrm{Im}(\mathbb{O})) = \{0\}$, so admissible connections on $\mathrm{Im}(\mathbb{O})$ are horizontal by Proposition \ref{prop:Cartan_Ehresmann}(1), while on full $\mathbb{O}$ the Lie nucleus is $\mathcal{N}_{\mathrm{Lie}}(\mathbb{O}) = \mathbb{R} \cdot 1$), and Theorem \ref{thm:Bianchi_gauge_EL}(2)--(3) shows that $d_D^{[\cdot,\cdot]_*} F_D = -\frac{1}{4}\mathrm{Assoc}_*(\vec{D}, \vec{D}, \vec{D}) \neq 0$ whenever $\vec{D}$ spans a non-associative 3-plane in $\mathrm{Im}(\mathbb{O})$ (for $n \ge 3$), whereas $d_D^{[\cdot,\cdot]_*} F_D = 0$ holds identically whenever $\dim M = 2$ or whenever $\vec{D}$ takes values in an associative quaternionic subalgebra $\mathbb{H} \subset \mathbb{O}$.
    \item \textbf{Solvable Lie Algebras and Affine Nil Algebras:} For a solvable Lie algebra $\mathfrak{g}$ of derived length $s$ ($\mathfrak{g}^{(s+1)} = 0$), Proposition \ref{prop:abelianization} shows that $D [\wedge]_{\mathfrak{g}} D \in \Omega^2(P; [\mathfrak{g}, \mathfrak{g}])$ shifts up the derived filtration and its iterated squares $(D [\wedge]_{\mathfrak{g}} D)^{[m]} \in \Omega^{2^{m+1}}(P; \mathfrak{g}^{(m+1)})$ vanish at step $m = s$, while confining the quartic interaction sector to $E_{[\mathfrak{g}, \mathfrak{g}]} \subsetneq E_{\mathfrak{g}}$. Finally, for any affine nil algebra $A$ of nilpotency degree $s$ \cite{Smoktunowicz}, $A$ is $(1,s)$-nil and $(2,s)$-nil, so Theorem \ref{thm:topological_obstruction}(1) implies that all internal topological Chern-Weil $A$-YMT functionals $S^A_{\mathrm{top}, r, \mathcal{P}}[D]$ vanish in dimensions $n \ge 2(s+1)$.
\end{enumerate}

\end{document}